\documentclass[a4paper,11pt]{article}

\pdfoutput=1

\usepackage[usenames,dvipsnames]{color}
\usepackage{jcappub}
\usepackage{epsfig}
\usepackage{amssymb}
\usepackage{amsmath}
\usepackage{graphicx}
\usepackage{multirow}
\usepackage{verbatim}
\usepackage{xspace}
\usepackage{placeins}
\usepackage{tabularx}
\usepackage{afterpage}
\usepackage{float}

\definecolor{darkblue}{rgb}{0,0,0.5}
\definecolor{darkgreen}{rgb}{0.1,0,0.3}
\definecolor{darkred}{rgb}{0.6,0,0}

\defcitealias{GS98}{GS98}
\defcitealias{AGSS}{AGSS09}
\defcitealias{AGS05}{AGS05}

\usepackage{xr}
\usepackage{amsmath}
\usepackage{graphicx}
\usepackage{amssymb}
\usepackage{parskip}
\usepackage[utf8]{inputenc}
\usepackage{tikz}
\usepackage{enumitem}
\newcommand{\ts}{\textsuperscript}
\usepackage{xcolor}
\newcommand*\circled[1]{\tikz[baseline=(char.base)]{%
            \node[shape=circle,fill=white,draw,inner sep=1pt] (char) {#1};}}
\usepackage{siunitx}
\usepackage{caption}
\DeclareMathOperator\erf{erf}
\renewcommand{\vec}[1]{\mathbf{#1}}

\begin{document}

%%%%%%%%%%%%%%%%%%%%%%%%%%%%%%%%%%%%%%%%%%%%%%%%%%%%%%%%%%%%%%%%%%%%%%%
\title{Simulation of energy transport by dark matter scattering in stars}

\author[1]{Hannah Banks,}
 \emailAdd{hmb61@cam.ac.uk}
 \affiliation[1]{Department of Applied Mathematics and Theoretical Physics, University of Cambridge, Wilberforce Road, Cambridge, CB3 0WA, UK }

\author[2]{Siyam Ansari,}
 \emailAdd{siyam-ans@hotmail.co.uk}
 \affiliation[2]{Department of Physics, Imperial College London, Blackett Laboratory, Prince Consort Road, London SW7 2AZ, UK}

\author[3,4,5]{Aaron C. Vincent,}
 \emailAdd{aaron.vincent@queensu.ca}
  \affiliation[3]{Department of Physics, Engineering Physics and Astronomy, Queen's University, Kingston ON K7L 3N6, Canada}
 \affiliation[4]{Arthur B. McDonald Canadian Astroparticle Physics Research Institute, Kingston ON K7L 3N6, Canada}
 \affiliation[5]{Perimeter Institute for Theoretical Physics, Waterloo ON N2L 2Y5, Canada}

\author[2,6]{and Pat Scott}
\affiliation[6]{School of Mathematics and Physics, The University of Queensland, St.\ Lucia, Brisbane, QLD 4072, Australia}
 \emailAdd{pat.scott@uq.edu.au}

\abstract{Asymmetric dark matter (ADM) that is captured in stars can act as an efficient conductor of heat. Small ADM-induced changes in a star's temperature gradient are known to alter neutrino fluxes and asteroseismological signatures, erase convective cores and modify a star's main sequence lifetime. The Sun's proximity to us makes it an ideal laboratory for studying these effects. However, the two formalisms commonly used to parametrize such heat transport were developed over 30 years ago, and calibrated with a single set of simulations. What's more, both are based on assumptions that break down at the Knudsen transition, where heat transport is maximized. We construct a Monte Carlo simulation to exactly solve the Boltzmann collision equation, determining the steady-state distribution and luminosity carried in stars by ADM with cross sections that depend on velocity and momentum. We find that, although the established (Gould \& Raffelt) formalism based on local thermal equilibrium does well for constant cross sections, the isothermal (Spergel \& Press) method actually performs better across all models with a simple, universal rescaling function. Based on simulation results, we provide recommendations on the parametrization of DM heat transport in stellar evolution models. }

\maketitle
%%%%%%%%%%%%%%%%%%%%%%%%%%%%%%%%%%%%%%%%%%%%%%%%%%%%%%%%%%%%%%%%%%%%%%%

\section{Introduction}
\label{sec:intro}
Despite the abundance of gravitational evidence supporting the presence of dark matter (DM) in the Universe, its fundamental nature and properties remain unknown \cite{Bertone2005ParticleConstraints, BertoneBook, Lisanti2016LecturesPhysics}.  If DM is able to interact with particles of the Standard Model (SM), for example via a weak-scale cross-section for elastic scattering with quarks, it could lose enough kinetic energy in collisions with nuclei to be gravitationally captured by astrophysical bodies such as the Sun and other stars \cite{Press1985CaptureParticles, Gould1987, Gould1987a}.

As long as the accumulated population of DM is not depleted by self-annihilation, evaporation or decay, trapped DM offers an alternative channel for macroscopic energy transport in the interiors of stars \cite{Steigman78,Spergel1985EffectInterior,Faulkner1985,Gould1990}. Such Asymmetric (A)DM particles can absorb energy in elastic interactions with nuclei near the core, and release it via collisions in the cooler outer layers. If a sufficient population of DM is able to accumulate in the star, the resulting heat redistribution can produce measurable modifications to the stellar structure and neutrino fluxes \cite{Lopes2001,Lopes2002SolarMatter,Bottino2002DoesParticles,Lopes2012,Lopes2014,Pantsy,Geytenbeek2016EffectInterior,Frandsen2010,Cumberbatch2010LightHelioseismology,Taoso2010EffectSun,Vincent14,Vincent2015,Vincent2016UpdatedMatter}.

\begin{figure}[t]
    \centering
    \hspace*{-1cm}\includegraphics[width=1.1\textwidth]{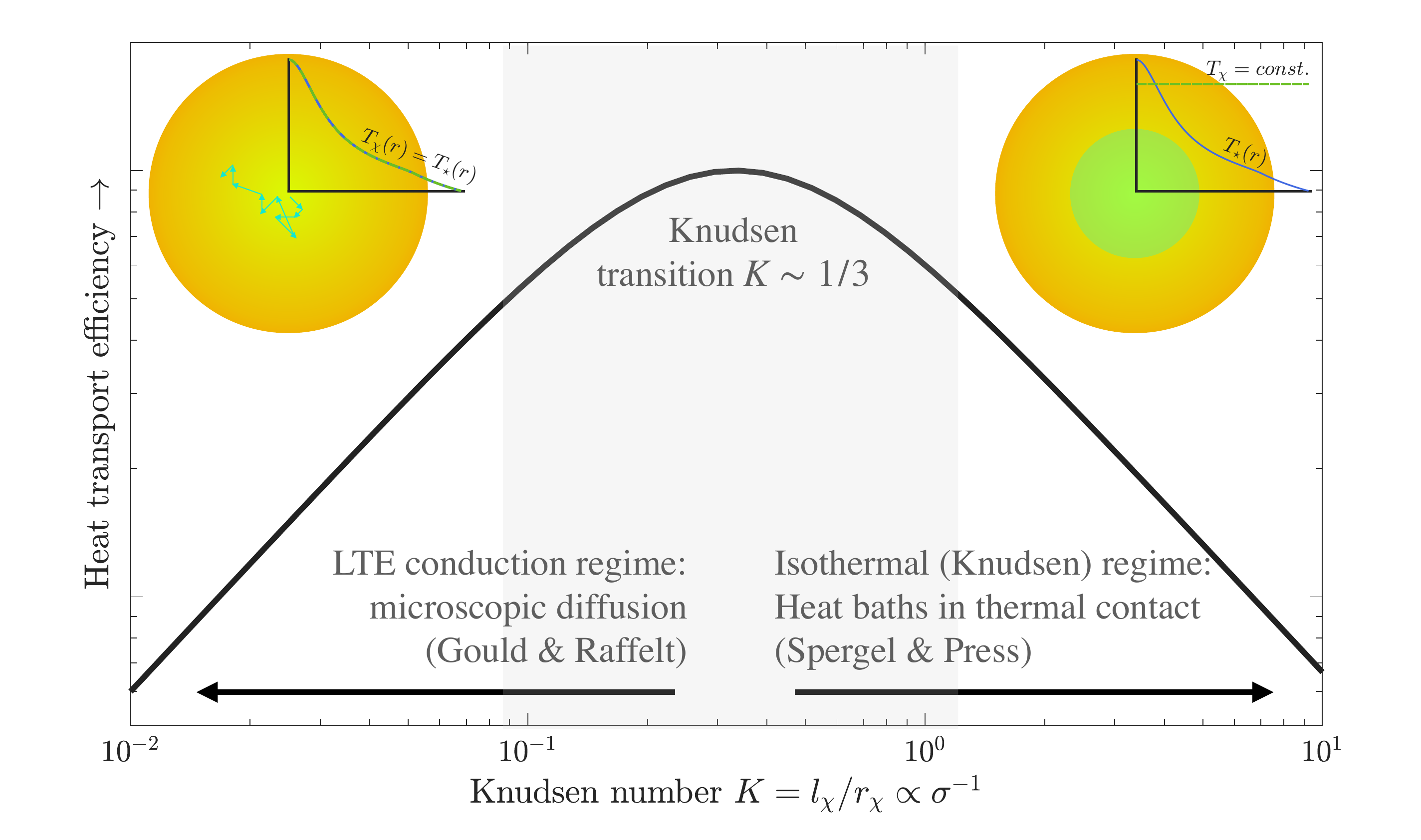}
    \caption{Schematic diagram showing the efficiency of heat transport in stars by a weakly interacting particle, as a function of the Knudsen number $K$. The two thermal conduction schemes commonly used to model heat transport are illustrated on either side. These rely on asymptotic behaviour with $K$ to obtain analytic solutions to the Boltzmann Collision Equation \eqref{eq:BCE}. The left of the figure shows the Local Thermal Equilibrium (LTE) regime, modelled by Gould \& Raffelt (GR, \cite{Gould1990}). The DM and plasma have the same temperature profile, and conduction takes place via a small dipole correction to the DM's Maxwell-Boltzmann distribution. Conduction is limited by the mean free path, and scales as $l_\chi \propto K$. The right side of the figure shows the Knudsen regime where the mean free path is large, modelled by Spergel \& Press (SP, \cite{Spergel1985EffectInterior}) using an isothermal DM distribution in thermal contact with the plasma. Conduction is limited by the interaction strength, and scales with the cross section $\sigma \propto K^{-1}$. The \textit{Knudsen transition} region in the centre of the figure can only be modelled via empirical calibration with simulations like those performed in this work.
    }
    \label{fig:KnudsenFigure}
\end{figure}

Heat transport by DM in stars is described by a Boltzmann Collision Equation (BCE), which depends on the microscopic details of the DM-nucleon interactions, the local gravitational potential $\phi(r)$ and the temperature $T(r)$, density and composition of the stellar plasma. The thermal behaviour of a captured DM population is commonly described by the Knudsen number
\begin{equation}
K = l_\chi/r_\chi,
\label{eq:K}
\end{equation}
 i.e.\ the ratio of the mean inter-scattering distance $l_\chi$ and the DM ``scale height'' within the star $r_\chi$, which is set by $T(r)$ and $\phi(r)$.
We provide a schematic illustration of the transport efficiency in terms of this parameter in Fig.\ \ref{fig:KnudsenFigure}. Approximate analytic solutions to the BCE have been obtained in the two limiting cases. In the \textit{Knudsen regime} of large mean free paths ($K \gg 1$), Spergel \& Press (SP) \cite{Spergel1985EffectInterior} modeled the DM as an isothermal heat bath, with heat transfer occurring due to the difference between the DM temperature and that of the local plasma temperature $T(r)$. In this regime, transport is limited by the weak interaction cross section, and scales as $K^{-1}$. In the opposite limit ($K \ll 1$), Gould \& Raffelt (GR) \cite{Gould1990} developed a Local Thermal Equilibrium (LTE)-based perturbative solution, modeling transport as caused by a dipolar perturbation to a local ($T_\chi = T(r)$) Maxwellian phase space distribution, with the perturbation linear in $l_\chi |\ln \nabla T|$. Here, transport is limited by the short inter-scattering distance, and scales as $K$. Both formalisms allow for general velocity or momentum-dependent interactions, but only constant cross sections were studied at the time.

These two regimes intersect at the \textit{Knudsen transition} at $K \simeq 1/3$, which has remained analytically intractable. In the Sun, this happens for DM-nucleon cross sections $\sigma_{\chi N}$ of about  $10^{-36}$ cm$^2$. Gould \& Raffelt \cite{Gould1990,Gould1990CosmionLimit} developed a Monte Carlo simulation designed to test these formalisms, using a simple setup in which the Sun is modelled as sphere of constant density with a linear temperature gradient. In the Knudsen regime, they found that the SP method overestimated transport by a factor of approximately two. In contrast, the GR solution performed well in the LTE limit, but required an additional ad-hoc ``radial suppression'' to deal with the breakdown of the dipole approximation as $r \rightarrow 0$.  Ref. \cite{Gould1990} also found that predicted luminosities could match simulation results across the Knudsen transition, given a constant rescaling depending only on $K$. Following this work, the GR solution, modified with radial and Knudsen corrections (as formalized in e.g.\ \cite{Scott2009}), became the accepted method for correctly modelling heat transport, though SP has been extensively used because of its relative simplicity.

Stellar simulations including heat transport by ADM, modelled using both GR and SP approaches, show that it may affect Hertzsprung-Russel (HR) diagram trajectories and extend the main sequence lifetimes of stars \cite{Iocco:2012wk,Lopes:2019jca,Raen:2020qvn}, affect asteroseismology \cite{Rato:2021tfc}, the luminosity of giant stars \cite{Lopes:2021jcy} and the shape of the neutrino spectrum from the Sun \cite{Lopes:2018wgp}. DM-mediated energy transport may also provide a means to alleviate the current tension between precision measurements of solar observables and their predictions from theoretical Standard Solar Models (SSMs) \cite{Frandsen2010,Cumberbatch2010LightHelioseismology,Taoso2010EffectSun,Vincent14,Vincent2015,Vincent2016UpdatedMatter}. Known as the solar abundance problem \cite{Asplund2009TheSun,Serenelli2009NewRevisited,Bergemann2014}, this constitutes a more than  $6\sigma$ mismatch between theory and observation. Refs. \cite{Vincent13,Pantsy} have further noted that nonstandard DM-nucleus interactions could yield measurable helioseismological effects whilst remaining beyond the reach of terrestrial direct detection experiments. As such experiments are sensitive only to the high-velocity tail of the DM halo's phase space distribution, they may be blind to important phenomenological effects at low velocity or momentum transfer.

Ref. \cite{Vincent13} extended the GR formalism to include velocity and momentum-dependent cross sections. Using these results, Refs. \cite{Vincent14,Vincent2015,Vincent2016UpdatedMatter} showed that some non-constant DM-nucleon cross sections could yield   changes in helioseismological observables that lead to a strong improvement over the SSM in terms of agreement with observations.

Although promising, the results of \cite{Vincent13} have never been verified by direct simulations. Given the empirical corrections required to model heat transport for constant cross sections, it remains a possibility that the approach does not generalize well to non-constant cross sections---and indeed, we will find this to be the case.

This leads to a more important concern regarding how heat transport is currently modeled. The GR approach is built on the assumption of local thermal equilibrium, and indeed locality is required at every step in deriving their solutions. However, cross sections that remain allowed by direct detection experiments (see e.g.\ \cite{PICO:2019vsc}) are well into the Knudsen regime. To-date, it has remained unclear whether the GR radial and Knudsen corrections work well deep into the Knudsen regime, especially in more realistic stellar environments than the initial toy model considered in Ref.~\cite{Gould1990CosmionLimit}.

Given that more than three decades have passed since the simulations of Gould \& Raffelt, more modern and extensive simulations are long overdue. Our aim in this work is thus to carry out a first-principles Monte Carlo simulation of energy transport in a star by DM with both constant and non-trivial couplings to the SM, i.e.\ depending on momentum transfer $q$ or relative velocity $v_\mathrm{rel}$. We perform simulations in both the idealized setup examined by Gould \& Raffelt, and in a more realistic stellar environment using a real Solar model, directly testing the predictions of the SP and GR approaches against simulation results. We restrict ourselves to scattering with a single nuclear species, equivalent to spin-dependent interactions with hydrogen. Following these extensive simulations, we find that heat transport turns out to be most robustly described by a modified version of the isothermal SP transport scheme.

We begin in Section \ref{sec:bg} by reviewing the physics of DM in the Sun, and the details of the SP and GR formalisms for describing its impact on solar energy transport.  We give the details of our Monte Carlo simulations in Section \ref{sec:methods} and present our results in Section \ref{sec:results}. Section \ref{sec:discussion} provides an updated, Monte Carlo-calibrated formalism that appears to accurately describe heat transport across all regimes, for constant and non-constant interaction cross sections, and across very different stellar structures. We summarise and conclude in Section \ref{sec:conclusions}.

%######################################################
\section{Background}
\label{sec:bg}
DM heat transport can affect stellar structure, observable via asteroseismology, stellar trajectories on the HR diagram, as well as main sequence stellar lifetimes. The closest, and easiest to study star is the Sun: the temperature, density and composition of the solar core can be probed using neutrino flux measurements and helioseismology.  The flux of ${}^{8}$B neutrinos provides a particularly stringent constraint on the core temperature $T_\mathrm{c}$, as it is proportional to $T_\mathrm{c}^{\beta}$ with $\beta \sim$ 20--25.  Helioseismology refers to the study of pressure wave propagation in the Sun. By inverting the frequencies of oscillations observed at the solar surface, precision determinations can be made of the sound speed profile, surface helium abundance and boundary of the convective zone \cite{Kosovichev2011}.

Early SSMs agreed with the solar observables, but this is no longer the case.  Upgrades from 1D hydrostatic to 3D hydrodynamic atmospheric models combined with improvements to the modelling of non-LTE line formation led to significant reductions in the solar metal content and thus the $ Z/X $ SSM calibration point \cite{CtoO,AspIV,AspVI,AspARAA,AGS05,ScottVII,Scott09Ni,Asplund2009TheSun,AGSS_NaCa,AGSS_FePeak,AGSS_heavy}. Whilst the consistency with observational neutrino fluxes was maintained, this revision brought the SSM into sharp contradiction with helioseismological observables \cite{Serenelli2009NewRevisited}. With the potential to reduce the core temperature and density of the Sun, conduction by gravitationally trapped DM has been suggested as a possible solution to this problem \cite{Frandsen2010,Cumberbatch2010LightHelioseismology,Taoso2010EffectSun,Vincent14,Vincent2015,Vincent2016UpdatedMatter}.

The potential impact of DM on stellar interiors can be predicted by incorporating DM capture and transport into evolutionary stellar models. To do this, quantitative descriptions of both of these processes are required.

For DM to become trapped in a star, collisions with stellar nuclei must result in sufficient momentum transfer to bring the DM velocity below the local escape velocity $v_\mathrm{esc}$. We will not reproduce the details of DM capture in stars, referring the interested reader to Ref.\ \cite{Vincent2015} for a treatment of capture of generalised velocity and momentum-dependent DM, and Ref.\ \cite{Catena2015FormTheories} for the corresponding treatment in terms of non-relativistic DM-nucleon effective operators. When interaction rates are high, capture via multiple interactions must be taken into account, either via a finite optical depth \cite{Gould1992,Busoni2017EvaporationSun} or by including multiple-scatter events \cite{Bramante:2017xlb,Ilie:2020vec}.

In order for DM to accumulate to levels at which the effects of additional energy transport become observable, the capture rate must dominate over annihilation and evaporation. For annihilation to be suppressed, the captured dark matter must posess some degree of asymmetry, either in its abundance (as in ADM), or in its interactions \cite{Clementz2015AsymmetricSun}. The evaporation rate depends on the DM mass, stellar mass, interaction type and cross section \cite{Busoni2017EvaporationSun,Garani:2021feo}. In practice, as long as the DM mass is higher than the evaporation threshold ($m_\chi \gtrsim 4$\,GeV in the Sun), evaporation can be neglected altogether. In this paper, we are only concerned with understanding the process of conductive energy transport, so we do not consider the effects of annihilation or evaporation, though the latter will be included in our simulations for self-consistency.

The capture rate becomes saturated if the total scattering cross-section reaches the cross-sectional area of the Sun (times a gravitational focussing factor), such that all incident DM is captured. This so-called geometric limit translates into a maximum possible solar DM population $N_{\chi}$, and hence a maximal possible impact of DM on the solar structure \cite{Vincent2015}. Based on the local density of DM in the Milky Way, the maximum number of DM particles that could have accumulated in the Sun to date is less than about 1 per $10^{10}$ baryons.

%############################################################################
\subsection{Energy transport by dark matter}
\label{sec:ET}

The impacts of a given DM population on a star can be fully described by the radial profile of its number density $n_{\chi} (r)$, and the effective luminosity it generates by scattering, annihilation and/or decay $L_{\chi} (r)$. Without annihilation or decay, these can be directly obtained from the DM phase space distribution, $F(\vec{v},\vec{r},t)$ \cite{Gould1990}, where $\vec{v}$ is the DM velocity.
The evolution of $F$ for a diffuse, weakly interacting species in the dense solar interior is formally governed by the Boltzmann collision equation (BCE) \cite{Gould1990,Vincent13}
\begin{equation}
\label{eq:BCE}
DF = l^{-1}_{\chi}CF,
\end{equation}
where $l_{\chi}$ is the typical inter-scattering length, $D$ is a differential (Liouville) operator representing the impact of diffusion and gravity and $C$ is the collision operator, which encodes microscopic scattering events. The length $l_{\chi}$ is formally defined as the inverse of the mean number of interactions per unit length.  It depends on the interaction cross-section $\sigma_{i}$ and number density $n_{i}$ of each distinct nuclear species with which the DM interacts, as
\begin{equation}
\label{eq:mfp}
l_{\chi}(\vec{v},\textbf{r}) =  \left[\sum_{i} n_{i}(r)\left\langle \sigma_{i}(\vec{v})\right\rangle\right]^{-1},
\end{equation}
where the sum is over all species with which the DM interacts, and the angular brackets denote the thermal average (i.e.\ over the nuclear velocity distribution).

$C$ depends on both the kinematic form of the DM-nucleus interaction and the distribution of target nuclei, and needs to be explicitly constructed for a given interaction law \cite{Vincent13}. As the population of DM is diffuse, DM-DM scattering events can be neglected relative to scattering from solar nuclei, and $C$ is linear \cite{Nauenberg1987,Gould1990}. It is typically assumed that the collisional time-scale is significantly shorter than that on which the solar structure changes, such that $C$ is time-independent and $F = F(\vec{v},\vec{r})$  is stationary \cite{Gould1990}.

DM-mediated energy transport can be broadly divided into two regimes, depending on the strength with which DM couples to the solar nuclei. These can be conveniently distinguished via the Knudsen number (Eq.\ \ref{eq:K}). Note in particular that $l_{\chi}$ depends on density and temperature, and is thus a function of radius $r$. We will follow the convention of defining $K$ at the centre of the star: $K = {l_{\chi}(r = 0)}/{r_{\chi}}$. The scale radius $r_\chi $ is conventionally set to the radius at which the distribution of DM confined to an isothermal, constant-density sphere with the same temparture $T_\mathrm{c}$ and density $\rho_\mathrm{c}$ as the solar core would peak. It is formally given by
\begin{equation}
\label{eq:scale}
 r_\chi=\sqrt{\frac{3k_\mathrm{B}T_\mathrm{c}}{2\pi G \rho_\mathrm{c} m_{\chi}}},
\end{equation}
where $m_{\chi}$ is the DM mass and $k_\mathrm{B}$ and $G$ are the Boltzmann and gravitational constants. As outlined in Sec.\ \ref{sec:intro} and Fig.\ \ref{fig:KnudsenFigure}, the two extremal regimes are thus:
\begin{enumerate}
    \item The \textit{Knudsen}, \textit{non-local} or \textit{isothermal} limit, where $K \gg 1$. The enhancement provided by longer inter-scattering distances is countered by a reduction in collision efficiency. Heat conduction is proportional to the cross section $\sigma \sim l_\chi^{-1}$.
    \item The \textit{conduction} or \textit{Local Thermal Equilibrium (LTE)} limit, where $K \ll 1$.  The comparatively strong DM-nucleus couplings result in a high collisional frequency, but suppress inter-scattering distances, meaning heat cannot be transported far. Conduction is proportional to the mean free path $l_\chi \sim \sigma^{-1}$.

\end{enumerate}
Optimal transport occurs at the so-called \textit{Knudsen transition} between these regimes, where transport is both non-local and efficient.  This occurs in the Sun for constant cross sections of $\sigma_{SD} \sim 10^{-36}$ cm$^2$, and $\sigma_{SI} \sim 10^{-37}$ cm$^2$, in the respective cases of spin-dependent and spin-independent interactions.

In the remainder of this section we provide the details of the existing formalisms used to describe these two regimes, and summarise existing numerical results aimed at linking them.

\subsubsection{Spergel \& Press: isothermal conduction}
\label{sec:SP}
In the non-local regime $K\gg $1, the DM population effectively has a single temperature $T_{\chi}$, which represents the average stellar temperature sampled by the DM in its interactions in the star, weighted by the radius-dependent collision rates. This situation was first modeled by Spergel and Press \cite{Spergel1985EffectInterior}, and relies on three assumptions: 1) the phase space distribution is Maxwellian,  2) the velocity distribution has a single constant temperature $T_\chi$ everywhere, and 3) the distribution is locally isotropic. In other words, $F(\vec v,\vec r,t)$ solves the collision\textit{less} BCE, $DF = 0$:
\begin{equation}
\label{eq:isothermaldist}
n(r)_{\chi,\mathrm{iso}} = N_\mathrm{iso}e^{-m_{\chi}\phi(r)/T_{\chi}},
\end{equation}
where $N_\mathrm{iso}$ is a normalisation constant, set by demanding that the integrated density matches the total DM population $N_\chi$.

The energy transferred per unit time per unit stellar mass is \cite{Spergel1985EffectInterior}
\begin{equation}
    \epsilon = \frac{1}{\rho(r)}\int d^3v f_\chi(v) \int d^3 u f_N(u) \sigma_\mathrm{tot} \langle \Delta E(u,v,\vartheta)\rangle |\vec v - \vec u|,
    \label{eq:SPint}
\end{equation}
where $\vec v$ and $\vec u$ are respectively the DM and nuclear velocities, and the average energy transferred per collision is
\begin{equation}
   \langle \Delta E \rangle = \frac{m_\chi m_n}{(m_\chi + m_N)^2} (1-Q) \left[ m_N v^2 - m_\chi u^2 + (m_\chi - m_N) uv \cos \vartheta\right].
\end{equation}
Here the angle $\vartheta$ between the incoming DM and nuclear velocity vectors is given by $\vec u \cdot \vec v \equiv uv \cos \vartheta$ and
\begin{equation}
    Q = \frac{\int \frac{d\sigma}{d\cos\theta_{\mathrm{cm}}}\cos\theta_{\mathrm{cm}} d \cos\theta_{\mathrm{cm}}}{\int \frac{d\sigma}{d\cos\theta_{\mathrm{cm}}} d \cos\theta_{\mathrm{cm}}},
\end{equation}
where $\theta_{\mathrm{cm}}$ is the scattering angle in the centre of momentum frame. Spergel \& Press \cite{Spergel1985EffectInterior} first derived the form of \eqref{eq:SPint}\footnote{ Spergel \& Press originally expressed this as the energy injected per unit volume, which is equal the definition that we use (Eq.\ \ref{eq:SP}) times $\rho(r)$.} for a constant DM-nucleus cross section:
\begin{equation}
    \epsilon_\mathrm{SP} = \frac{8}{\rho(r)} \sqrt{\frac{2}{\pi }} \frac{m_\chi m_N}{(m_\chi + m_N)^2} n_\chi(r) n_N(r) \sigma_\mathrm{tot}k_B\left(T_\chi-T(r)\right) \left(\frac{k_BT(r)}{m_N}+\frac{k_BT_\chi}{m_\chi}\right)^{\frac{1}{2}}.
  \label{eq:SP}
\end{equation}
where $\sigma_\mathrm{tot} = \int (d\sigma/d\cos\theta) d\cos\theta = 2\sigma$.
The DM temperature $T_\chi$ can be found iteratively through the luminosity condition:
\begin{equation}
    L_\mathrm{tot} = 4\pi \int_0^{R_\odot} \rho(r) r^2 \epsilon_\mathrm{SP}(r,T_\chi) dr = 0,
    \label{eq:lumcondition}
\end{equation}
\textit{i.e}. that the DM is neither a net source nor a net sink of energy.

We may generalize this formalism to account for velocity and momentum-dependent cross sections. Including a velocity-dependent cross section amounts to including additional factors of $|\vec v - \vec u|$ in the integrand. Eq.\ \eqref{eq:SPint} can then be evaluated as described in Ref.\ \cite{Spergel1985EffectInterior}, by first integrating over all angles except $\vartheta$, and then splitting the final velocity integrals into the $u - v > 0$ and $u - v < 0$ cases, with the change of variables $z dz = -uvd\cos\vartheta$ (taking care to avoid typos in the integration limits of Eq. A.5 of \cite{Spergel1985EffectInterior}).

For cross sections proportional to $v^{2n}$ or $q^{2n}$, this integral yields
\begin{equation}
    \epsilon = \frac{A_{2n}}{\rho(r)} \sqrt{\frac{2}{\pi }} \frac{m_\chi m_N}{(m_\chi + m_N)^2} n_\chi(r) n_N(r) (1-Q) \sigma_\mathrm{tot}k_B\left(T_\chi-T\right) \left(\frac{k_BT}{m_N}+\frac{k_BT_\chi}{m_\chi}\right)^{\frac{1}{2}+n}
    \label{eq:SPgeneral}
\end{equation}
Where
\begin{equation}
A_{-2} = 2, \qquad A_{0} = 8, \qquad A_{2} = 48, \qquad A_{4} = 384.
\end{equation}
For $q$-independent cross sections, $(1-Q)\sigma_\mathrm{tot} = 2\sigma_0/v_0^{2n}$. For $q$-dependent cross sections, this becomes
\begin{equation}
    (1-Q)\sigma_\mathrm{tot} = B_{2n}\frac{2^n m_\chi^{2n}\sigma_0}{(1+\mu)^{2n} q_0^{2n}},
    \label{eq:SPQintegral}
\end{equation}
where $\mu = m_\chi/m_N$, and $B_2 = 8/3$, $B_4 = 4$, while $B_{-2} = 2$, the value for $q$-independent scattering, as we are using the momentum-transfer cross section that eliminates the forward-scattering divergence for $q^{-2}$.

Gould \& Raffelt \cite{Gould1990CosmionLimit} have pointed out that all three assumptions in the SP approach (that the DM velocity distribution is uniform, isotropic and Maxwellian) are violated in the case of DM transport in a star, leading to a formally incorrect solution. In the Knudsen limit, they found that the DM temperature was never perfectly uniform, instead lying somewhere between the DM temperature $T_\chi$ found with Eqs. (\ref{eq:isothermaldist}--\ref{eq:lumcondition}) and the local stellar temperature $T(r)$. Earlier simulations by Nauenberg \cite{Nauenberg1987} concluded that SP overestimated DM luminosity by a factor of $\sim 3$; Gould \& Raffelt refined this factor to $\sim$ 2.

\subsubsection{Gould \& Raffelt: LTE conduction}

Refs \cite{gilliland1986solar,Nauenberg1987} first derived expressions for heat transport in the LTE  $K \ll 1$ limit, based on conduction theory. These were found to be inconsistent, and only valid in the small DM mass limit $m_\chi/m_N \ll 1$ \cite{Gould1990}. An essentially analytical treatment of the LTE limit was established by Gould and Raffelt \cite{Gould1990}, who solved the BCE by means of a first order perturbation expansion of $F$ in the small parameter $\varepsilon \equiv l_\chi |\nabla \log  T|$. Assuming the heat transport was purely due to a dipole correction to the locally Maxwellian distribution, they showed that the radial dependence of the DM number density and luminosity can be expressed in terms of dimensionless diffusion ($\alpha$) and  conduction ($\kappa$) coefficients, as
\begin{equation}
\label{eq:conductiondist}
n_{\chi,\mathrm{LTE}}(r) = n_{\chi}(0)\left[\frac{T(r)}{T(0)}\right]^{3/2}\exp \left[-\int_{0}^{r}dr'\frac{k_\mathrm{B}\alpha(r')\frac{dT(r')}{dr'}+m_{\chi}\frac{d\phi(r')}{dr'}}{k_\mathrm{B}T(r')}\right]
\end{equation} and
\begin{equation}
\label{eq:LLTE}
L_{\chi,\mathrm{LTE}}(r) = 4\pi r^{2}\kappa(r)n_{\chi}(r)l_{\chi}(r)\left[\frac{k_\mathrm{B}T(r)}{m_{\chi}}\right]^{1/2}k_\mathrm{B}\frac{dT(r)}{dr},
\end{equation}
resulting in a transported energy per unit stellar mass of
\begin{equation}
\epsilon_{\rm LTE} =  \frac{1}{4\pi r^2 \rho(r)} \frac{dL_\chi}{dr}.
\label{eq:ELTE}
\end{equation}

 In this scenario the DM shares the temperature of the local background nuclei, as it is in local thermal equilibrium with them.  The quantities $\alpha$ and $\kappa$ are formally calculated via inversion of the collision operator, and are thus particular to the kinematic form of the DM-nucleus interaction. For interactions with nuclei of a given mass, these quantities are a function of the DM-nucleus mass ratio $\mu = m_{\chi}/m_{N}$ only. For scattering with multiple elements, effective values of $\alpha$ and $\kappa$ must be calculated by taking a weighted average of the contributions of each species, according to their local abundances. This causes $\alpha$ and $\kappa$ to become functions of radius.  Values for $\alpha$ and $\kappa$ for various $\mu$ were computed by Gould \& Raffelt for the case of constant DM-nucleon cross-sections \cite{Gould1990}. This treatment was extended to DM-nucleon couplings with non-trivial kinematic dependences in Ref.~\cite{Vincent13}, where the values of $\alpha$ and $\kappa$ for cross-sections scaling with various powers of the relative velocity or momentum transfer were computed.

\subsubsection{Previous Monte Carlo results and the Knudsen transition}

Since a formal solution to the full BCE remains elusive, Monte Carlo simulation is the only way to directly verify the approximation schemes described above, and to access the Knudsen transition regime in which both approximations fail. As the BCE is linear in the dilute limit, time-averaging the properties of a single particle is equivalent to computing the instantaneous ensemble averages of the same properties for a population of DM particles \cite{Nauenberg1987,Gould1990}. In this way the desired number density and luminosity profiles can be computed entirely independently of the BCE.

This approach was first proposed and implemented by Nauenberg \cite{Nauenberg1987}, and was later employed by Gould and Raffelt \cite{Gould1990CosmionLimit,Gould1990} to investigate DM-mediated energy transport in a simple solar model for the case of constant DM-nucleus couplings. Gould and Raffelt found that the luminosity for values of $K$ covering the full range of the Knudsen transition is well-described by the LTE solution, \eqref{eq:LLTE}, with two important modifications:
\begin{enumerate}
    \item A Knudsen suppression function $\mathfrak{f}(K)$ must be used to to suppress the total luminosity found in the $K=0$ LTE limit.  The same function can also be used to interpolate between the LTE and isothermal density distributions $n_{\chi}(r)$ \cite{Scott2009}.
    \item A radial suppression factor $\mathfrak{h}(r)$ must be included, as the LTE solution overestimates the conduction rate near the core. This is expressed succinctly \cite{Gould1990}: ``Since the energy transported by the WIMPs is generated typically at a distance $l_\mathrm{eff}$ ``upstream'', all the luminosity at $r (r < l_\mathrm{eff})$ is generated within a sphere of volume $(4\pi/3)r^3$, as opposed to a shell with volume $4\pi r^2 l_\mathrm{eff}$ for $r \gg l_\mathrm{eff}$.''
\end{enumerate}
These suppression/interpolations factors are well fit by \cite{Scott2009,Bottino2002DoesParticles}
\begin{equation}
\label{eq:fgothhgoth}
\mathfrak{f}(K) = 1- \left[1+\left(\frac{K}{K_{0}}\right)^{1/\tau}\right]^{-1} \quad\text{and}\quad \mathfrak{h}(r,t) = 1 + \left[\frac{r-r_{\chi}(t)}{r_{\chi}(t)}\right]^{3},
\end{equation}
with $\tau \simeq 0.4$ and $K_{0}\simeq 0.5$. Energy transport at any Knudsen number can then be expressed in terms of the LTE and isothermal behaviour as
\begin{equation}
\label{eq:8}
n_{\chi} = \mathfrak{f}(K)n_{\chi,\mathrm{LTE}}+[1-\mathfrak{f}(K)]n_{\chi,\mathrm{iso}}
\end{equation}
and
\begin{equation}
\label{eq:LGR}
L_{\chi,\mathrm{total}}(r,t) = \mathfrak{f}(K)\mathfrak{h}(r,t)L_{\chi,\mathrm{LTE}}(r,t).
\end{equation}

Given the absence of any analytical framework with which to determine the density and energy transport at the Knudsen transition, these relations represent the only quantitative handle on this regime. Despite having been widely used in the literature, these results, based on a handful of simulations, have remained virtually uninterrogated for over three decades. This is especially important now, as direct detection limits push allowed DM-nucleon cross sections well into the Knudsen regime. Given that the widely-accepted scheme (GR with a Knudsen correction) fundamentally rests on a short mean-free-path approximation that is explicitly broken, the time is ripe to re-examine these results.  We thus come to the main goal of this work: to explore in more detail the results of the earlier Monte Carlo simulations, and to update them to include the effects of non-constant interaction cross sections. No such study has ever been completed for momentum or velocity-dependent couplings.

%#########################################################################

\section{Methodology}
\label{sec:methods}

\subsection{Dark matter-nucleus interactions}
\label{sec:DMmodels}
In this work we consider DM-nucleon couplings that are either constant, or depend on some power of the momentum $q$ exchanged during interactions, or the DM-nucleon relative velocity $v_\mathrm{rel}$. Though our results should generalise straightforwardly to spin-independent interactions, we will focus for simplicity on spin-dependent couplings---which in this context means scattering with hydrogen only.

We parameterise the DM-nucleon differential cross-section in the centre-of-mass frame relative to a reference constant cross-section $\sigma_{0}$ as
\begin{equation}
\label{eq:cs1}
\frac{d\sigma}{d\cos \theta_\mathrm{cm}}(v_\mathrm{rel}) = \sigma_{0} \Big(\frac{v_\mathrm{rel}}{v_{0}}\Big)^{2n},
\end{equation}
and
\begin{equation}
\label{eq:cs2}
\frac{d\sigma}{d\cos\theta_\mathrm{cm}}(q) = \sigma_{0} \Big(\frac{q}{q_{0}}\Big)^{2n}.
\end{equation}
Note that whilst $v_\mathrm{rel}$ is an initial-state quantity and therefore has nothing to do with the centre-of-mass scattering angle $\theta_\mathrm{cm}$, the momentum transfer $q$ is a function of both $v_\mathrm{rel}$ and $\theta_\mathrm{cm}$,
\begin{equation}
\label{eq:q}
q^2 = (1 - \cos\theta_\mathrm{cm})v_\mathrm{rel}^{2}\frac{2m_{\chi}^2}{(1+\mu)^2}.
\end{equation}
Eq.\ \eqref{eq:cs1} is therefore isotropic, but Eq.\ \eqref{eq:cs2} retains an implicit angular dependence.  The values of $v_0$ and $q_0$ are arbitrary, and set the relative velocity and momentum transfer at which the reference cross-section is defined. For our realistic simulations, we will choose $v_0= 220$\,\si{km.s^{-1}}, which is the typical velocity of DM in the Milky Way halo, and $q_0= 40$\,\si{\mega \electronvolt}, a typical recoil energy in terrestrial direct detection experiments. Apart from the standard case of a constant cross-section ($n = 0$), we consider $n \in \left\{-1, 1, 2\right\}$, all of which arise as the leading-order DM-nucleon coupling in concrete theoretical models.

When considering DM in the Sun, it is the DM-\textit{nucleus} cross-section which governs energy transport. DM-nucleus interactions are typically classified as spin-dependent (SD, which traditionally refers specificially to the coupling of DM to nuclear spin), or spin-independent (SI, where the DM couples coherently to every nucleon). In the SI case, the cross-section, $\sigma_{0,i}$, for scattering with the $i\ts{th}$ nuclear species of mass $m_i$ and nucleon number $A_i$ is related to that for scattering with a single nucleon (i.e. a hydrogen nucleus) $\sigma_0 \equiv \sigma_{0,H}$, by (see e.g.\ \cite{Akrami11DD} and references therein)
\begin{equation}
\label{eq:SI}
\sigma_{0,i} = \sigma_{0}A_i^{2}\Big(\frac{m_i}{m_p}\Big)^2\Big(\frac{m_\chi + m_p}{m_{\chi} + m_i}\Big)^2,
\end{equation}
and in the SD case,
\begin{equation}
\sigma_{0,i} = \sigma_{0}\frac{4(J_{i}+1)}{3J_{i}}\left|\langle S_{p,i} \rangle + \langle S_{n,i} \rangle \right|^2 \Big(\frac{m_i}{m_p}\Big)^2 \Big(\frac{m_{\chi}+m_p}{m_{\chi}+m_i}\Big)^2,
\label{eq:SD}
\end{equation}
where $m_p$ is the proton mass, $\langle S_{p,i} \rangle $ and  $\langle S_{n,i} \rangle $ are the expectation values of the spins of the proton and neutron subsystems, and $J_{i}$ is the nuclear spin. Being by far the most abundant element in the Sun with spin, here we only consider SD interaction of DM with hydrogen.

Note that Eq.\ \ref{eq:SI} only holds at low $q$; at large $q$ DM no longer couples coherently to every nucleon, and the cross-section is suppressed. This is characterised by an additional multiplicative, isotope- and momentum-dependent nuclear form factor---or \textit{nuclear response function}---which accounts for the internal structure of nuclei. We do not consider the nuclear response in this paper, as most of the DM scattering events contributing to energy transport in the Sun occur with hydrogen and helium at low momentum transfer. Where scattering from heavier elements does occur, it has been shown that the suppression of the cross-section produced by introducing the form factor is insignificant at the relevant values of the momentum transfer \cite{Vincent13}.  Of course, such form factors become much more important in the context of DM capture.

Recent work \cite{Fitzpatrick2013TheDetection,Catena2015FormTheories} has formalized an equivalent parameterisation, based on \textit{non-relativistic effective operators} (NREOs), which decompose interactions into a basis of scalar products of the identity, DM spin $\vec S_\chi$, nucleon spin $\vec S_N$, exchanged momentum $\vec q$ and $\vec v_\perp$, the component of $\vec v_{\mathrm{rel}}$ orthogonal to $\vec q$. Nuclear response functions corresponding to each of the 14 operators identified for spin-1/2 DM have been computed and tabulated in \cite{Catena2015FormTheories}, for the 16 most abundant elements in the Sun. In the language of NREOs, the constant SI and SD interactions respectively correspond to $\mathcal{O}_1 = \mathbb{I}$ and $\mathcal{O}_4  = \vec S_N \cdot \vec S_\chi$. Other operators can be straightforwardly translated to and from the form adopted here (\ref{eq:cs1}-\ref{eq:cs2}) via the corresponding DM response functions $R(q,v_\perp)$ presented in Ref. \cite{Catena2015FormTheories}. Such a refactoring can be found e.g. in the public \texttt{Captn' General} code \cite{Kozar:2021iur}.

\subsection{Simulations}

We simulate energy transport by considering 2-body elastic collisions between a single DM particle and a thermal distribution of nuclei, confined within the simple harmonic oscillator gravitational potential of a uniform sphere of constant density $\rho_\mathrm{sho}$:
\begin{equation}
\label{eq:pot}
V(r) = \frac{2 \pi G \rho_\mathrm{sho} r^2}{3}.
\end{equation}
This allows for analytic expressions for the orbits followed by individual particles (Appendix \ref{sec:app}). We track the DM particle on a collision-by-collision basis, recording its position, velocity and collisional energy transfer. We assume the nuclei to be in local thermal equilibrium, with velocities described by a Maxwell-Boltzmann distribution at a local temperature $T(r)$.  We work in a Cartesian coordinate system with origin at the centre of the Sun.

Although the realistic mass distribution of the Sun is more complicated, the benefit of having analytical orbits here is substantial, both in terms of computational time and complexity. Given that trapped DM is predicted to cluster in the core of the Sun, where the density is approximately constant, this is not an unreasonable way to model energy transport by DM scattering.  Note also that whilst we employ a constant $\rho_\mathrm{sho}$ for determining the gravitational potential experienced by DM particles between scattering events, this is entirely independent of the density profile that we employ for scattering nuclei in the simulation. We will consider two scenarios, one with a constant density of scatterers \textit{\`a la} Gould and Raffelt, and one that will include a radially-dependent density distribution $n(r)$.

We simulate the random walk of a DM particle undergoing scattering events in the Sun according to the following procedure:

\begin{enumerate}[label=\protect\circled{\textbf{\arabic*}},leftmargin=*]
\setlength{\itemsep}{20pt}

\item \textbf{Selection of the initial position and velocity}\label{rep:zero}\vspace{1mm}\\
Using the rejection method, we select the initial height of the DM particle in the Sun from a distribution
\begin{equation}
\label{exdis}
P(r) \varpropto r^2 e^{-\left(\frac{r}{r_\mathrm{sho}}\right)^2},
\end{equation}
where $r_\mathrm{sho}$ is the effective scale length of the distribution obtained by evaluating Eq.\ \ref{eq:scale} with $\rho_c = \rho_\mathrm{sho}$. We draw the corresponding angular co-ordinates $\theta$ and $\phi$ isotropically from uniform distributions.  We subsequently draw the three Cartesian components of the initial velocity vector independently from 1D Maxwell-Boltzmann distributions characterised by the local temperature. We found that the results of the simulation were largely independent of the form of the initial sampling distributions, so the assumption of local thermal equilibrium for obtaining the starting velocity does not impact the results.

\item \textbf{Determination of the next point of collision}\label{rep:first}\vspace{1mm}\\
Following each collision, we draw a number of optical depths over which the DM particle will travel before interacting again.  We select this number from an exponentially falling distribution proportional to $e^{-\tau}$. The optical depth $\tau$ relates a time interval $t$ to the sum of the total rates $\omega_{i}$ with which a DM particle of velocity $\vec{v}$ scatters with the $i\ts{th}$ nuclear species, via the differential equation
\begin{equation}
\label{eq:dtau}
d\tau = \sum_{i} \omega_{i}(\vec{v}) dt.
\end{equation}
The rate at which a DM particle of velocity $\vec{v}$ interacts with particles of number density $n_{i}(r)$, each moving with a velocity $\vec{u}$, is given by the product of the total interaction cross-section,  $\sigma_{i}$, with the flux, $n_{i}(r)\left| \vec{v} - \vec{u} \right|$. In the Sun, where the velocities of a given nuclear species follow a distribution $f(\vec{u};r)$, the quantity $\omega_i$ is constructed by averaging the rates for each possible $\vec{u}$ weighted by $f(\vec{u};r)$. This leads to the general result that
\begin{equation}
\label{eq:rate}
\omega_{i} = \int d^3 \vec{u} \sigma_{i}(\left|\vec{v}-\vec{u}\right|)n_{i}(r)|\vec{v} - \vec{u}|f(\vec{u};r).
\end{equation}
As the nuclei are in LTE, $f(\vec{u};r)$ takes the form of a Maxwell-Boltzmann distribution at the local temperature $T(r)$. In general, the total cross-section $\sigma_{i}$ depends on the  DM-nucleus relative velocity $v_\mathrm{rel} \equiv |\vec{v} - \vec{u}|$. This is most clearly seen from its relation to the differential cross-section
\begin{equation}
\label{eq:sigma}
\sigma_{i} = \int_{-1}^{1} d\cos\theta_\mathrm{cm} \frac{d\sigma_{i}}{d\cos\theta_\mathrm{cm}}(v_\mathrm{rel},q).
\end{equation}
Whilst we parameterise the differential cross-sections in terms of $v_\mathrm{rel}$ and $q$, the latter can itself be expressed in terms of $v_\mathrm{rel}$ and $\theta_\mathrm{cm}$ (Eq.\ \ref{eq:q}). After integrating over the scattering angle, the total cross-section therefore depends solely on the relative velocity of the colliding particles.

For cross-sections proportional to $q^{-2}$, the proportionality of the differential cross-section to $(1-\cos\theta_{\mathrm{cm}})^{-1}$ arising from Eq.\ \ref{eq:q} causes the total cross-section to diverge at $\theta_\mathrm{cm}$ = 0. This scattering angle corresponds to forward scattering, which for elastic interactions does not result in momentum transfer. Given that momentum transfer is necessary for energy transport, forward scattering is not of any relevance to this discussion. We thus regulate the divergent behaviour by using the ``momentum transfer cross-section" $\sigma_{T}$,
\begin{equation}
\frac{d\sigma_{T}}{d\cos\theta_\mathrm{cm}}(q,\theta_\mathrm{cm}) \equiv (1-\cos\theta_\mathrm{cm}) \frac{d\sigma}{d\cos\theta_\mathrm{cm}},
\end{equation}
which considers only those interactions that result in momentum transfer \cite{Krstic,Tulin:2013teo,Vincent13}.

The analytical expressions that we obtain for $\omega_i$ by computing Eq.\ \ref{eq:rate} for each specific DM-nucleus interaction are given in Table \ref{tab:omega_i}.

We find the time corresponding to the desired value of $\tau$ by integrating Eq.\ \ref{eq:dtau} along the DM trajectory using a 5$\ts{th}$ order Runge-Kutta algorithm with an embedded $4\ts{th}$ order step.
This is an iterative technique, which combines 6 evaluations of the integrand at successive times within a small interval $\delta t$ in two different ways to give two independent evaluations of $\tau (t+\delta t)$. The difference between these evaluations is used to estimate the truncation error and adjust the step size accordingly. As $t$ is increased, $\tau$ increases monotonically. We stop iterating when the desired value of $\tau$ is reached and interpolate back to find the corresponding time value.  Once the collision time is known, we obtain the DM position and velocity using Eqs.\ \ref{eq:traj} and \ref{eq:velt}.

\begin{table}
\begin{center}
\begin{tabular}{c |c} \hline
\multicolumn{1}{c|}{Interaction Type}& \multicolumn{1}{c}{$\omega_{i}$} \\
\hline

$\sigma  = const.$ & $2\sigma_{0}n_iv_{T}\sqrt{\mu}\left[\big(y+ \frac{1}{2y})\erf{(y)}+\frac{1}{\sqrt{\pi}}\exp{(-y^2)}\right]$ \\[8pt]

$\sigma  \propto v_\mathrm{rel}^2$ &
$2\sigma_{0}n_i\mu^{3/2}v_{T}\big(\frac{v_{T}}{v_0}\big)^2 \left[\big(\frac{3 + 12y^2 + 4y^4}{4y})\erf{(y)}+\frac{5 + 2y^2}{2\sqrt{\pi}}\exp{(-y^2)}\right]$ \\[8pt]

$\sigma  \propto q^2$ &
$\frac{4\sigma_{0}n_i\mu^{3/2}v_{T}m_{\chi}^{2}}{(1+\mu)^2}\big(\frac{v_T}{q_0}\big)^2\left[\big(\frac{3 + 12y^2 + 4y^4}{4y})\erf{(y)}+\frac{5 + 2y^2}{2\sqrt{\pi}}\exp{(-y^2)}\right]$ \\[8pt]

$\sigma  \propto v_\mathrm{rel}^4$ &
$2\sigma_{0}n_i\mu^{5/2}v_{T}\big(\frac{v_{T}}{v_0}\big)^4 \left[\big(\frac{15 + 8y^6 + 60y^4+90y^2}{8y})\erf{(y)}+\frac{33 + 4y^4 +28y^2}{4\sqrt{\pi}}\exp{(-y^2)}\right]$ \\[8pt]

$\sigma  \propto q^4$ &
$\frac{32\sigma_{0}n_i\mu^{5/2}v_{T}m_{\chi}^{4}}{3(1+\mu)^4}\big(\frac{v_T}{q_0}\big)^4\left[\big(\frac{15 + 8y^6 + 60y^4+90y^2}{8y})\erf{(y)}+\frac{33 + 4y^4 +28y^2}{4\sqrt{\pi}}\exp{(-y^2)}\right]$ \\[8pt]

$\sigma  \propto v_\mathrm{rel}^{-2}$ &
$2\sigma_{0}n_i\mu^{-1/2} \big(\frac{v_{T}}{v_0}\big)^{-2}  v_{T }y^{-1}\erf{(y)}$\\[8pt]

$\sigma  \propto q^{-2}$ &
$\frac{\sigma_{0}n_i\mu^{-1/2} q_{0}^{2}(1+\mu)^2}{v_T m_{\chi}^{2}}y^{-1}\erf{(y)}$\\[8pt]

\hline
% \hline
\end{tabular}
\caption{The analytical form of the total rate $\omega_{i}$ at which a DM particle of velocity $\vec{v}$ scatters with the $i\ts{th}$ nuclear species. For ease of mathematical notation we define $v_T^2 = 2k_\mathrm{B}T/m_{\chi}$ and $y^2 = |\vec{v}/v_T|^2/\mu$. \label{tab:omega_i}}
\end{center}
\end{table}

\item \textbf{Selection of a Scattering Nucleus}\vspace{1mm}\\
The probability of interaction with the $i\ts{th}$ nuclear species is proportional to $\omega_{i}$. This depends on both the velocity of the DM particle and its position. In the simulations that we present here, we only consider spin-dependent scattering on hydrogen; in more general simulations, one would randomly select the nuclear species involved in each individual interaction, with the relative probabilities of different species given by their respective interaction rates $\omega_i$.

We randomly generate an incoming velocity vector of the scattering hydrogen nucleus, according to the probability distribution for the local rate of DM interaction with hydrogen. The nuclear velocity $\vec{u}$ enters this rate calculation both directly via its velocity distribution $f(\vec{u};r)$, and also in the DM-nucleus relative speed
\begin{equation}
v_\mathrm{rel} = |\vec{v} - \vec{u}| = (v^{2} + u^{2} - 2vu \cos\vartheta)^{1/2},
\end{equation}
where $\vartheta$ is the angle between the incoming velocities $\vec{v}$ and $\vec{u}$ of the DM and the nucleus respectively, in the frame of the Sun.

The 2D distribution for the nuclear speed and scattering angle thus takes the form
\begin{equation}
\label{eq:2drej}
P(u,\vartheta) \varpropto u^{2}(v^{2} + u^2 - 2uv\cos\vartheta)^{1/2 + n} \exp\left[-\frac{u^{2}}{v_{T}^{2}\mu}\right],
\end{equation}
where $v_T$ is defined as $\sqrt{2k_\mathrm{B}T/m_{\chi}}$.
Here, the additional factor of  $u^{2}$ compared to Eq.\ \ref{eq:rate} arises from the conversion of the (three-dimensional) nuclear velocity distribution $f(\vec{u},r)$ to the corresponding (one-dimensional) Maxwell-Boltzmann speed distribution. The additional factors of the relative velocity come from the form of the total cross-section.
We employ a 2-dimensional rejection technique to draw from this distribution, simultaneously considering candidate values for both $u$ and $\cos\vartheta$. These two factors completely specify the DM-nucleus relative velocity.

\item \textbf{Perform the Collision in the Centre of mass frame}\label{rep:last}\vspace{1mm}\\
We simulate the DM-nucleus interaction in the centre-of-mass (CoM) frame. The DM velocity $\vec{t}$ in this frame is related to the velocity $\vec{v}$ in the solar (lab) frame as
\begin{equation}
\vec{t} = \vec{v} - \vec{s},
\end{equation}
where the velocity of the CoM of the system in the solar frame is
\begin{equation}
\vec{s} \equiv \frac{m_{\chi}\vec{v}+ m_{N}\vec{u}}{m_{\chi}+m_{N}}.
\end{equation}
Energy and momentum conservation require that in CoM frame, the magnitude of each particle's momentum, and hence speed, are unchanged by the interaction. The remaining two parameters required to describe the outgoing system, namely the polar angles $\theta_\mathrm{cm}$ and $\phi$ through which the DM particle is scattered, are unconstrained, and thus governed by the probabilities of quantum mechanics.

The angle $\theta_\mathrm{cm}$ corresponds to the angle between the incoming and outgoing DM velocity vectors, and is thus referred to as the centre-of-mass scattering angle.  The probability distribution for this angle is dictated by the form of the DM-nucleus interaction.  The azimuthal angle $\phi$ is arbitrary, depending on neither the form of the interaction nor the kinematics.  It therefore follows a uniform distribution.

To describe a collision in this CoM frame, we first construct an orthonormal basis with unit vectors $\{\hat{a_1},\hat{a_2},\hat{a_3}\}$, taking $\hat{a_{3}}$ to be the unit vector in the direction of $\vec{t}$. The outgoing velocity of the DM particle $\vec{t'}$ in this frame  is thus
\begin{equation}
\vec{t'} = t\,\sin\theta_\mathrm{cm} \,\cos\phi \, \vec{\hat{a_1}} + t\,\sin\theta_\mathrm{cm} \,\sin\phi \,\vec{\hat{a_2}} + t\,\cos\theta_\mathrm{cm} \, \vec{\hat{a_3}}.
\end{equation}
This is illustrated in Fig.\ \ref{fig:axis}.

\begin{figure}
  \centering
     \includegraphics[width=0.8\columnwidth]{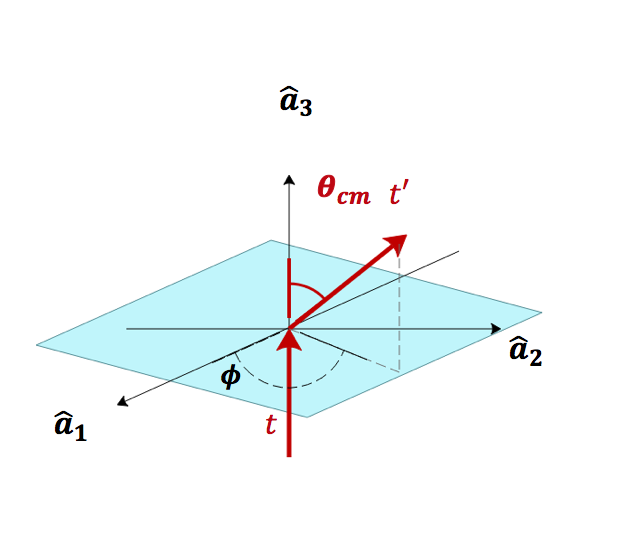}
     \caption{A schematic of the collision in the Centre of Momentum (CoM) frame. The red arrows $\vec{t}$ and $\vec{t'}$ represent the ingoing and outgoing DM velocity vectors. The angle marked in red, $\theta_\mathrm{cm}$, is the CoM scattering angle. It is defined such that $\theta_\mathrm{cm}$ = 0 corresponds to forward scattering.
     \label{fig:axis}
     }
\end{figure}

We draw values for $\theta_\mathrm{cm}$ based on the form of the differential cross section for a given interaction. In the case of velocity-dependent interactions, the distribution in $\cos \theta_\mathrm{cm}$ is simply uniform over the interval $[-1,1]$. For $q$-dependent interactions, the differential cross-section can be expressed in terms of the CoM scattering angle via Eq.\ \ref{eq:q}. We thus select $\theta_\mathrm{cm}$ from a distribution proportional to $(1-\cos\theta_\mathrm{cm})^{n}$.

Once the outgoing velocity has been calculated in the CoM frame, we then transform it back to the lab frame by adding $\vec{s}$ (which remains constant during the collision).

\item \textbf{Record and repeat}\vspace{1mm}\\
After each collision, we record the position of the DM particle and its outgoing velocity, and use them to determine its new trajectory. We also log the energy transferred to the nucleus, given by the difference between the DM kinetic energy immediately before and after the collision.

We then repeat steps \ref{rep:first} -- \ref{rep:last} until the particle has undergone a large number of interactions, typically between $10^6$ and $10^7$.

In addition to the data that we record at each collision point, we also log the position and velocity of the DM at regular time intervals over the course of the simulation, in order to allow us to construct stationary distributions of relevant properties of the DM population.

\end{enumerate}

%#######################################################################

\subsection{Treatment of particles leaving the Sun}

If at any point in its trajectory a DM particle reaches the solar surface, it leaves the Sun.  If the velocity on exit is less than the surface escape velocity, it does not escape to infinity, but will return at some later time, to participate in further scattering events. Outside the Sun, the orbits of such particles become entirely Keplerian. We use the exit position and velocity of the DM particle to calculate the form of this trajectory, and by finding its intersection points with the boundary of the Sun, compute the position and velocity at which it will re-enter the Sun. We use the re-entry position and velocity to find the new simple-harmonic trajectory of the DM particle, and then continue from Step \ref{rep:first}. We compute the time that the DM particle spends outside the Sun from the ratio of the area of its orbit swept out whilst absent to the total orbital area, and add it to the total simulation time.

If the DM particle reaches the boundary of the Sun with a velocity exceeding the surface escape velocity, we terminate the simulation and begin a new one from Step \ref{rep:zero}. This corresponds to an evaporation event. We observe a negligible number of such cases throughout our simulations.

\subsection{Obtaining distributions}
\label{sec:dists}

The luminosity transported by DM to a height $r$ in the Sun is
\begin{equation}
\label{eq:Lum}
L(r) =\int dL = 4\pi \int_{0}^{r} r'^{2} \rho (r') \epsilon_{\chi}(r') dr',
\end{equation}
where $\rho(r)$ is the total matter density, and $\epsilon (r)$ is the energy injected into the stellar plasma by DM scattering, per unit time and mass of solar material. The integrand $dL$ corresponds to the total rate of energy transfer in a spherical shell centred at radius $r$. To estimate $dL$ from $E_i$, the set of energy transfer values in collisions in our simulation, in a given radial bin we compute
\begin{equation}
\label{eq:E}
\langle dL \rangle = \sum_{i} \frac{E_{i}}{t_\mathrm{sim}},
\end{equation}
where $t_\mathrm{sim}$ is the total simulation time. This corresponds to the mean rate of energy transfer by a single DM particle across the entire shell. The total luminosity at radius $r$ can then be determined by simply extending the shell the entire way from the centre of the Sun to $r$.  To map this result to the expected luminosity for an entire population of DM in the Sun, Eq.\ \eqref{eq:E} must be multiplied by $\eta N_\mathrm{B}$, where $N_\mathrm{B} \equiv M_{\odot}/m_p$ is the number of baryons in the Sun, and $\eta = N_\chi/N_\mathrm{B}$ is the DM fraction accumulated in the Sun.

In the latter parts of this paper, we will use the maximum value of the luminosity, i.e.\ $L_{max} \equiv \max\limits_r |L(r)|$, as a measure of the transport efficiency of a given DM model---the absolute value here is because the luminosity is negative everywhere by convention. Explicitly, we take $L_{max}$ and its associated error, to be the luminosity and error of the bin with highest luminosity.

The standard error on a mean of some quantity $x$ estimated from Monte Carlo simulation can be estimated as
\begin{equation}
\label{eq:error}
\sigma_\mathrm{SE}^{2} = \frac{\langle x_{i}^{2} \rangle - \langle x_i \rangle ^{2}}{N},
\end{equation}
where $N$ is the number of samples used to construct the mean. The standard error on our estimate of the luminosity is therefore
\begin{equation}
\sigma_{\mathrm{SE}} = \frac{1}{t_{sim}} \sqrt{\sum_{i = 1}^N E_{i}^2 - \frac{1}{N}\left(\sum_{i=1}^N E_i\right)^2 },
\end{equation}
where for the error on $dL$ in a shell, the sum extends over all events in the radial bin, and for the error on $L(r)$, over all events at radii less than $r$.

\section{Results}
\label{sec:results}
In this section we present the main results of our Monte Carlo simulations. We will focus on two large sets of simulations: 1) an \textit{idealized} setup identical to what was used by Gould \& Raffelt, presented in Sec.\ \ref{sec:r1}, and 2) a \textit{realistic} setup using realistic DM and nuclear masses, with temperature and density data from a Standard Solar Model, in Sec \ref{sec:real}.

\subsection{\textit{Idealized} simulations}
\label{sec:r1}
In order to verify our Monte Carlo procedure, we initially study energy transport under the same conditions as Gould \& Raffelt \cite{Gould1990,Gould1990CosmionLimit}. Here, we model the star as a uniform sphere of constant density $\rho = \rho_\mathrm{sho}$, consisting of a single nuclear species with the same mass as the conducting DM particles. The physical scales of this system follow a simple SI-based unit system, with $m_p = m_\chi$ = 1 kg, T = 1 K and a constant density $\rho_\mathrm{sho} = \sqrt{3 k_B/2\pi G} = 10^{-13}$ kg m$^{-3}$.
We take the radius of the Sun to be 2.5\,m such that the Solar temperature remains above zero everywhere. We note that this, and our corresponding procedure to deal with DM particles leaving and re-entering the Sun, marks the only difference between our simulation and that produced of Gould \& Raffelt, who did not any such radial boundary.

The solar temperature profile is taken to be a simple linear gradient of the form
\begin{equation}
\label{temp:grad}
  T(r) = (1.65 - 0.65 r) \,\mathrm{K}.
\end{equation}
This is constructed such that $r = 1$ corresponds to the radius at which $T = T_\chi$, the isothermal (Spergel \& Press) DM temperature for this system.

Adopting such a simple model is advantageous for three reasons. First, it allows direct comparison to the results of the numerical simulations performed by Gould \& Raffelt \cite{Gould1990CosmionLimit}; second, the simple temperature and density profiles mean that comparing with the analytical distributions outlined in Section \ref{sec:ET} is particularly straightforward; and third, the large difference in length, density and gravitational scales between the \textit{idealized} and \textit{realistic} approaches will provide a strong test of the robustness of any heat conduction approximation scheme.

To verify this setup, we first ``switch-off'' the temperature gradient,and follow the evolution of DM in a sphere of nuclei at a uniform temperature $T=1$~K. We note that in this specific situation, $r_{\chi}$ = 1\,m and the Knudsen number $K$ is numerically equal to the DM mean-free path, which in this instance is constant throughout the Sun.

\begin{figure}
  \centering
     \hspace*{-1cm}\includegraphics[width=0.55\textwidth]{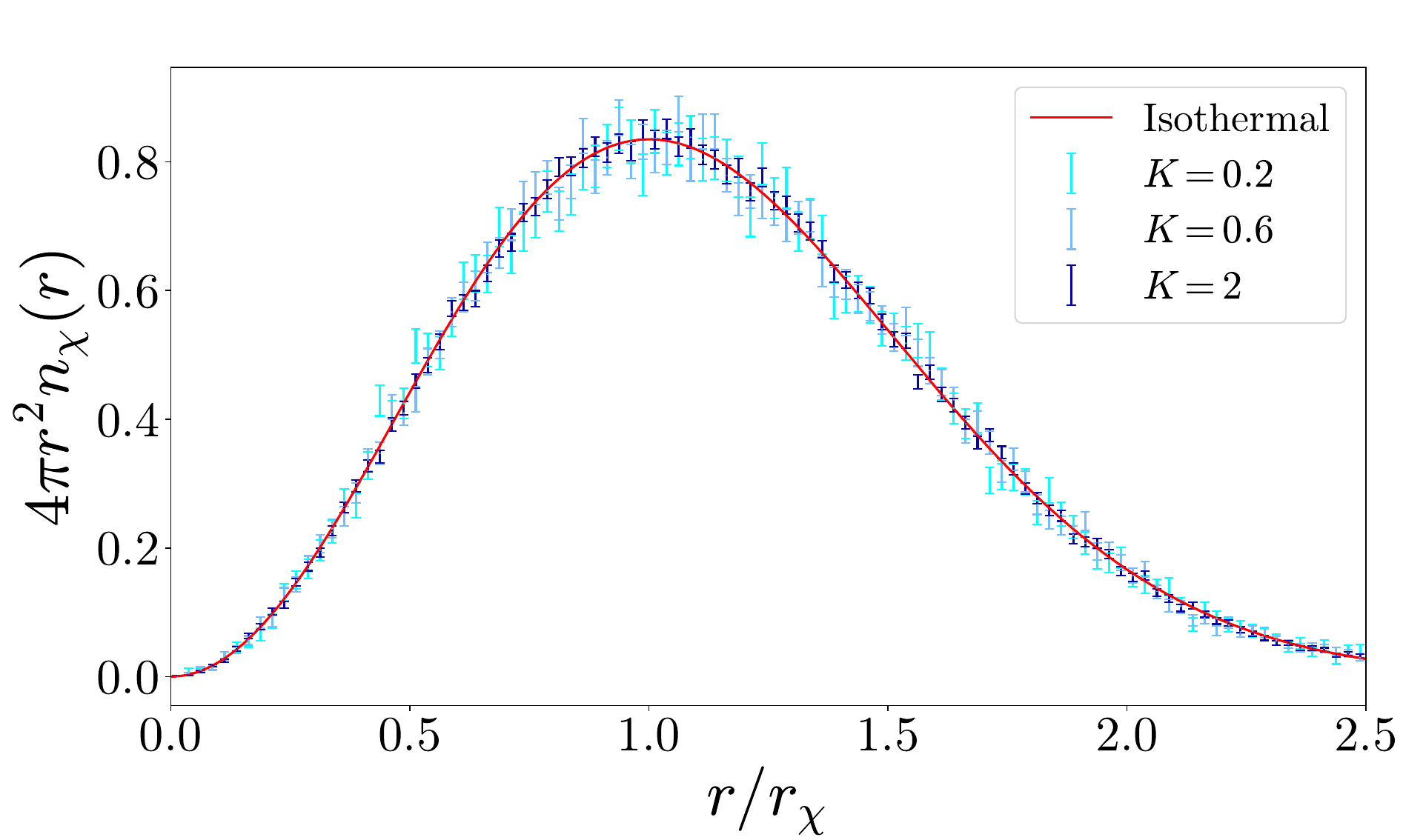}\includegraphics[width=0.55\textwidth]{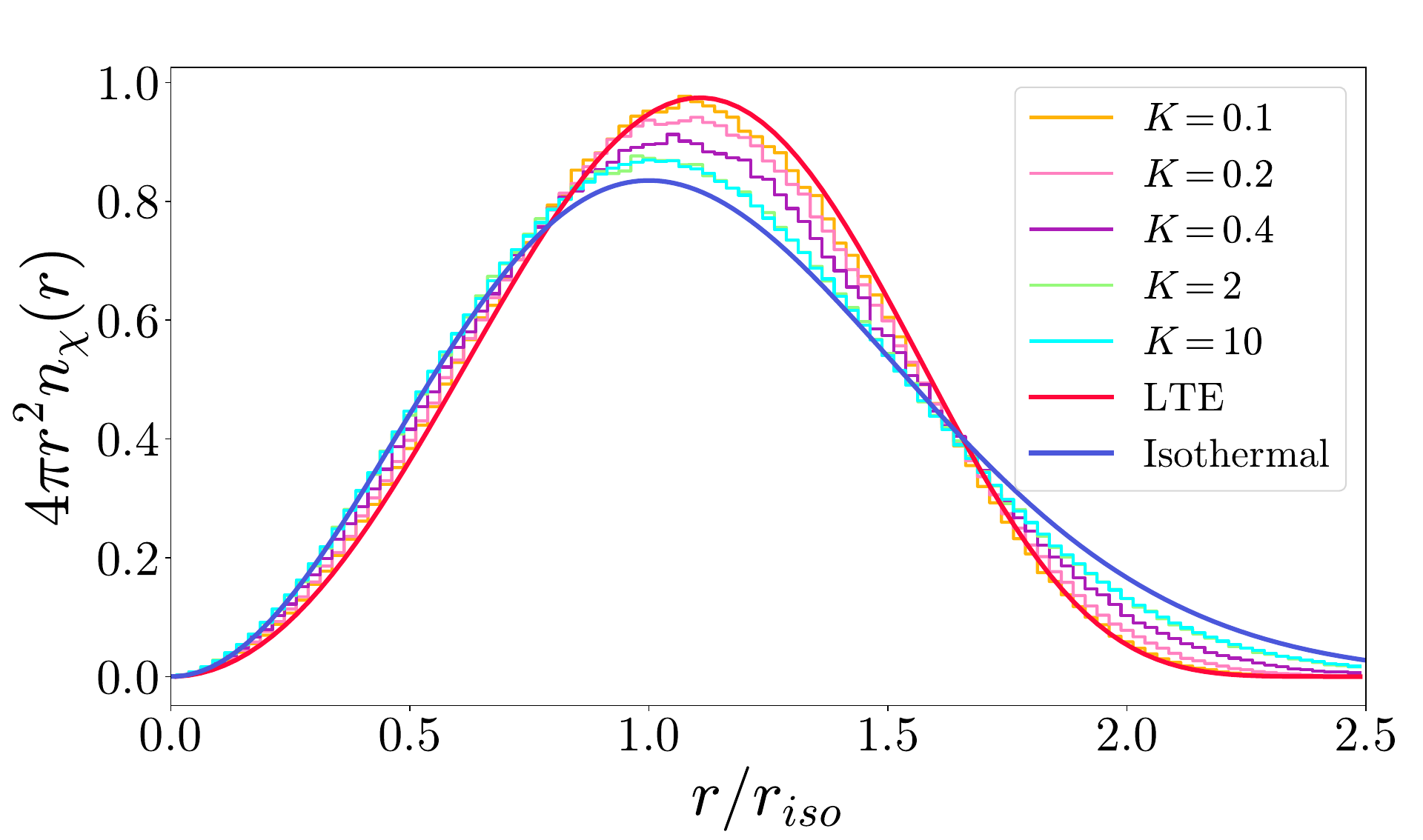}

        \caption{ Radial distribution of DM obtained at various Knudsen numbers, for velocity- and momentum-independent interactions in a constant density sphere, with a uniform temperature of 1\,K (\textit{left}) and the linear temperature gradient defined in Eq.\ \eqref{temp:grad} (\textit{right}).  This latter case is what we denote the \textit{idealized} set-up. The analytic LTE and isothermal profiles are shown as solid lines.
                \label{fig:CT}
               }
\end{figure}

For this scenario, the DM radial distribution should follow
\begin{equation}
\label{eq:dist}
n_\chi(r) = \frac{e^{-\big(\frac{r}{r_{\chi}}\big)^{2}}}{r_\chi^3 \pi^{3/2}}  .
\end{equation}
The function $4\pi r^2 n_\chi(r)$  peaks at a radius of $r$ = 1\,m, as per the definition of the scale length.  The distribution obtained for three different values of $K$ using our Monte Carlo program is shown in the left hand panel of Fig.\ \ref{fig:CT}, demonstrating excellent agreement with the analytical form. As expected, we observe an identical distribution over a wide range of $K$ values.

\begin{figure}
  \centering
     \hspace*{-1.2cm}\includegraphics[width=1.2\textwidth]{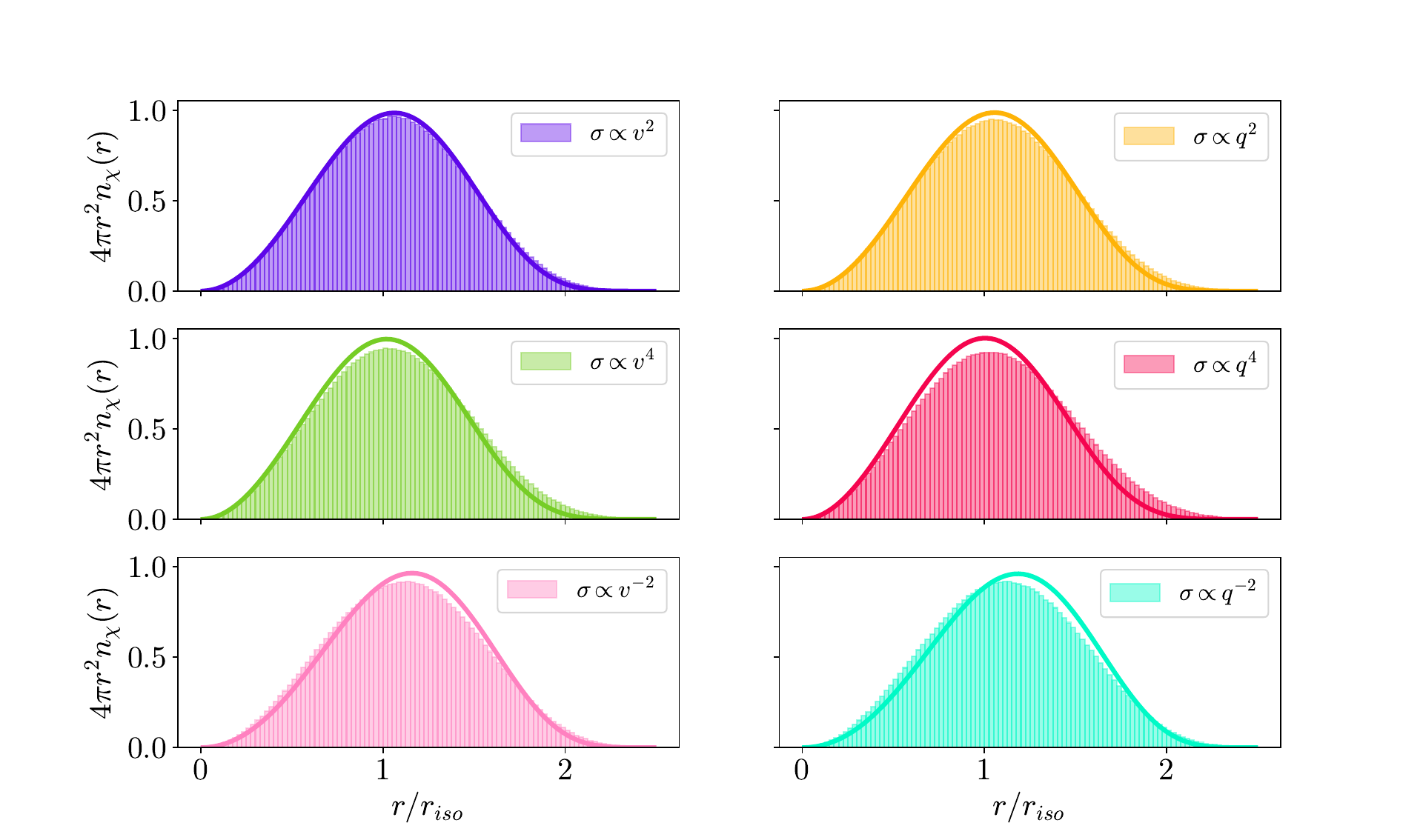}
     \caption{Radial distribution of DM with non-constant cross sections obtained from \textit{idealized} simulations (uniform temperature gradient), for $K = 0.2$. \textit{Left}: $\sigma \propto v^{2n}$. \textit{Right}: $\sigma \propto v^{2n}$. The expected form of the distribution is shown as a solid line.
     \label{fig:distributionsvnqn}
     }
\end{figure}

We then turn to simulations in the \textit{idealized} setup, with a constant density profile and a linear temperature gradient identical to that used by Gould \& Raffelt \cite{Gould1990CosmionLimit} as given in Eq.\ \eqref{temp:grad}.

For non-uniform temperature profiles, the DM radial distribution is dependent on $K$, with the $K \ll 1$ and $K\gg 1$ limits corresponding to the conduction and isothermal regimes respectively. We perform simulations at different values of $K$ by varying the interaction parameter $\sigma_0$, which implicitly defines the inter-scattering distance, $l_{\chi}$ in Eq.\ \eqref{eq:K}.
At $K$ = 0.1, transport is local and the conduction approximation should be expected to be fairly accurate. Since all of the scattering nuclei are identical in this model, the diffusion coefficient $\alpha$ is independent of radius. The analytical expression for the number density (Eq.\ \ref{eq:conductiondist}) thus simplifies to
\begin{equation}
\label{eq:approximate}
n_{\chi}(r) = n_{\chi}(0)\left[\frac{T(r)}{T(0)}\right]^{3/2 - \alpha} e^{\Big(- \int_{0}^{r}dr'\frac{ m_{\chi}\phi}{k_\mathrm{B} T(r')}\Big)}.
\end{equation}
As the temperature profile applied in this situation is analytic, the integral in equation Eq.\ \eqref{eq:approximate} can be evaluated to give a closed form expression for $n_{\chi}$. 

As $K$ increases, we observe a smooth transition from the local regime to the isothermal one. The Monte-Carlo distributions for a range of $K$ between 0.1 and 10 are are shown in the right-hand panel of Fig.\ \ref{fig:CT}, where the red curve is the theoretically-expected LTE distribution \eqref{eq:approximate}, and the blue one is the isothermal distribution \eqref{eq:isothermaldist}. Performing a least-squares fit of $\alpha$ in Eq.\ \eqref{eq:approximate} to the Monte Carlo radial distribution for $K$=0.1, we find $\alpha = 2.316 \pm 0.005$. This is in direct accordance with the value of 2.3190 computed analytically by Gould \& Raffelt \cite{Gould1990} for  $\mu = 1$. 

Fig.\ \ref{fig:distributionsvnqn} shows the distributions obtained for non-constant cross sections with $K = 0.2$, overlaid with the predicted LTE distribution \eqref{eq:conductiondist} using the values of $\alpha$ computed in Ref.\ \cite{Vincent13}, and without the Knudsen-dependent correction $\mathfrak{f}(K)$ \eqref{eq:fgothhgoth} applied. The slight departure from the LTE prediction is similar in magnitude to the $K = 0.2$ constant case seen in the right hand panel of Fig.\ \ref{fig:CT}.

\begin{figure}
  \centering
      \hspace*{-1.5cm}\includegraphics[width=1.2\textwidth]{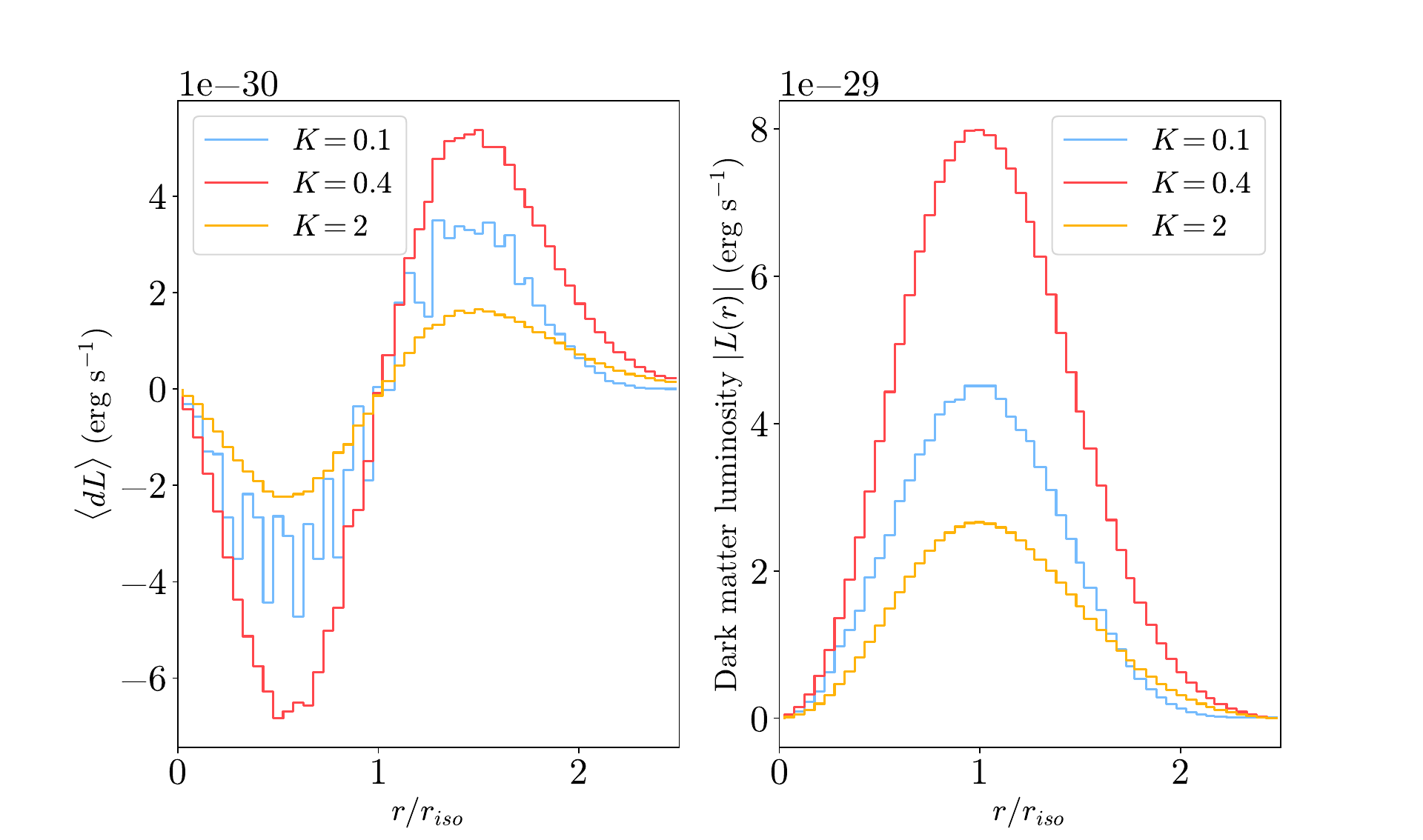}

        \caption{\textit{Left}: Total energy transferred to the nuclei per unit time in 100 radial bins for momentum- and velocity-independent interactions at three different Knudsen numbers $K$, in the \textit{idealized} simulations. A negative energy corresponds to energy being absorbed by the DM, (removed from the plasma) and positive energy represents energy deposition. \textit{Right}: luminosity $L(r)$ carried by DM at each radius.
                \label{fig:enTran}
           }
\end{figure}

We now turn to transported heat: Fig.\ \ref{fig:enTran} shows the radial luminosity $L(r)$ and energy transport profiles for three different momentum- and velocity- independent simulations in the \textit{idealized} (linear temperature gradient) setup. These correspond to transport in the local regime ($K = 0.1$; blue line), Knudsen transition ($K = 0.4$; red line), and the non-local/Knudsen regime ($K = 2$; yellow line). As expected, transport is suppressed in both the LTE limit due to the short mean free path, and the Knudsen limit due to the low interaction rate. The lower $K$ cases are noisier; this is because far more simulation steps are required to adequately sample the full stellar distribution, since $l_\chi$ is small and particles travel only a short distance between collisions.

\begin{figure}
  \centering
      \includegraphics[width=1.0\columnwidth]{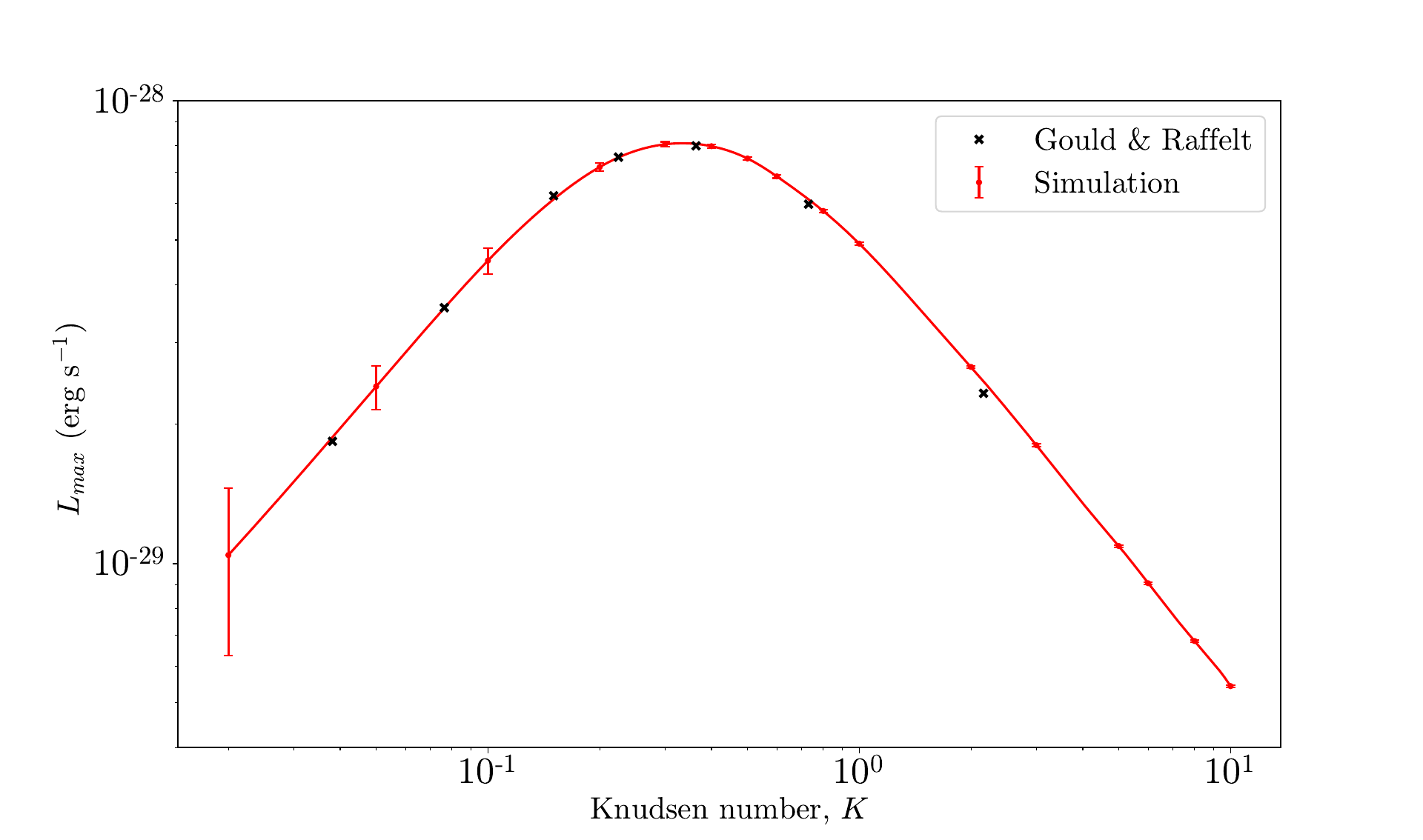}

        \caption{The transition from the conduction regime to the Knudsen limit. Each red point corresponds to the peak of the radial luminosity $L_\mathrm{max} \equiv \max\limits_r |L(r)|$ (Eq.\ \ref{eq:Lum}) obtained following $\sim$ \SI{1e7} simulated interactions. The solid red line is a cubic spline through this data set. The black points correspond to the results of an identical study by Gould \& Raffelt \cite{Gould1990CosmionLimit}, scaled to our units.
                \label{fig:GR}
               }
\end{figure}

Our results for velocity- and momentum-independent interactions are in agreement with Monte Carlo simulations performed by Gould \& Raffelt \cite{Gould1990CosmionLimit}. Each red point in Fig.\ \ref{fig:GR} is obtained from a separate simulation at that value of $K$  and shows the maximum luminosity $L_{max}$ found over 100 radial bins. The solid red line is a cubic spline through the results.  Leftwards of the peak corresponds to the conduction regime (where errors are larger for the reasons pointed out above), whereas rightwards of the peak represents the Knudsen limit of non-local transport.  We show the results of Gould \& Raffelt \cite{Gould1990CosmionLimit}, upon which this simulation is based, as black crosses on the same figure. In doing this, we have converted from the units used for mean-free-path by Gould \& Raffelt to the Knudsen number. The Gould \& Raffelt values of luminosity, originally presented in arbitrary units, have all been scaled by the same constant factor to match our normalisation.

\begin{figure}
    \centering
    \hspace*{-0.8cm}\includegraphics[width=0.55\textwidth]{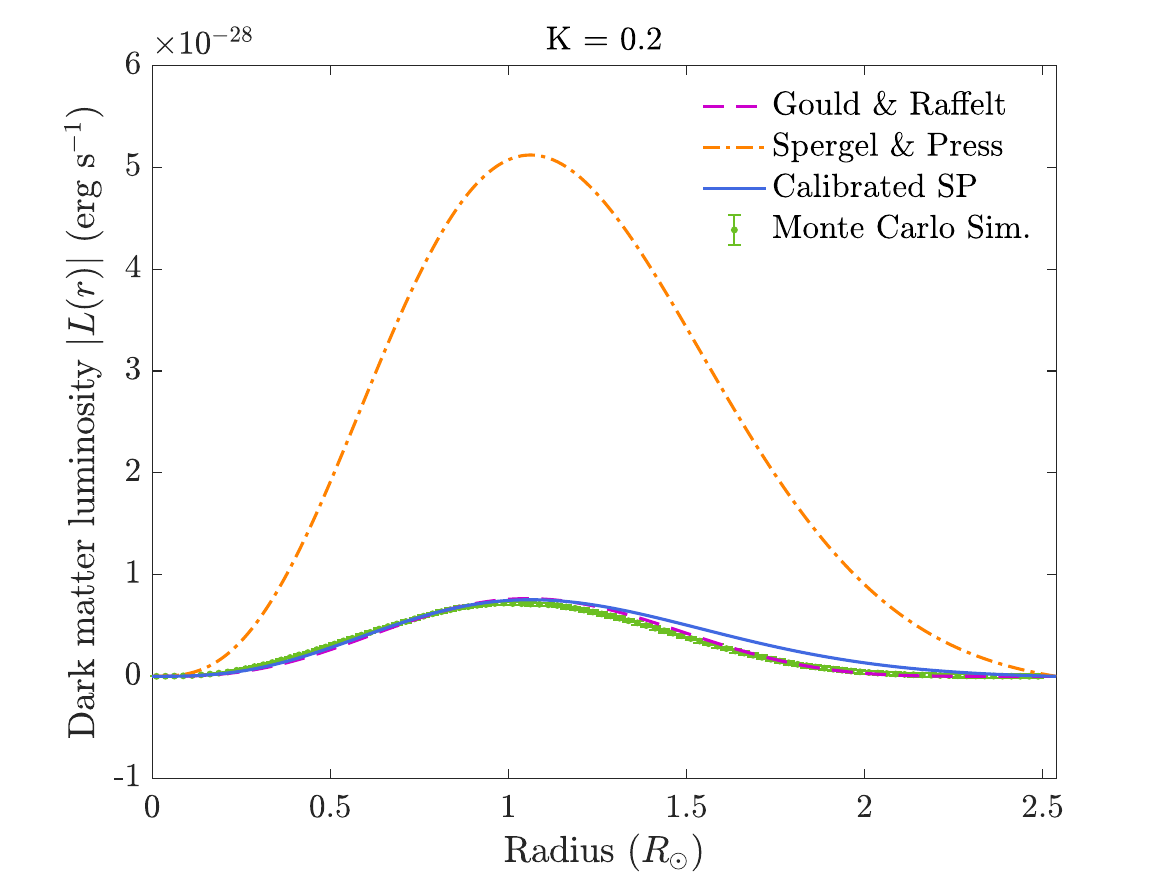}\includegraphics[width=0.55\textwidth]{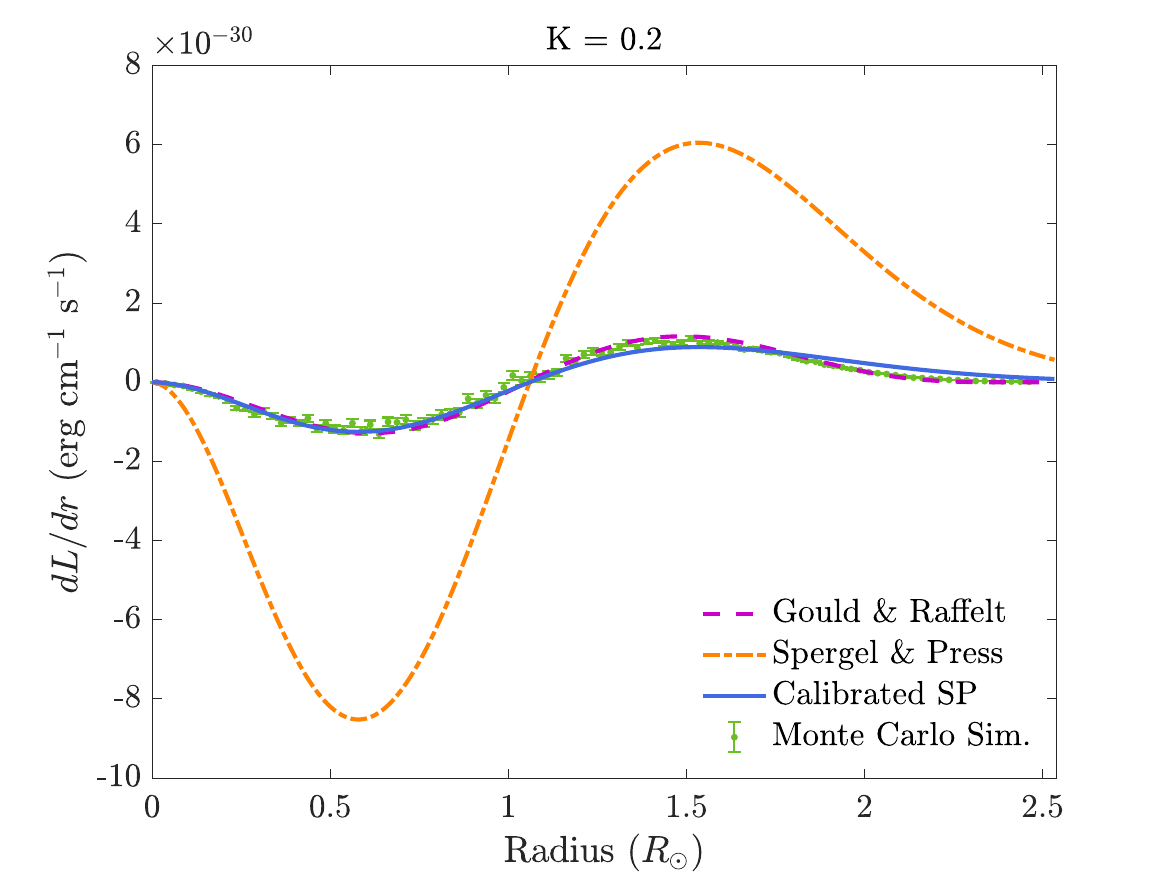}\\
    \hspace*{-0.8cm}\includegraphics[width=0.55\textwidth]{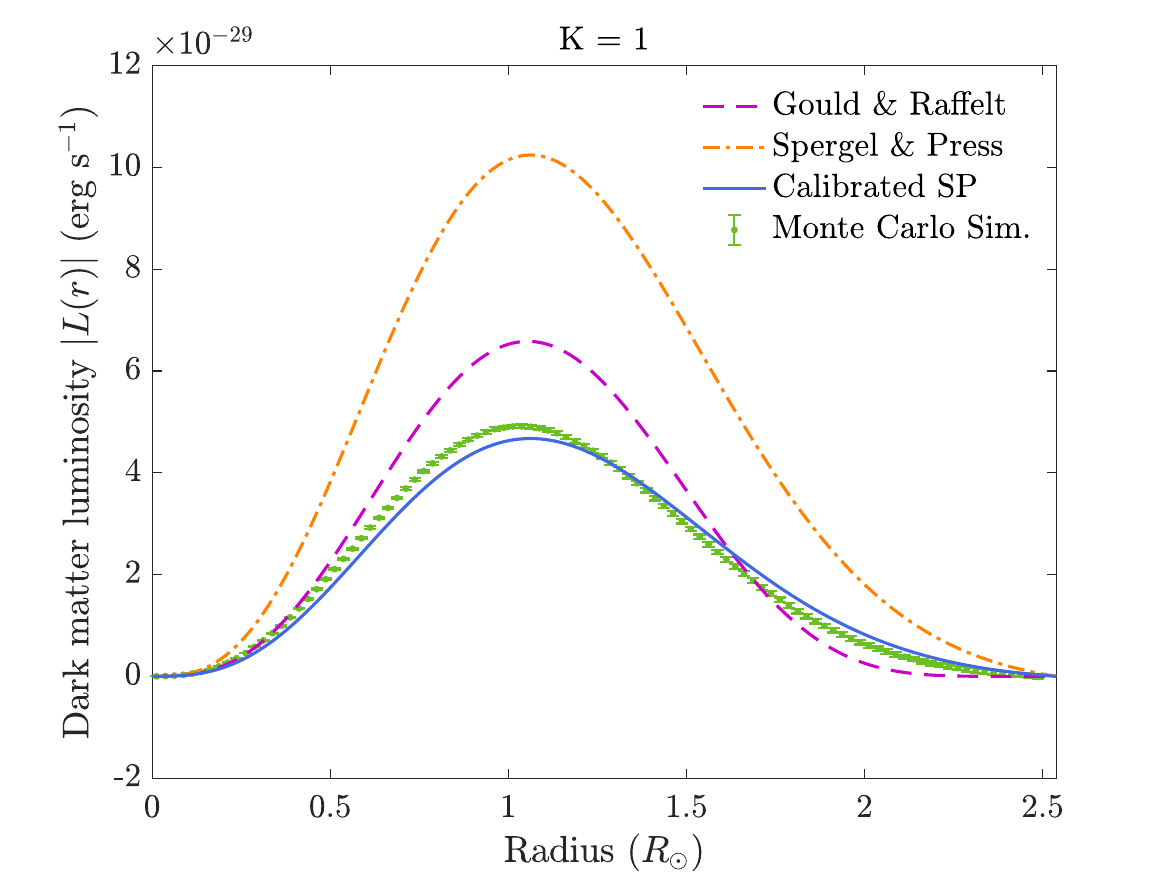}\includegraphics[width=0.55\textwidth]{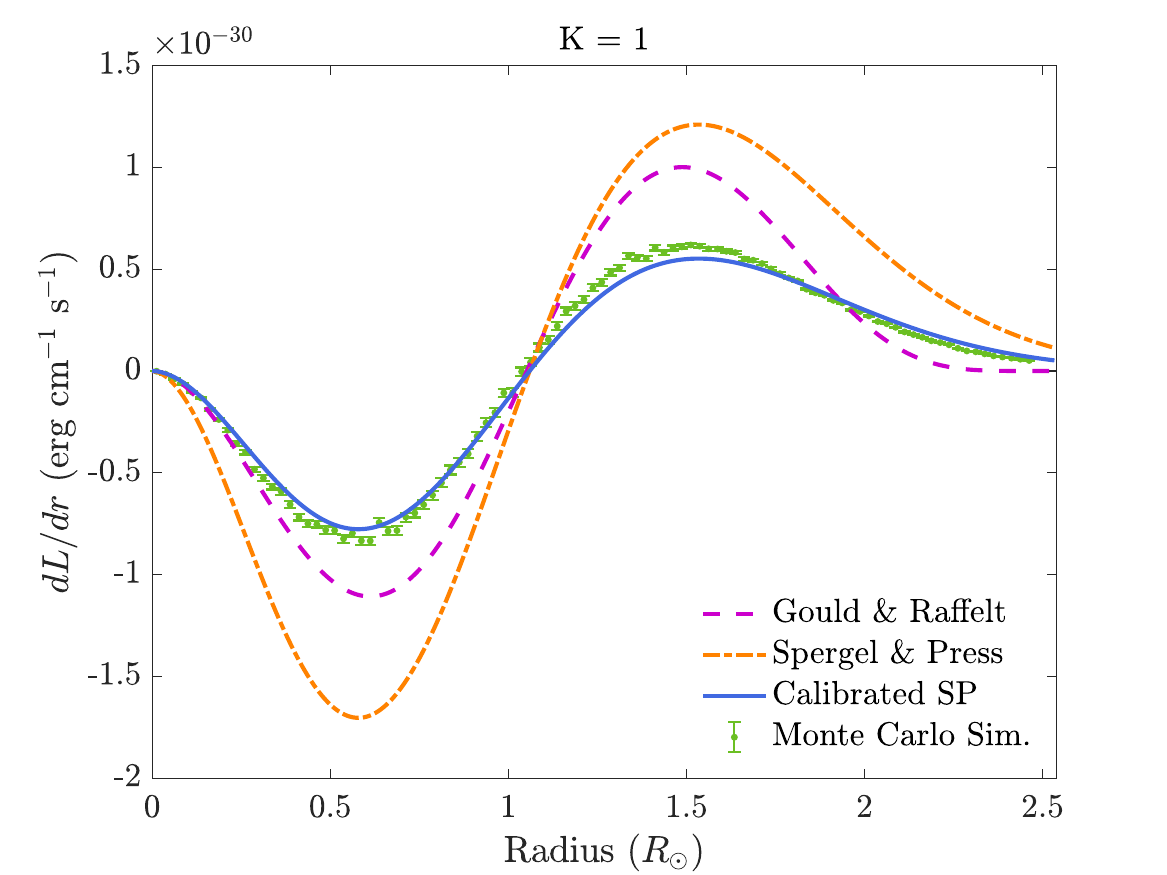}\\
    \hspace*{-0.8cm}\includegraphics[width=0.55\textwidth]{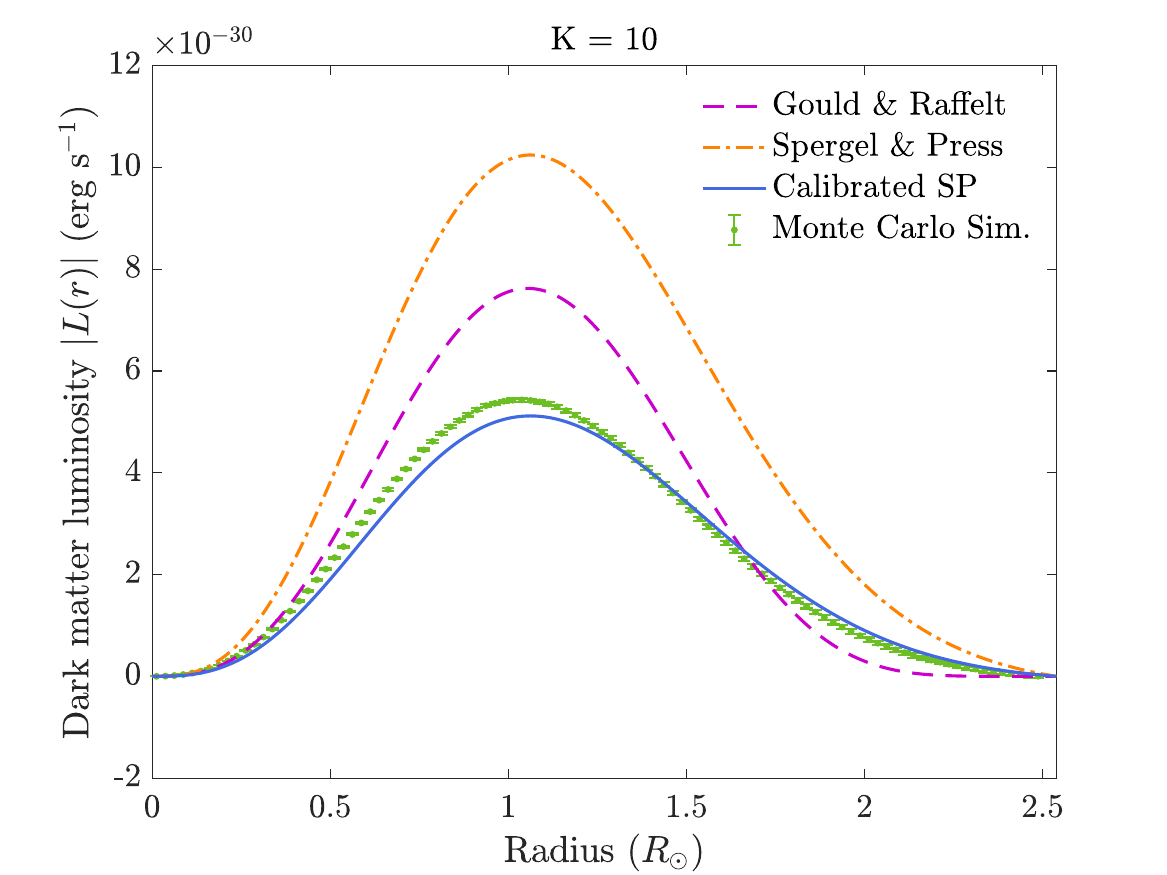}\includegraphics[width=0.55\textwidth]{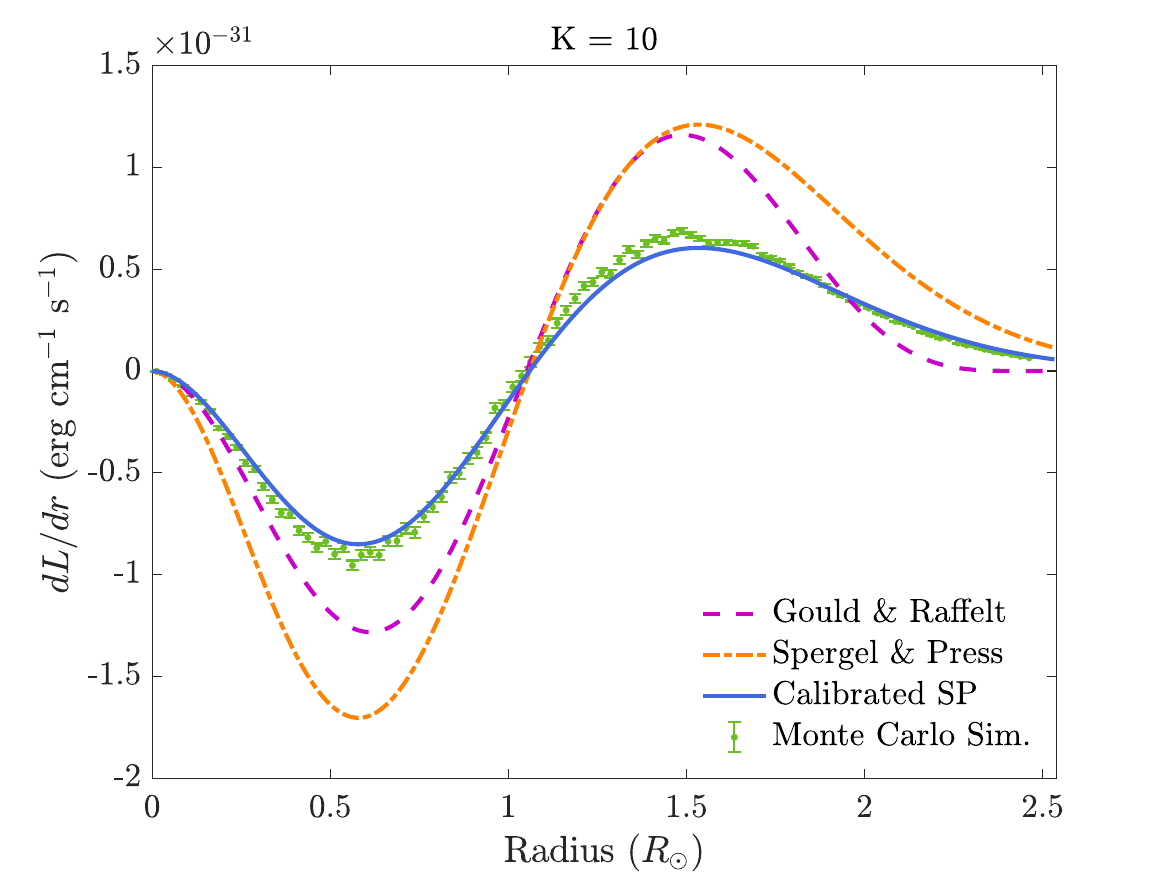}
    \caption{Transported heat in the \textit{idealized} simulation (simple harmonic oscillator potential, constant density and linear temperature gradient) for a constant elastic scattering cross section and $m_\chi = m_p$. Results are shown for the two commonly-used frameworks (Spergel \& Press \cite{Spergel1985EffectInterior}, Gould \& Raffelt \cite{Gould1990,Scott2009,Vincent13}), as well as the `Calibrated SP' method developed here. Green data points represent results of our Monte Carlo simulations. Cross sections and corresponding Knudsen numbers $K$ are given for each of the three plotted cases. }
    \label{fig:idealconduction_constant}
\end{figure}

Radial distributions of luminosity $L(r)$ for DM with velocity- and momentum-independent interactions are shown for three different values of $K$ in the left-hand panels of Fig.\ \ref{fig:idealconduction_constant}. Results from our Monte Carlo simulations are shown in green. The right-hand panel shows $dL/dr$, which is proportional to the transported energy per unit volume \eqref{eq:ELTE}. Negative values of $dL/dr$ correspond to heat being removed from the stellar material, whereas positive $dL/dr$ corresponds to heat deposition. The three values of $K$ roughly span the LTE regime ($K = 0.2$, top), near the Knudsen transition ($K =1$, middle), and the Knudsen regime ($K = 10$, bottom). Dashed and dot-dashed curves in Fig.\ \ref{fig:idealconduction_constant} represent the predicted heat transport under the two heat transport schemes described in Sec.\ \ref{sec:ET}. The isothermal (Spergel \& Press; SP) scheme is shown in orange, whereas the LTE (GR) solution, including the $\mathfrak{f}(K)$ and $\mathfrak{h}(r)$ Knudsen and radial suppression factors (Eq.\ \ref{eq:fgothhgoth}), is shown in magenta. We will return to the solid blue line, labeled ``Calibrated SP'', in Sec.\ \ref{sec:discussion}. Predictably, the GR solution fares well in the LTE regime, but despite the correction factors, tends to overestimate the peak luminosity into and above the Knudsen transition. The SP method, by comparison, is systematically too high by a factor of two in the Knudsen regime (as also noticed by Gould \& Raffelt \cite{Gould1990CosmionLimit}), and deviates substantially from the direct simulation in the LTE regime, as conduction begins to scale with $K$ rather than with $K^{-1} \propto \sigma$. Both approaches correctly identify the radius $r_\chi$ at which heat transport is maximal, i.e.\ where $dL/dr = 0$ and energy goes from being removed at smaller radii to being deposited at larger radii.

\begin{figure}
    \centering

    \hspace*{-0.8cm}\includegraphics[width=0.55\textwidth]{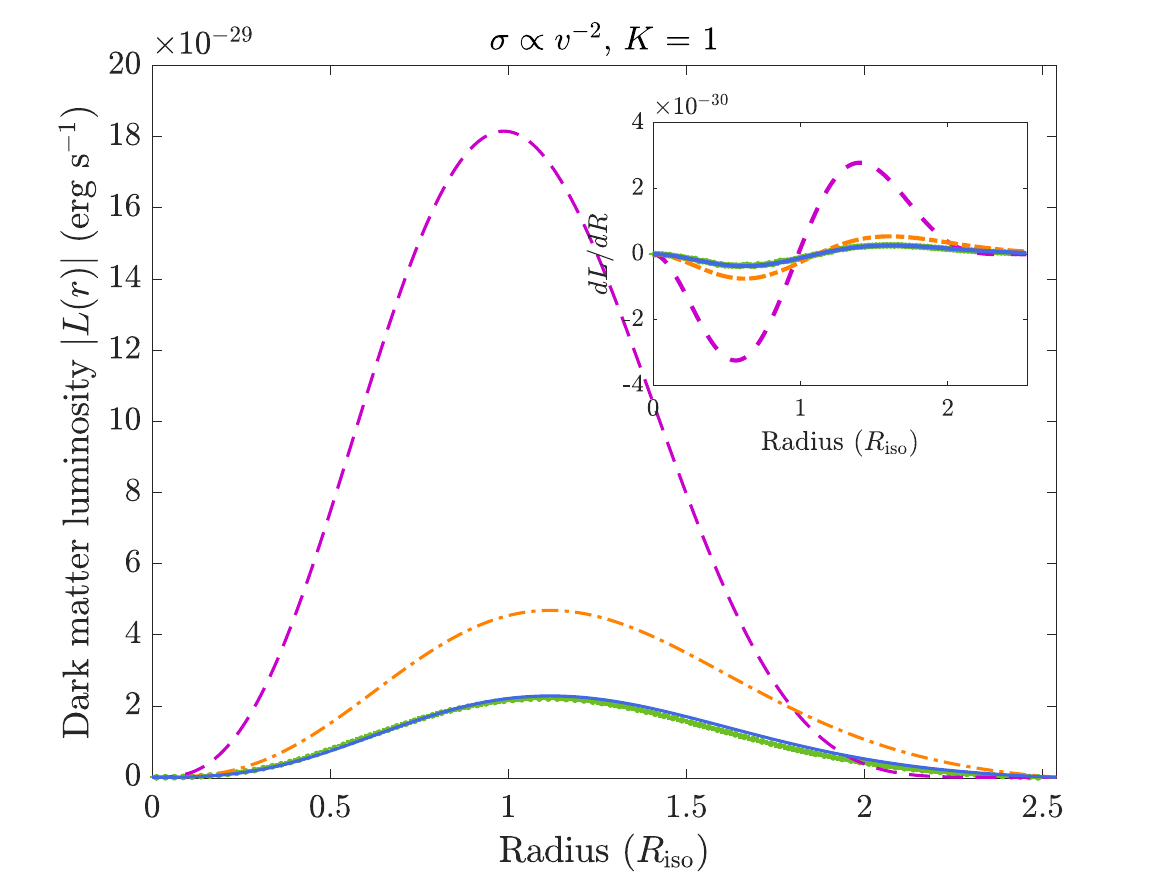}\includegraphics[width=0.55\textwidth]{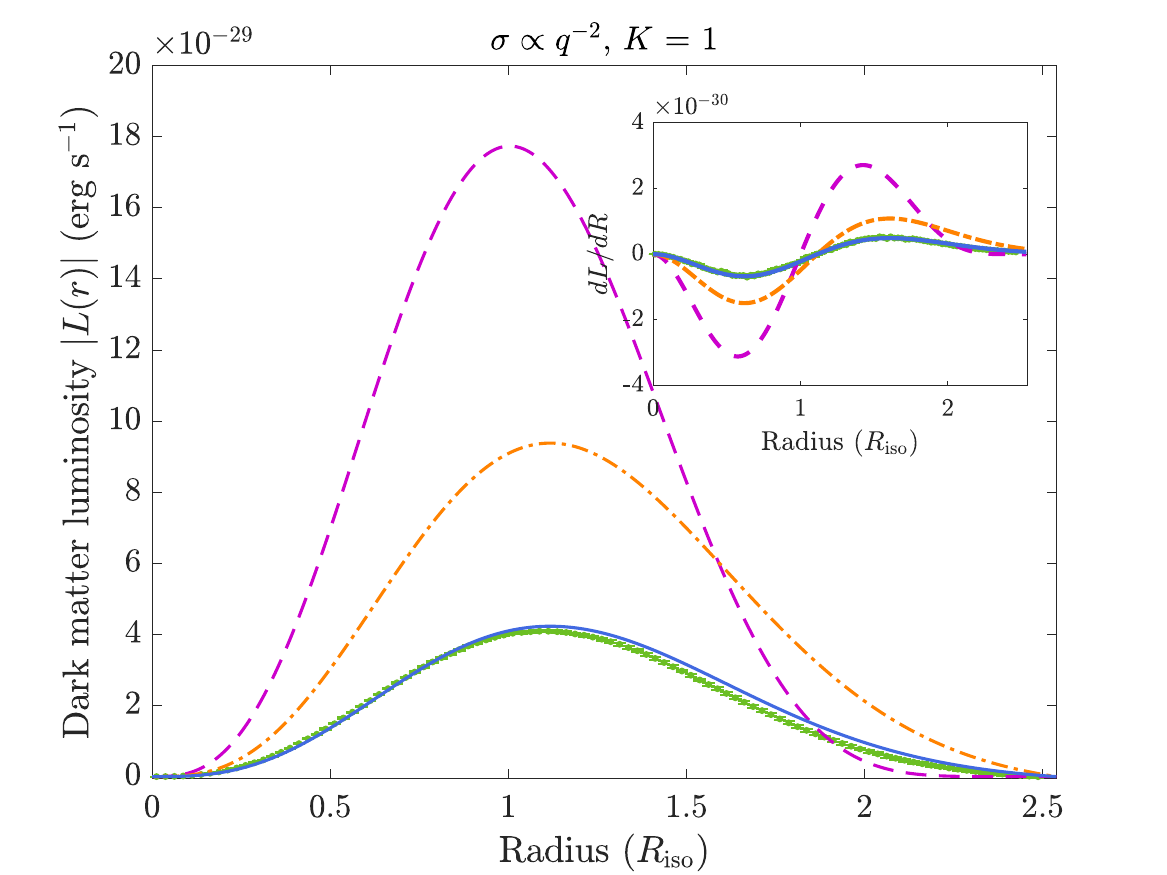}\\
    \hspace*{-0.8cm}\includegraphics[width=0.55\textwidth]{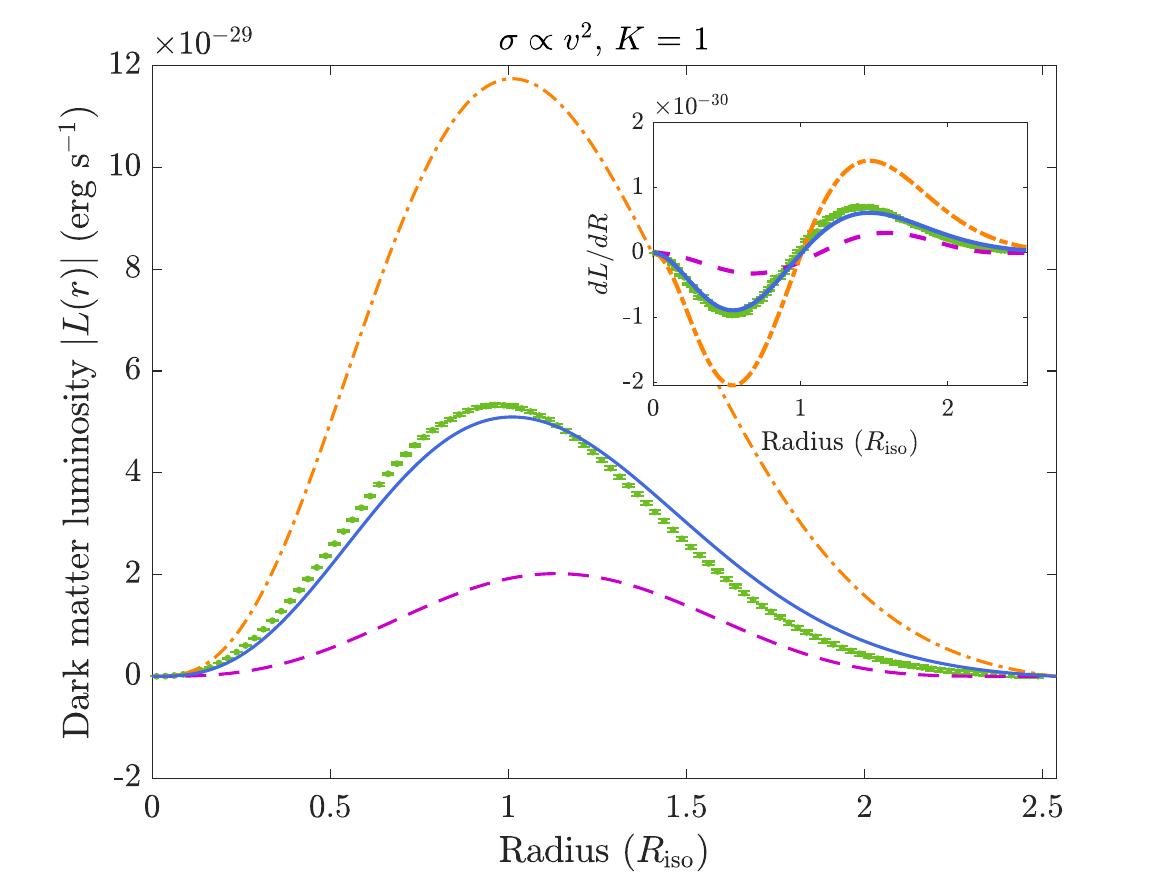}\includegraphics[width=0.55\textwidth]{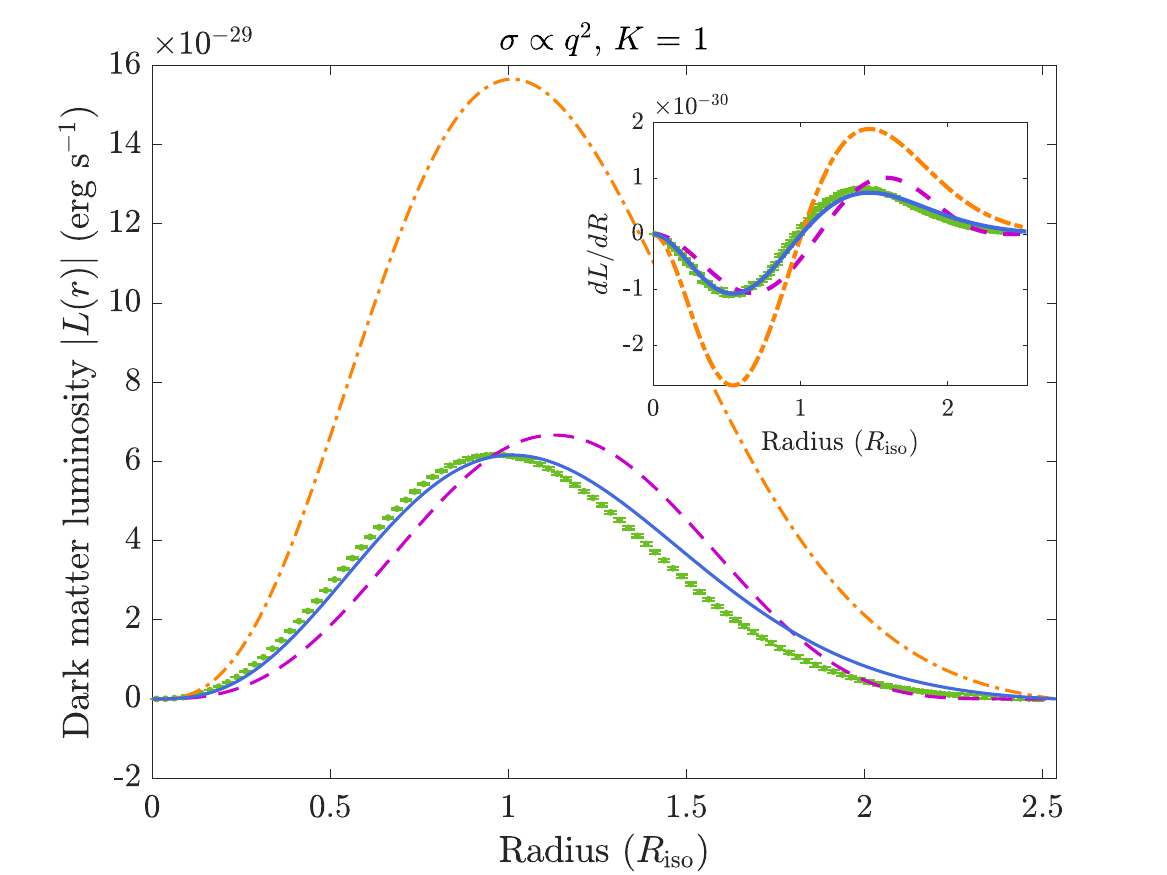}\\
    \hspace*{-0.8cm}\includegraphics[width=0.55\textwidth]{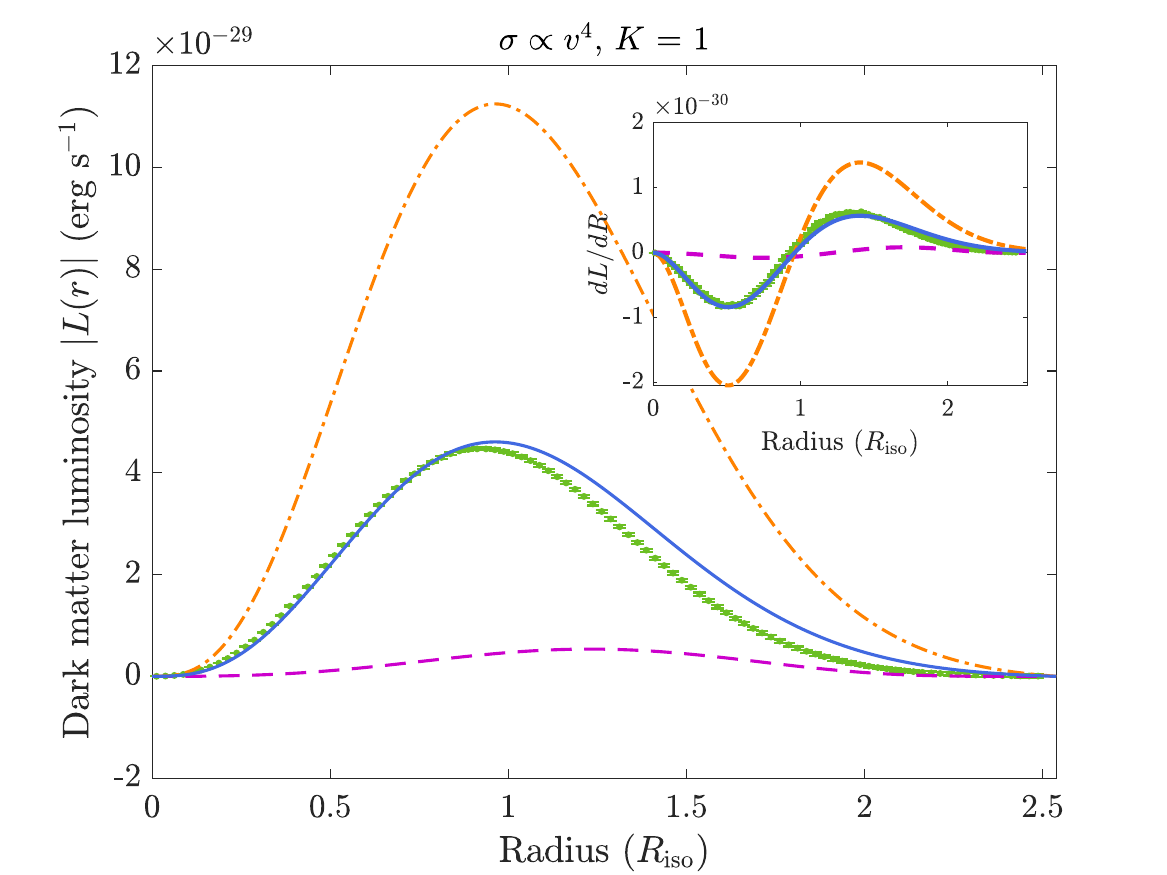}\includegraphics[width=0.55\textwidth]{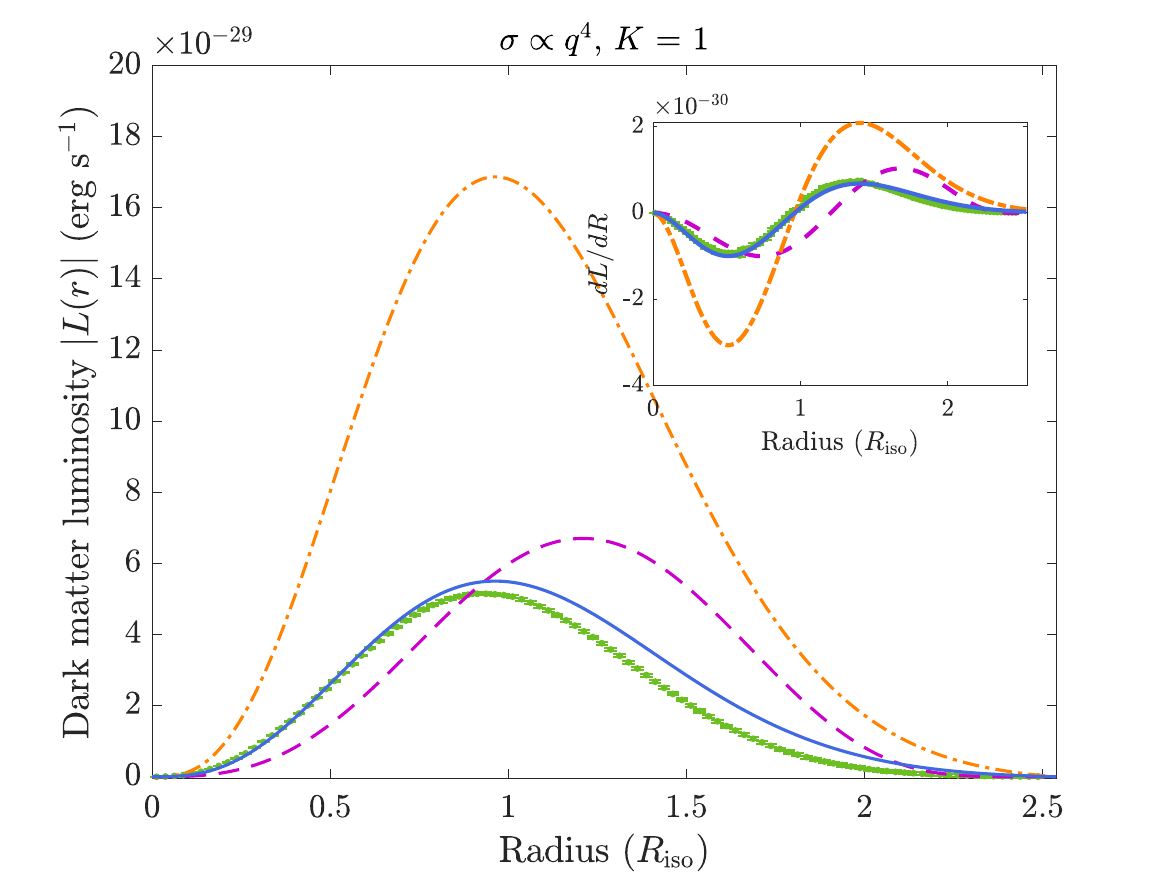}\\
     \includegraphics[width=0.8\textwidth]{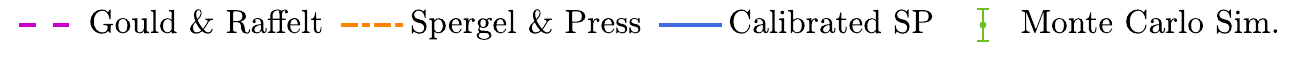}\\
    \caption{Transported heat in the \textit{idealized} simulation (simple harmonic oscillator potential, constant density and linear temperature gradient) for elastic scattering cross sections that depend on $v^{2n}$ (left) or $q^{2n}$ (right). Line styles as in Fig.\ \ref{fig:idealconduction_constant}. }
    \label{fig:idealconduction_vnqn}
\end{figure}

\afterpage{\clearpage}

Fig.\ \ref{fig:idealconduction_vnqn} is similar to Fig.\ \ref{fig:idealconduction_constant}, except that it shows $L(r)$ for $K=1$ in every panel, for cross sections that scale as $v^{2n}$ (left), and $q^{2n}$ (right).  Rows from top to bottom respectively show $n = -1, 1$ and 2. The derivatives $dL/dr$ are shown as insets, in the same units as in the right-hand panels of Fig.\ \ref{fig:idealconduction_constant}. Predicted curves from each theoretical heat transport scheme are again shown, with each curve appropriately adapted to the different cross section models following Eqs.\ (\ref{eq:SPgeneral}--\ref{eq:SPQintegral}) for SP, and using the results of Ref.\ \cite{Vincent13} for GR. The SP behaviour is very similar to the constant case, overestimating transport by a factor of two in the Knudsen regime, and failing to account for local transport for $K \ll 1$. Some of the GR curves match results relatively well (e.g. $q^2$), but they are generally not a good match to the explicit simulations, sometimes over or under-predicting the luminosity by several orders of magnitude, and generally failing to accurately predict the location of $r_\chi$.

\begin{figure}[p!]
  \centering
      \includegraphics[width=0.995\textwidth]{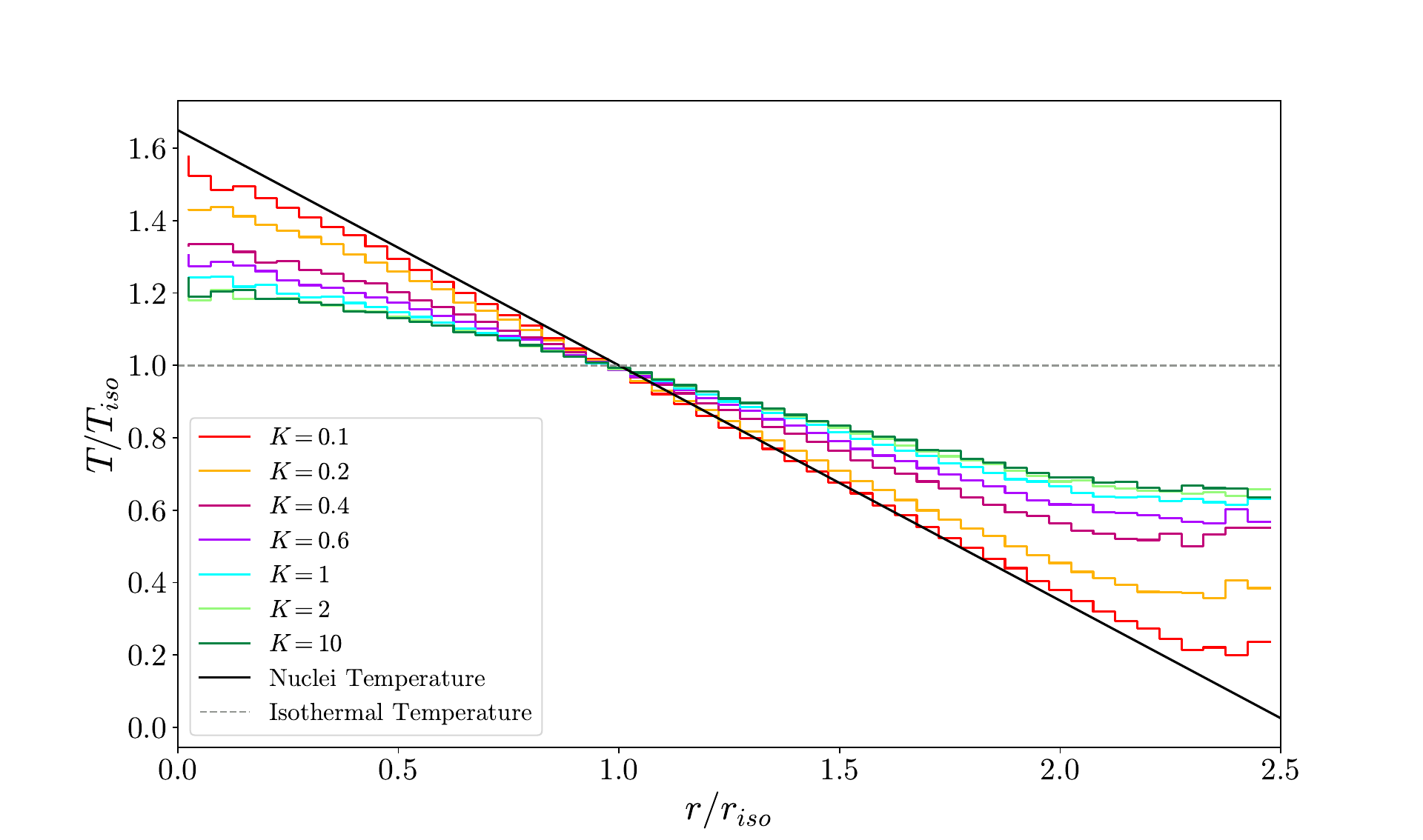}
        \caption{Effective temperature $T_\mathrm{eff}$ (Eq.\ \ref{eq:Teff}) of DM with velocity- and momentum- independent interactions as a function of radius for different Knudsen numbers in the \textit{idealized} simulations. For low $K$, the DM approaches local thermal equilibrium with the nuclei. At large $K$, the DM tends towards the isothermal distribution assumed in the Spergel \& Press formalism, but never actually reaches it.
        \label{fig:T}
        }
\end{figure}

In order to understand the behaviour of DM energy transport in our simulations, it is instructive to consider the effective DM temperature $T_\mathrm{eff}$, given by
\begin{equation}
T_\mathrm{eff} = \frac{2\langle E_{k}\rangle}{3k_\mathrm{B}},
\label{eq:Teff}
\end{equation}
where $\langle E_{k}\rangle $ is the time-averaged DM kinetic energy, with angled brackets representing an average within each radial bin. Figure \ref{fig:T} shows $T_\mathrm{eff}$ as a function of radius for different Knudsen numbers $K$, for $n=0$ (constant cross-section). At small $K$, as expected, the DM temperature closely follows that of the background nuclei, approaching local thermal equilibrium. Here, the DM interacts a sufficient number of times to thermalise before it diffuses to a region of different temperature. As $K$ is increased, the increase in inter-scattering distance causes the effective DM temperature to deviate from that of the nuclei, tending towards a constant temperature corresponding to the statistically-averaged isothermal temperature $T_\mathrm{iso}$, which can be obtained by solving Eq.\ \ref{eq:lumcondition}, and for this system, has a value of $\sim~1$K.

\begin{figure}[p!]
    \centering
    \includegraphics[width = 0.995\textwidth]{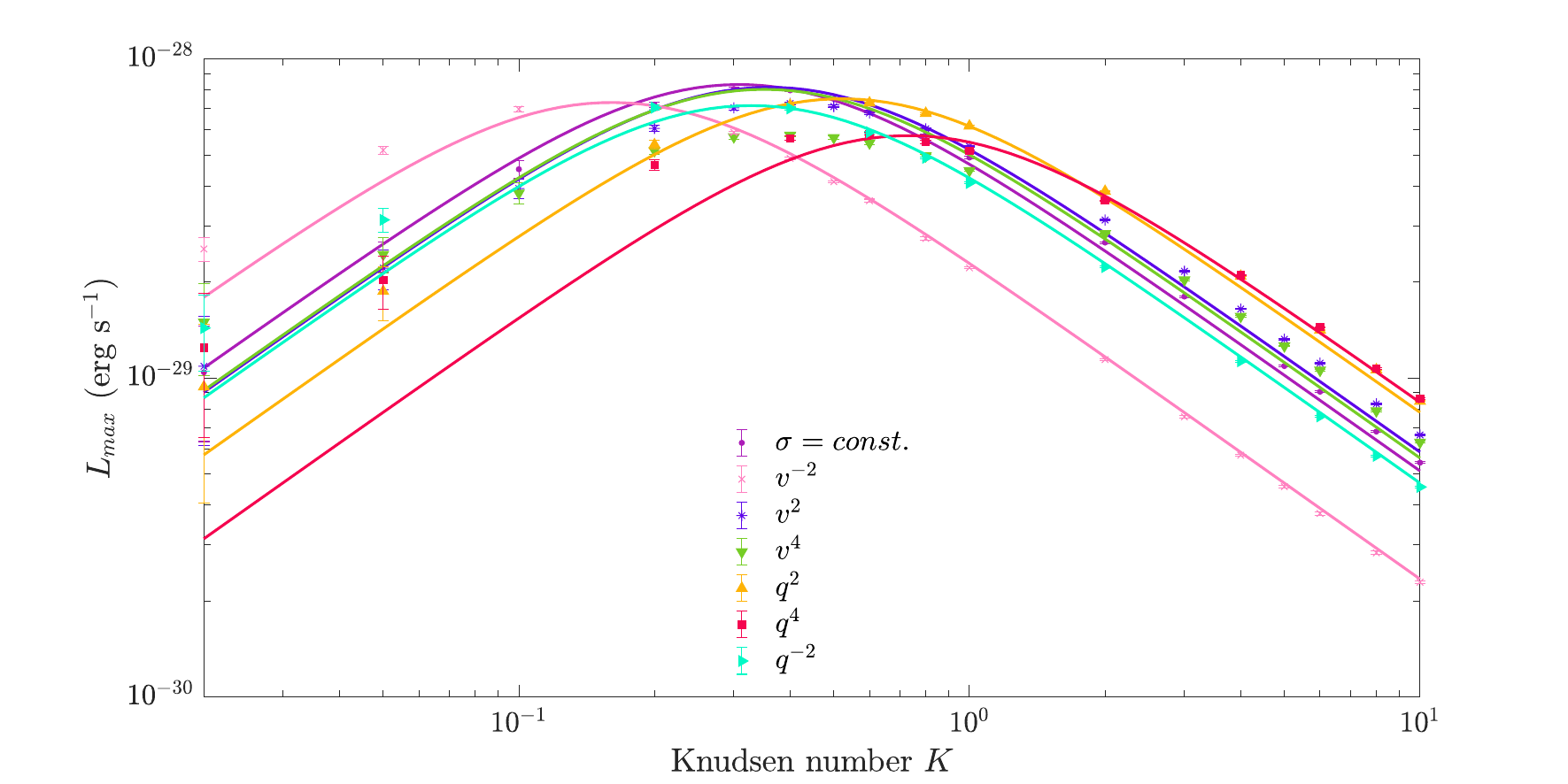}
    \caption{Maximum transported luminosity in the \textit{idealized} simulations, with constant stellar density and $m_\chi = m_p$, as a function of the Knudsen number $K$.  Here we show results for constant dark matter-nucleon cross sections as well as those proportional to $v^{2n}$ and $q^{2n}$. Data points with error bars are results of our simulations, and curves show maximum luminosities predicted by the calibrated SP transport scheme \eqref{eq:SPrescale}. $K_0$ is given in Table \ref{tab:K0real}.}
    \label{fig:maxlumGR}
\end{figure}

\afterpage{\clearpage}

Fig.\ \ref{fig:T} provides some insight into the level of accuracy that we can expect from our conduction formalism. As $K$ rises above $\sim$$0.2$, the average DM temperature departs noticeably from $T(r)$, indicating a breakdown of the conditions behind the GR perturbative expansion of the BCE \eqref{eq:BCE}. As noted in Ref.\ \cite{Gould1990CosmionLimit}, $T_\mathrm{eff}$ does not converge to the isothermal limit, but rather asymptotes to an intermediate value at large $K$, indicating that the isothermal assumption is never strictly realized. This should not be too surprising, as the isothermal solution $f(E) \propto e^{E/T_{\rm iso}}$  with  constant $T_{\rm iso}$ is the exact solution to the Collision\textit{less} Boltzmann Equation $DF = 0$ , whereas $T_{\rm iso}$ in the SP approach is defined by the (nonzero) collision operator on the right-hand side of the BCE.

To end this section, we show in Fig.\ \ref{fig:maxlumGR} the maximum luminosity across all $\sigma \propto v^{2n}$, $q^{2n}$ models. We find that they all show very similar behaviour with $K$, scaling proportionately to $K$ in the LTE regime, and to $K^{-1}$ for large $K$. In contrast to what is expected on the basis of a dipole-based perturbative solution to the BCE \cite{Vincent13}, the peak luminosities are very close to one another, suggesting that heat transport is limited by kinematics, and not by the form of the interaction cross section. We also find that the Knudsen transition occurs at slightly different values of $K$ for each model. This could be because we chose to define $K$ using the mean inter-scattering distance at $r = 0$, as velocity dependence introduces an implicit temperature (and therefore radial) dependence to $l_\chi$, even when the density of targets is constant.

\subsection{\textit{Realistic} simulations}
\label{sec:real}

We now turn to a more realistic setup for our Monte Carlo simulations.
We  study energy conduction in the Sun using the temperature profile and chemical composition of the present-day AGSS09ph standard solar model \cite{Serenelli2009NewRevisited}.\footnote{See \url{https://www.ice.csic.es/personal/aldos/Solar_Models.html} for up-to-date SSM data files.} In order to keep the orbital trajectories analytically tractable, we keep the simple harmonic oscillator (SHO) gravitational potential, rather than computing it self-consistently from the stellar density. We set the scale of the gravitational potential, $\rho_\mathrm{sho}$, to 148.9 \si{g.cm^{-3}}. This corresponds to the density in the core of the Sun and amounts to approximately 100 times the average solar density. Because of the high core density, most of the DM in a real star will be localized very near the core. For DM masses greater than $\sim 10$\,GeV, both the SP and GR distributions predict that 99.9\% of the DM is confined below $0.1 R_\odot$, where the potentials differ by $\lesssim 20\%$. For $m_\chi = 10$\,GeV, the predicted luminosity using the true Solar potential is 12\% higher, and peaks at 0.43 $R_\odot$, rather than 0.41 $R_\odot$ for the SHO potential. The two models deviate more from one another for lower DM masses, as the distribution more easily samples higher radii, where the two potentials begin to differ.
We will consider DM masses of $5-20$\,GeV.  These are above the threshold for evaporation, which is about 4\,GeV in the Sun, but should be larger for the potential that we consider here. Although our simulation is capable of dealing with evaporation events, we recorded none in the duration of the realistic study. We again consider scattering with a single isotope, which we model by assuming spin-dependent (SD) scattering (Eq.\ \ref{eq:SD}) on hydrogen only.

\begin{figure}[p]
    \centering
    \includegraphics[width=0.5\textwidth]{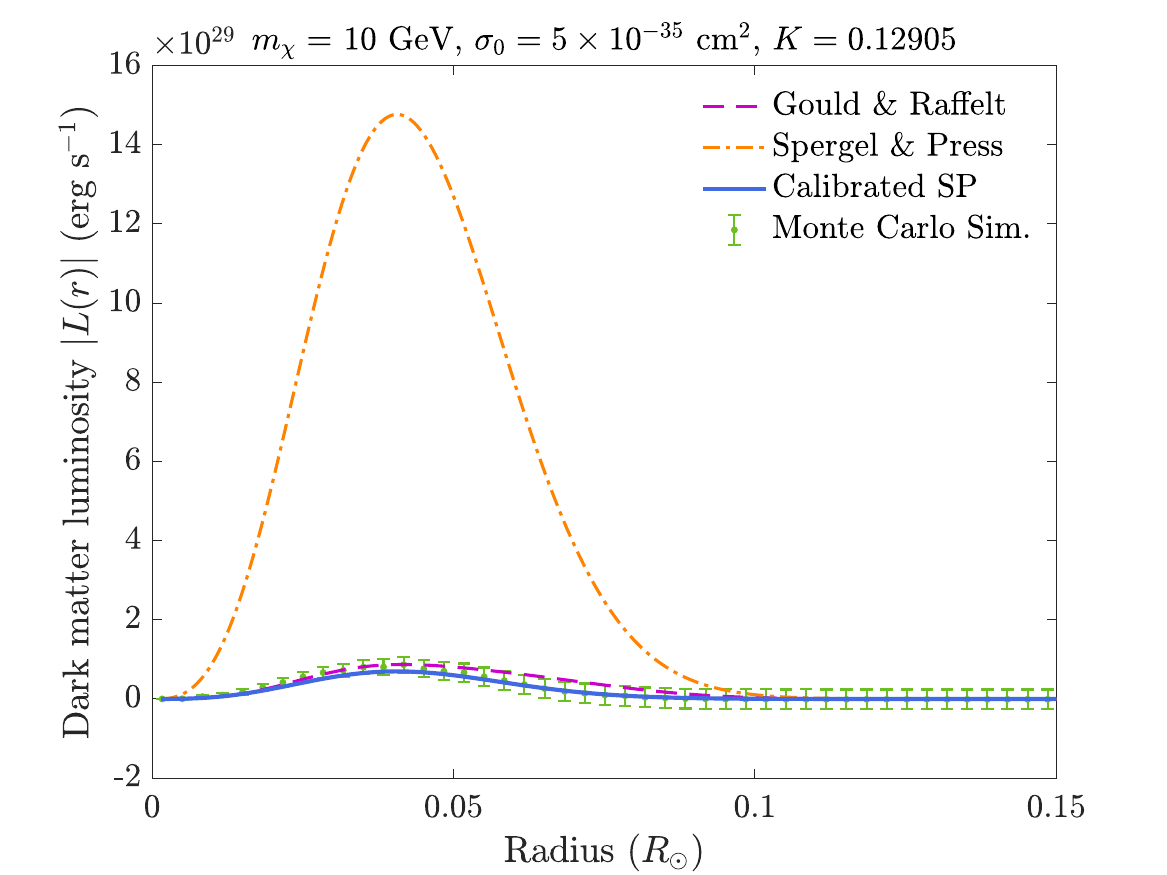}\includegraphics[width=0.5\textwidth]{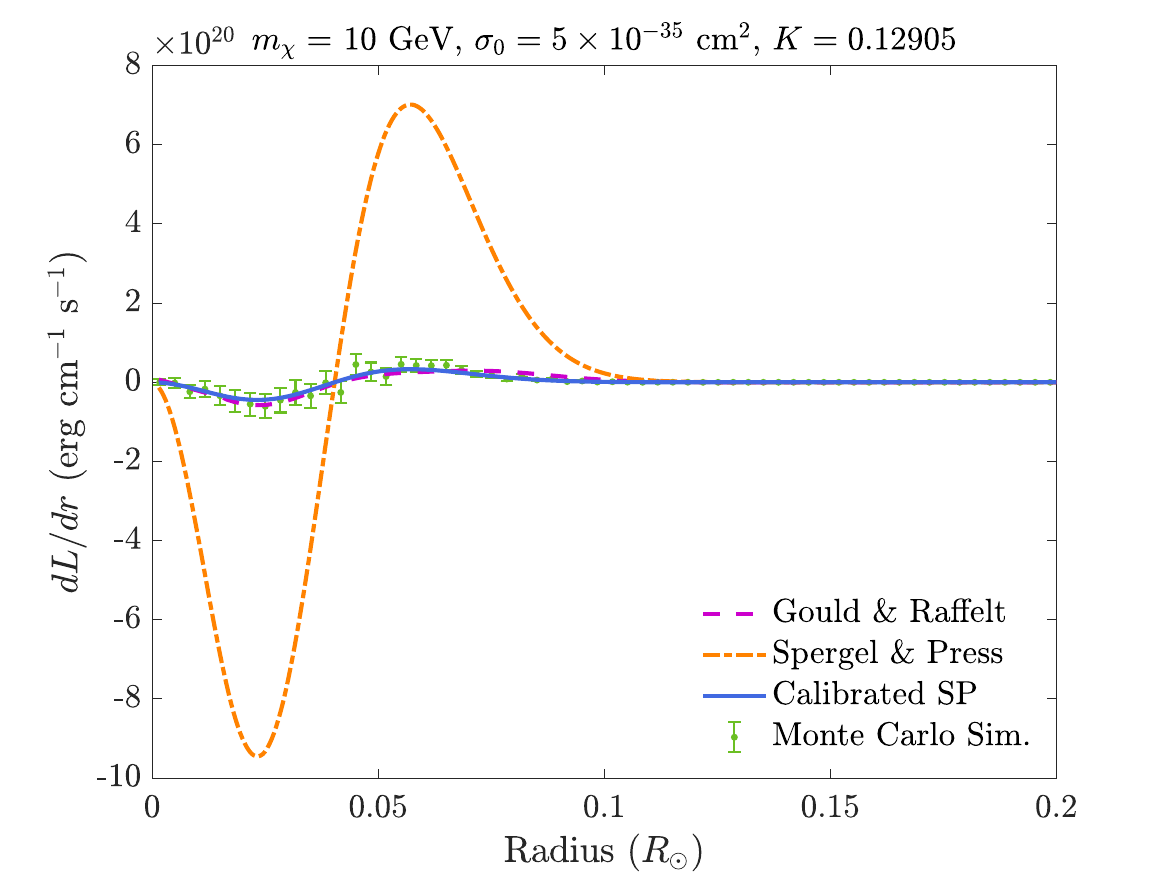}\\
    \includegraphics[width=0.5\textwidth]{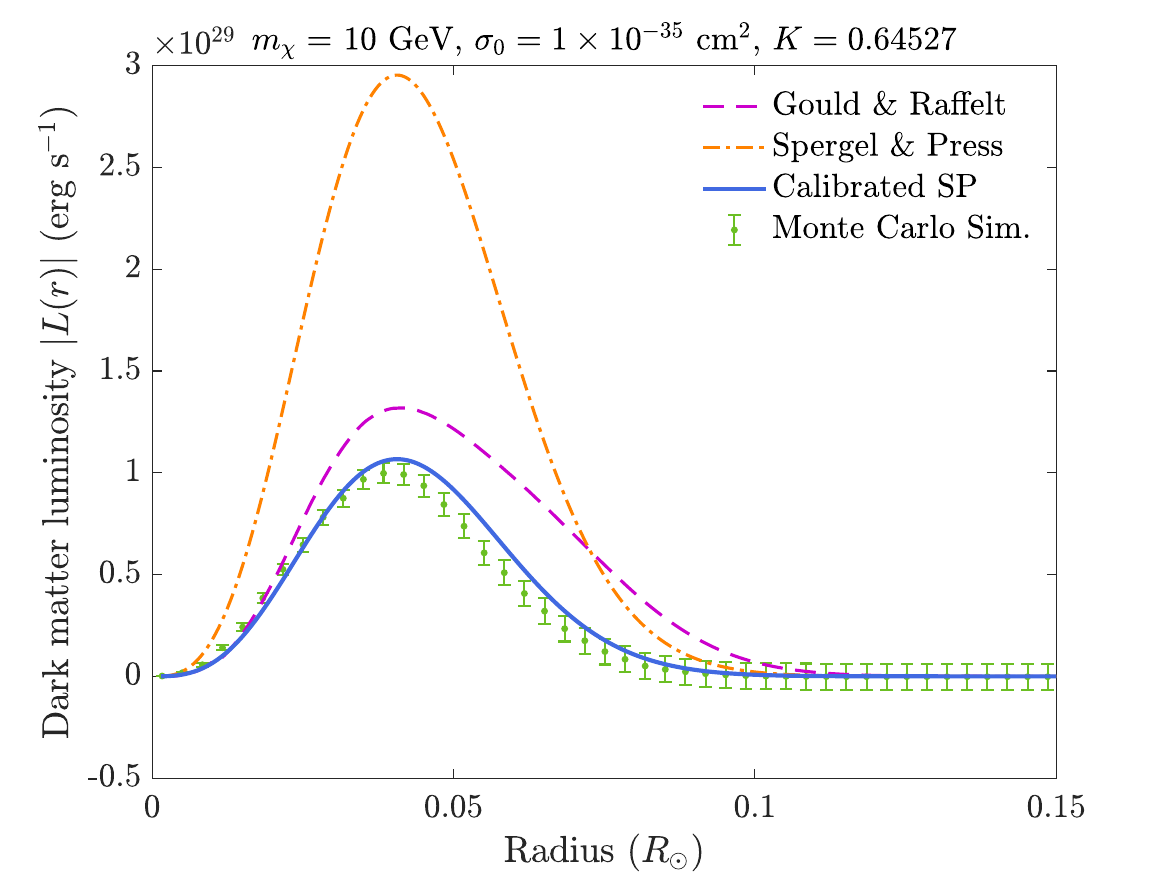}\includegraphics[width=0.5\textwidth]{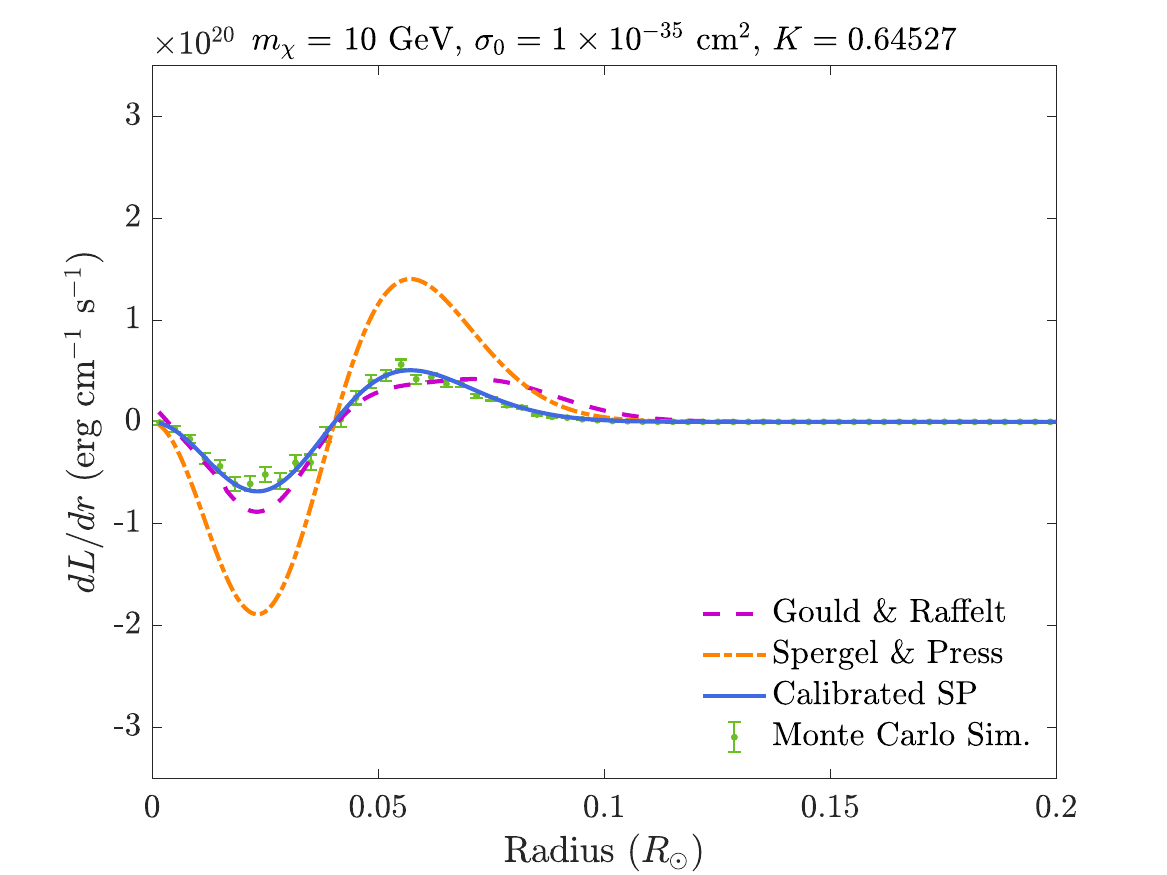}\\
    \includegraphics[width=0.5\textwidth]{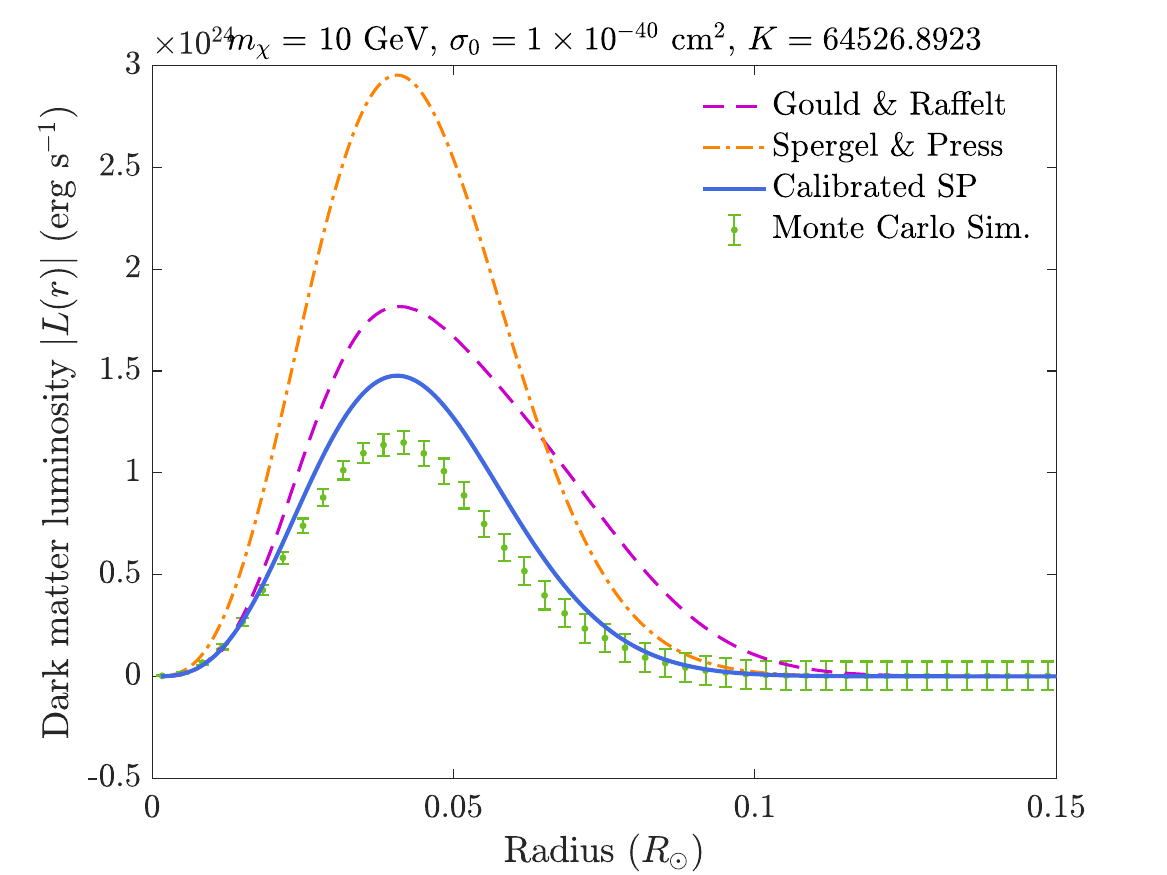}\includegraphics[width=0.5\textwidth]{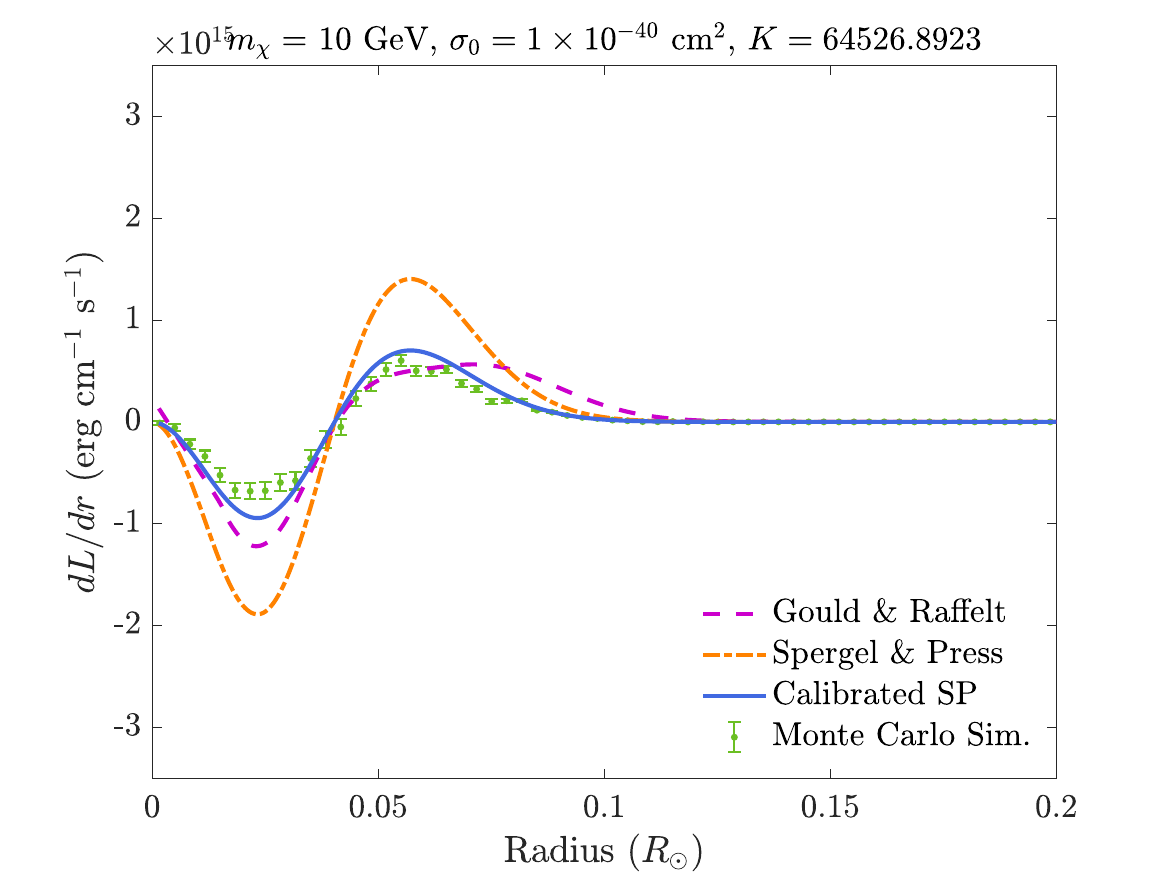}\\
    \caption{Transported heat in the \textit{realistic} solar model (simple harmonic oscillator potential, using SSM profiles for target density and temperature profiles), for a constant elastic scattering cross section, $m_\chi = 10$\,GeV and $n_\chi/n_b = 10^{-15}$. Results are shown for the two commonly-used frameworks (Spergel \& Press \cite{Spergel1985EffectInterior}, Gould \& Raffelt \cite{Gould1990,Scott2009,Vincent13}), as well as the `Calibrated SP' method developed here. Green data points represent results of our Monte Carlo simulations. Cross sections and corresponding Knudsen numbers $K$ are given for each of the three plotted cases.  }
    \label{fig:realconduction_constant}
\end{figure}

In Fig.\ \ref{fig:realconduction_constant} we show the radial luminosity profiles (left) and $dL/dr$ (right) for a selection of constant cross-section values. These range from conduction near the LTE limit ($\sigma_{0} = 5\times 10^{-35}$ cm$^2$, $K = 0.12$), through the transition ($\sigma_{0} = 10^{-35}$ cm$^2$, $K = 0.65$) and well into the Knudsen regime at a cross-section value near to the current experimental limit ($\sigma_{0} = 10^{-40}$ cm$^2$, $K = 64 \times 10^4$). As in Figs.\ \ref{fig:idealconduction_constant} and \ref{fig:idealconduction_vnqn}, we also include predictions from the two theoretical heat transport schemes. The GR method does as well as the \textit{idealized} setup, remaining within error bars in the LTE regime.  In the Knudsen limit, GR overestimates the peak luminosity by only about $10$--$20\%$, but fares far worse at large radii, substantially overpredicting the effect of DM heat transport outside the stellar core. In contrast, the SP method behaves much as in the \textit{idealized} cases, overestimating the luminosity by a factor of $\sim$$2$ in the Knudsen limit, and becoming steadily more inaccurate as $K$ decreases. However, the shape predicted by the SP formalism is in better general agreement with the results of the simulations than is the shape predicted by GR.

\begin{figure}[p]
    \centering
    \includegraphics[width=0.5\textwidth]{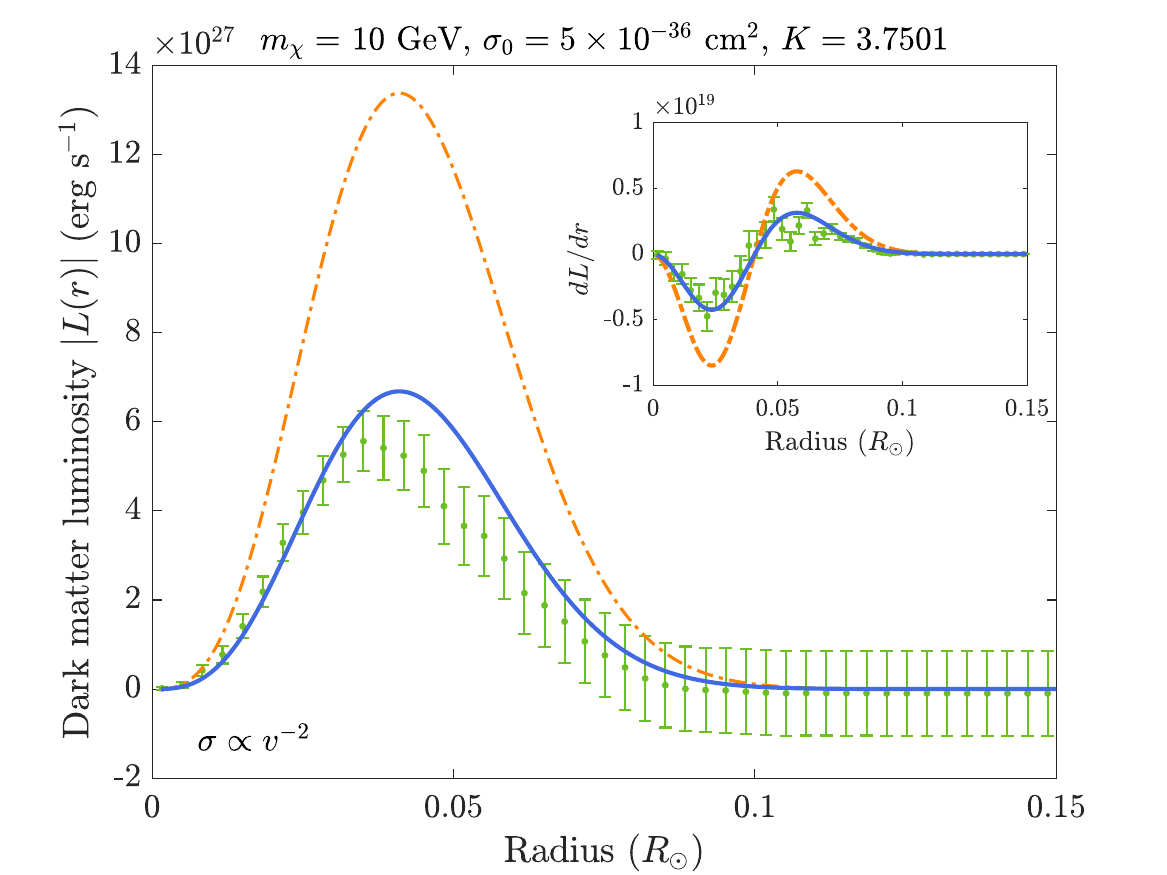}\includegraphics[width=0.5\textwidth]{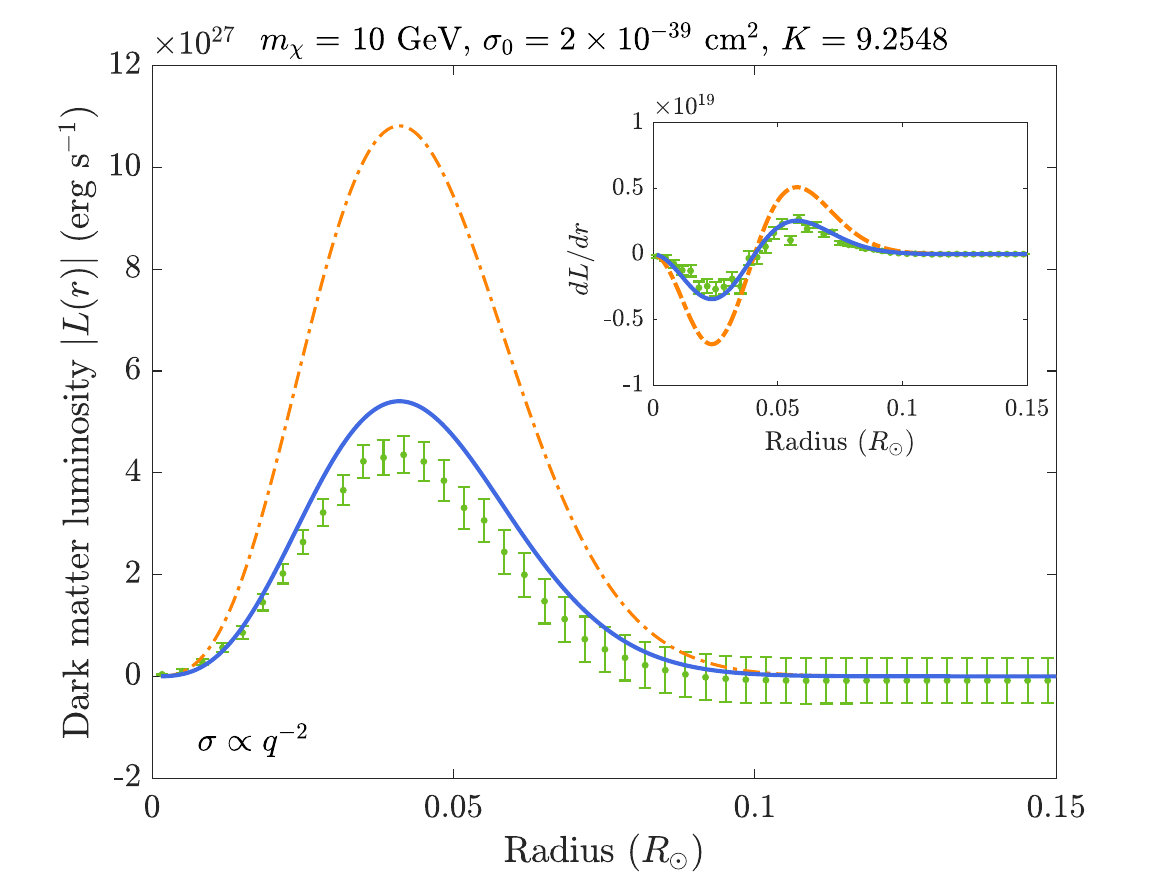}\\
    \includegraphics[width=0.5\textwidth]{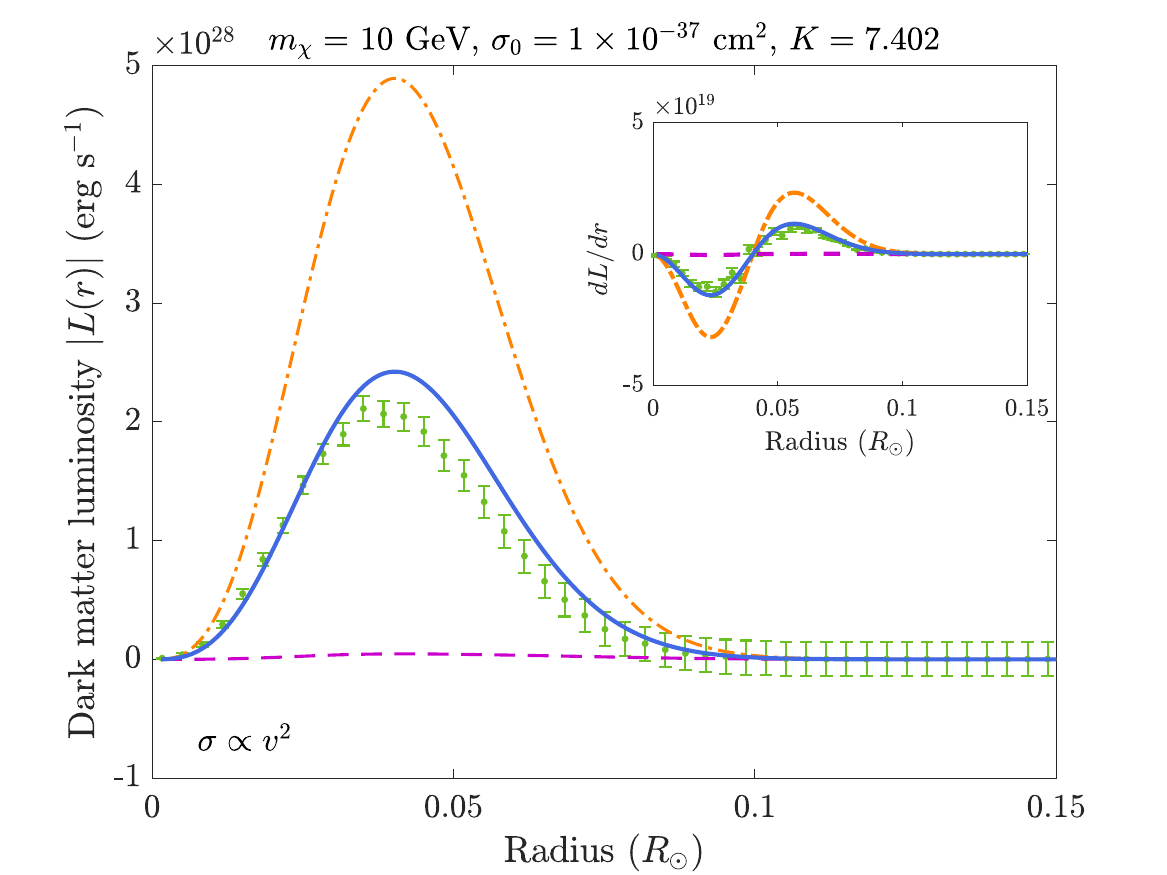}\includegraphics[width=0.5\textwidth]{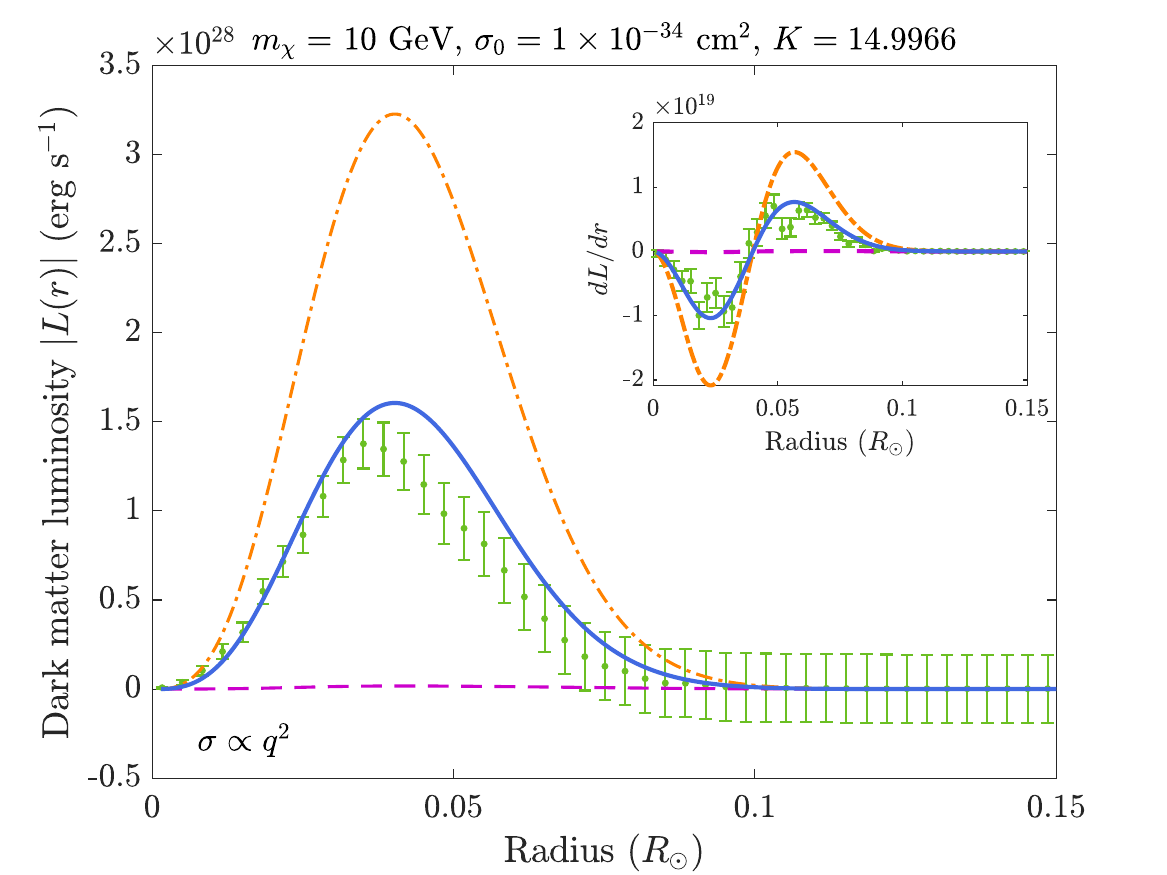}\\
    \includegraphics[width=0.5\textwidth]{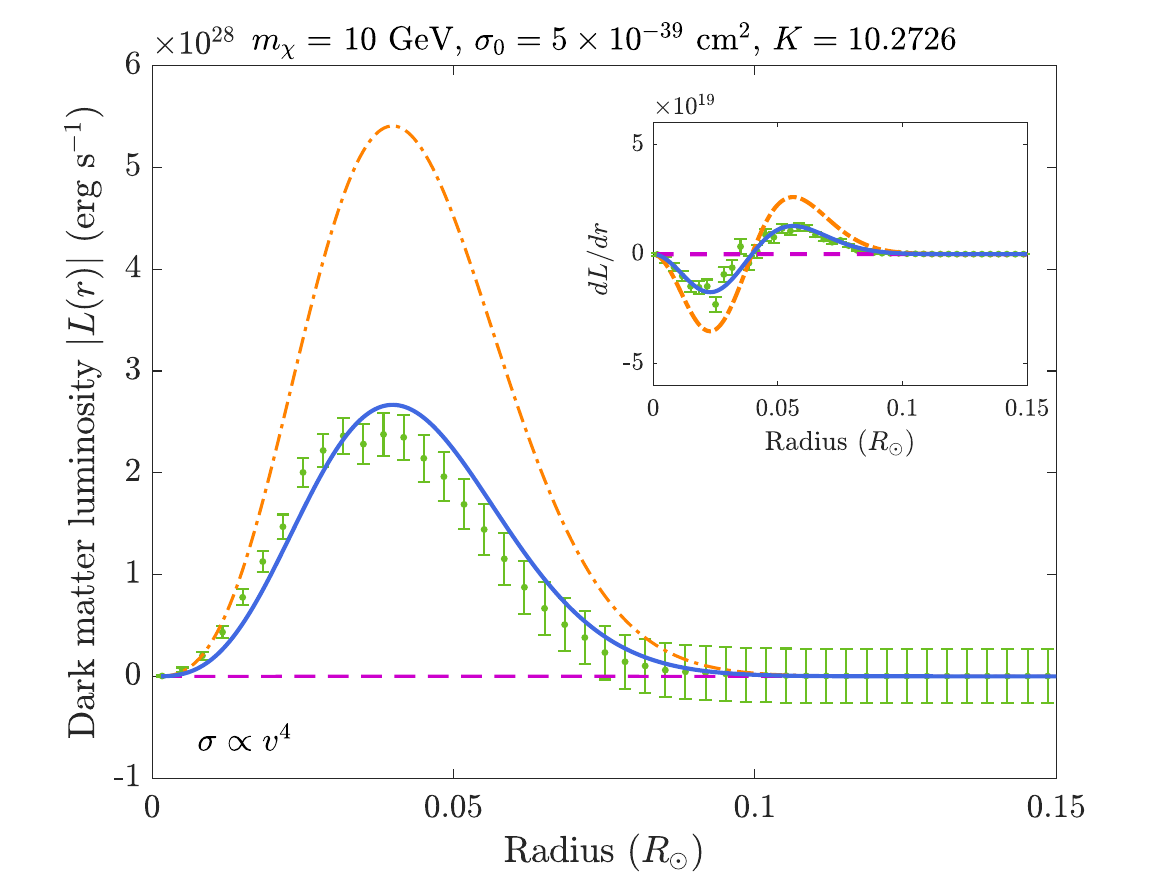}\includegraphics[width=0.5\textwidth]{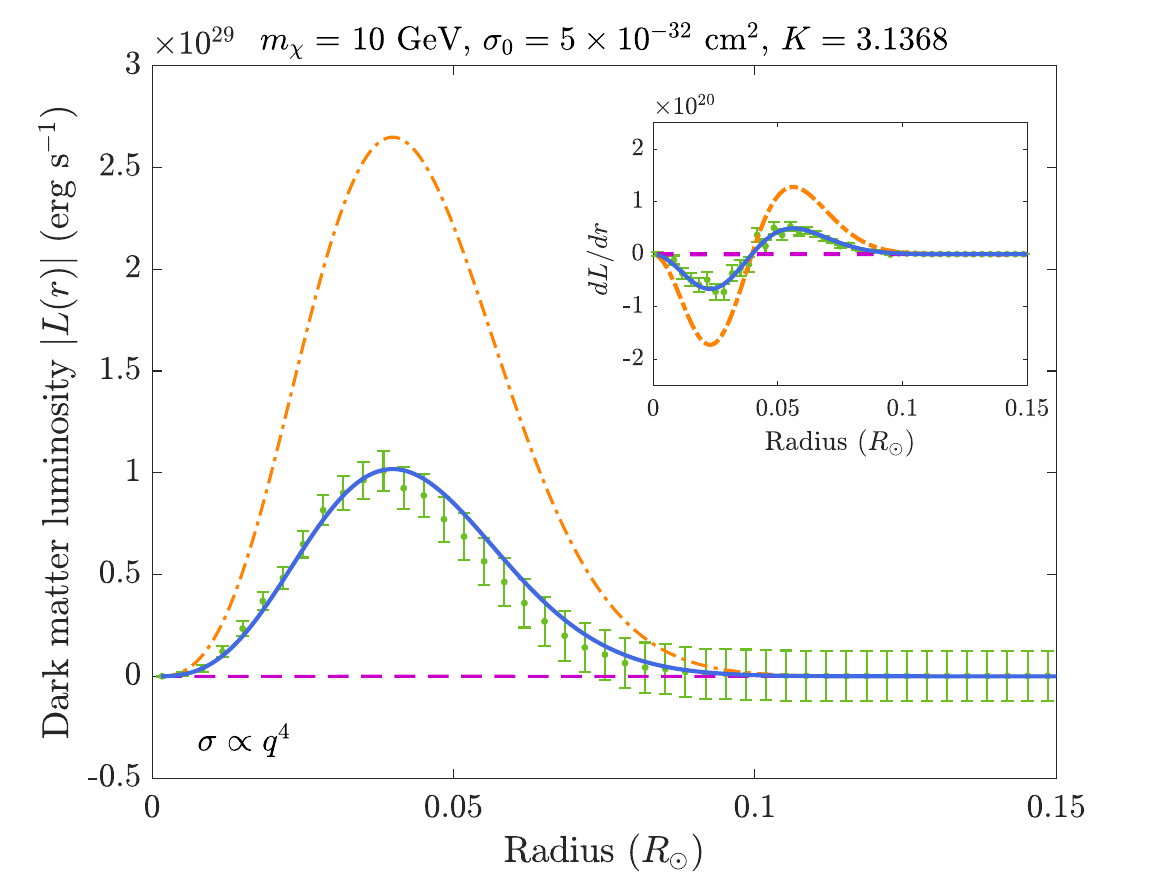}\\
     \includegraphics[width=0.8\textwidth]{figures/legend.pdf}\\
    \caption{
    Transported heat in the \textit{realistic} solar model (simple harmonic oscillator potential, using SSM profiles for target density and temperature profiles) for elastic scattering cross sections that depend on $v^{2n}$ (left) or $q^{2n}$ (right).  As in Fig.\ \ref{fig:realconduction_constant}, $m_\chi = 10$\,GeV and $n_\chi/n_b = 10^{-15}$. Line styles are as in Fig.\ \ref{fig:realconduction_constant}. The GR line has been removed from the top panels, as it overestimates conduction by an order of magnitude. }
    \label{fig:realconduction_vnqn}
\end{figure}

\begin{figure}[t]
    \centering
    \hspace*{-1.6cm}\includegraphics[width=1.2\textwidth]{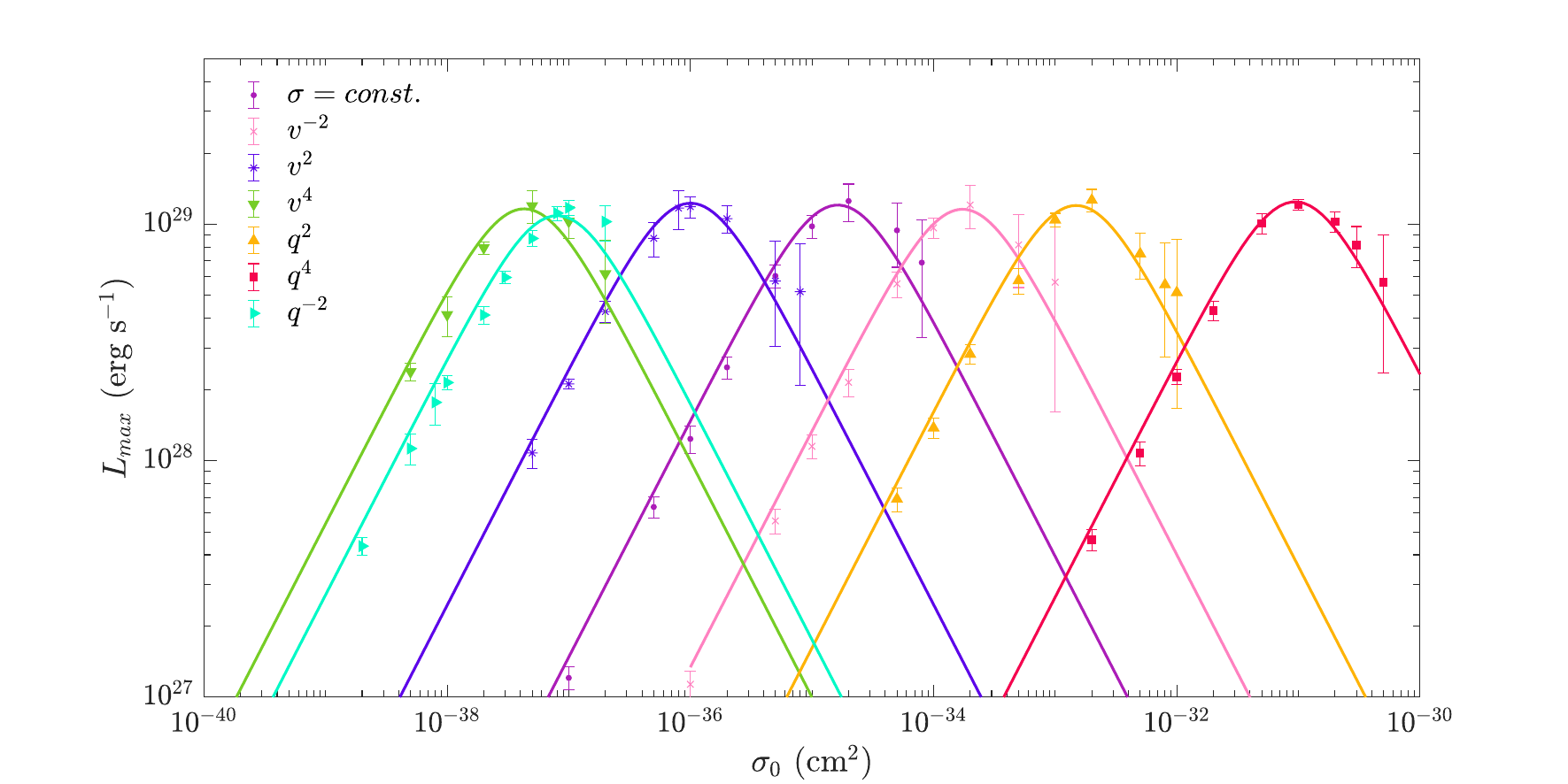}
    \caption{Maximum luminosity as a function of $\sigma_0$ in the \textit{realistic} simulations, for spin-dependent interacting DM with $m_\chi = 10$\,GeV. Data points are maximum luminosities estimated from Monte Carlo simulations and solid curves are luminosities computed with Eq.\ \eqref{eq:SPrescale}, where the value of $K_0$, provided in Table \ref{tab:K0real} have been fitted. }
    \label{fig:maxlumReal}
\end{figure}

\begin{figure}[!ht]
    \centering
    \includegraphics[width=0.5\textwidth]{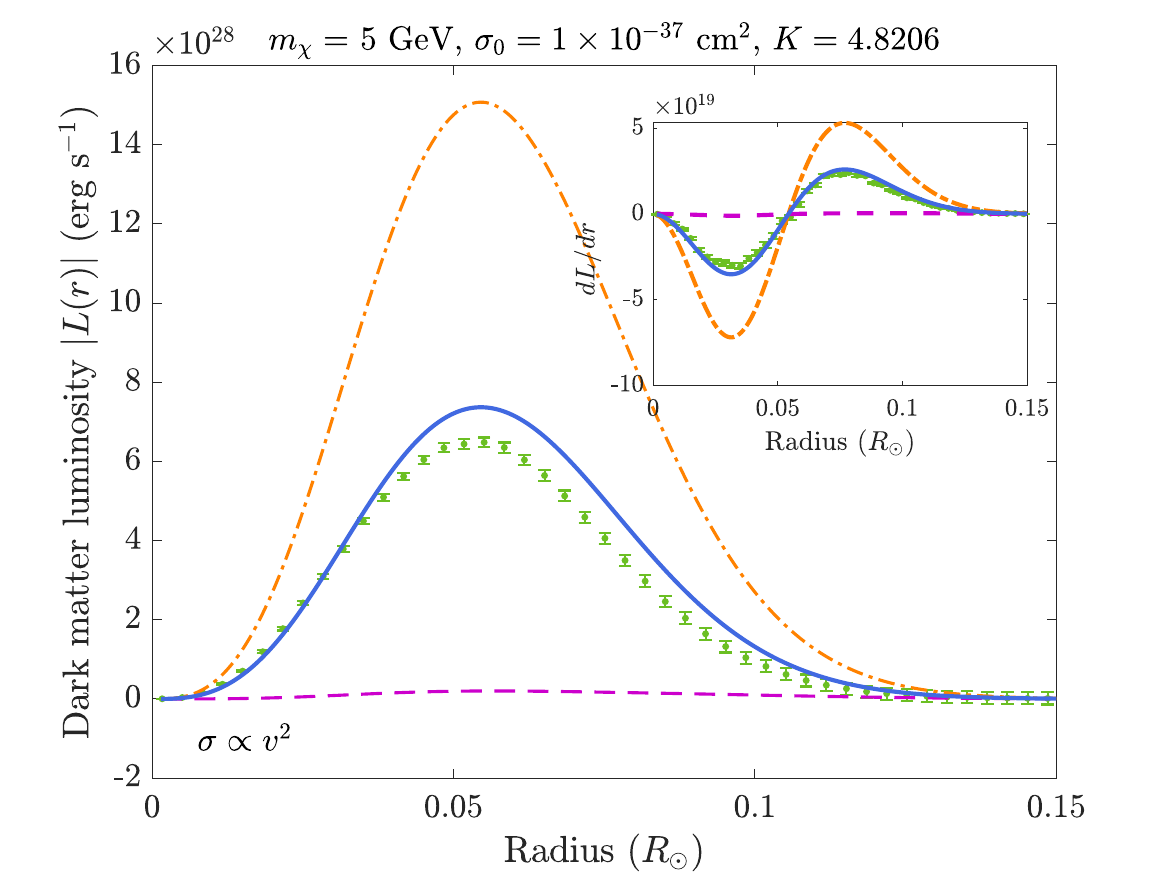}\includegraphics[width=0.5\textwidth]{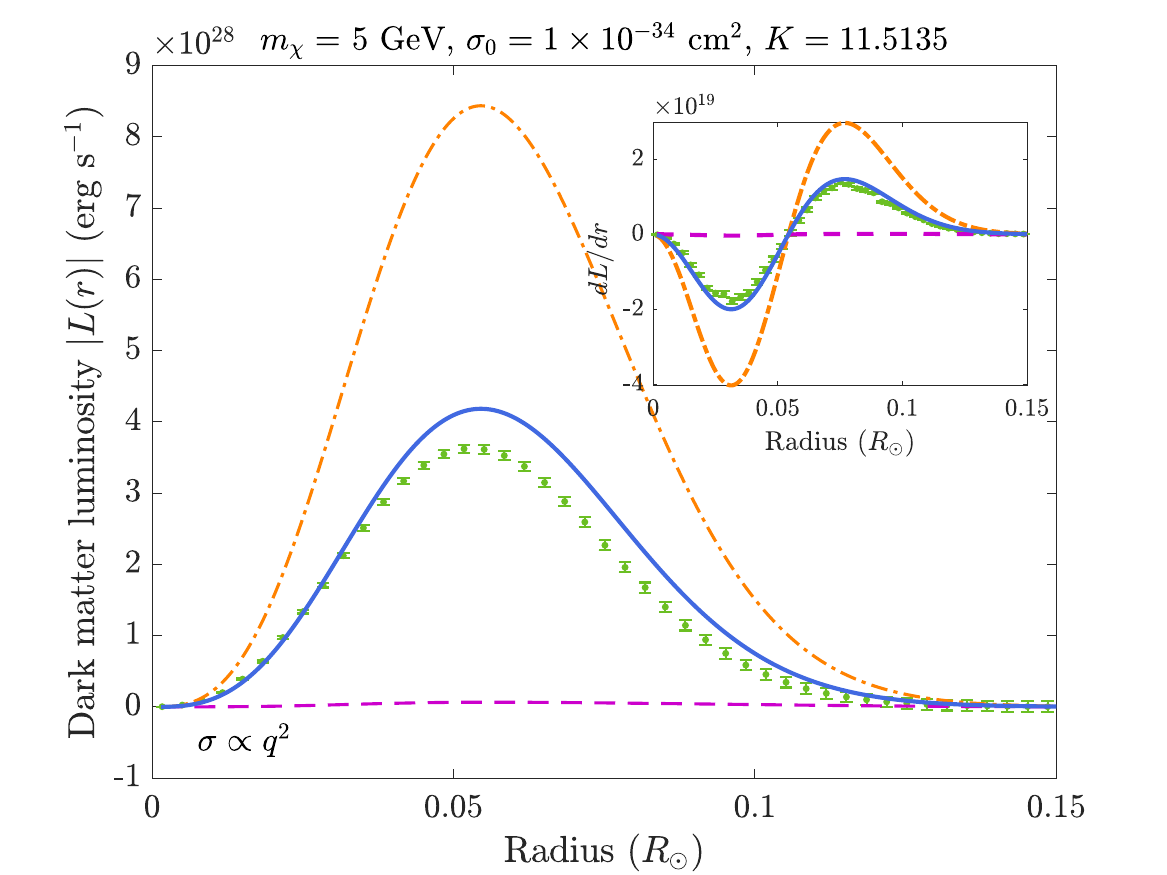}\\
    \includegraphics[width=0.5\textwidth]{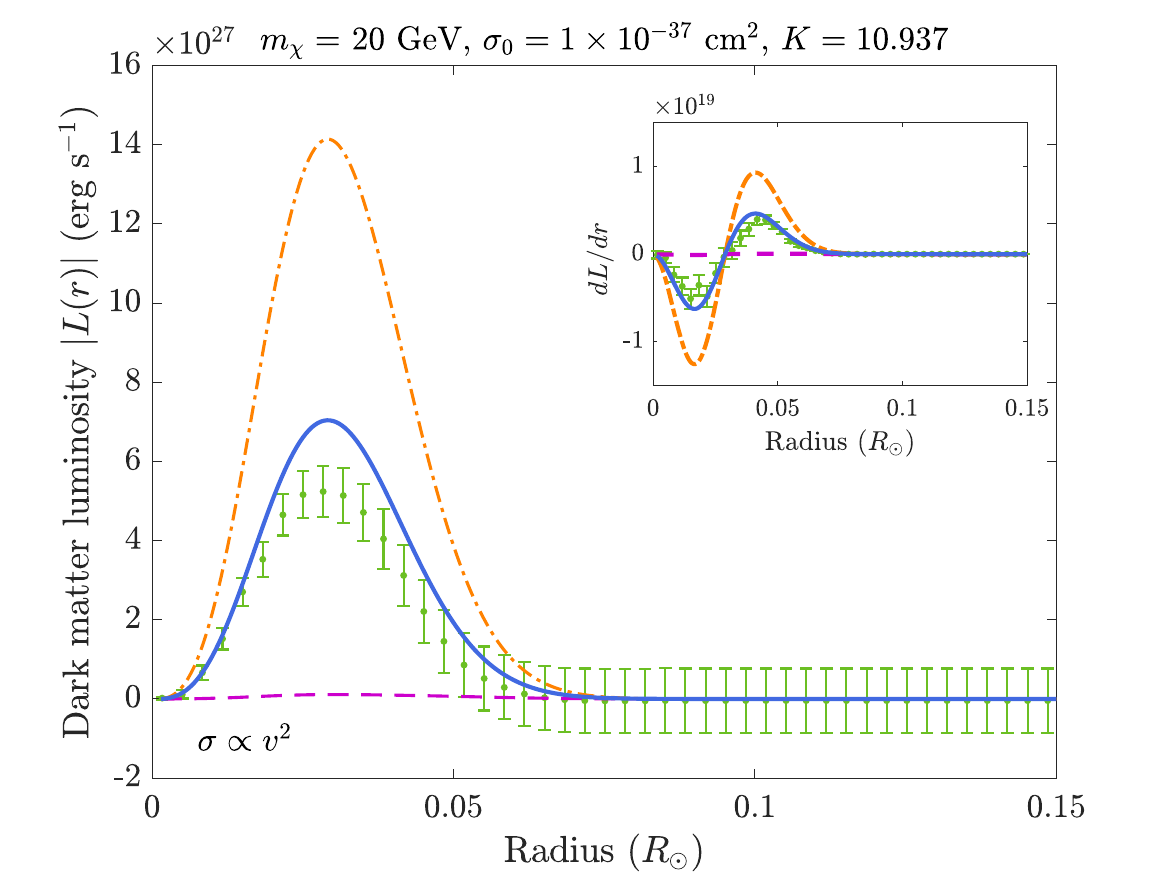}\includegraphics[width=0.5\textwidth]{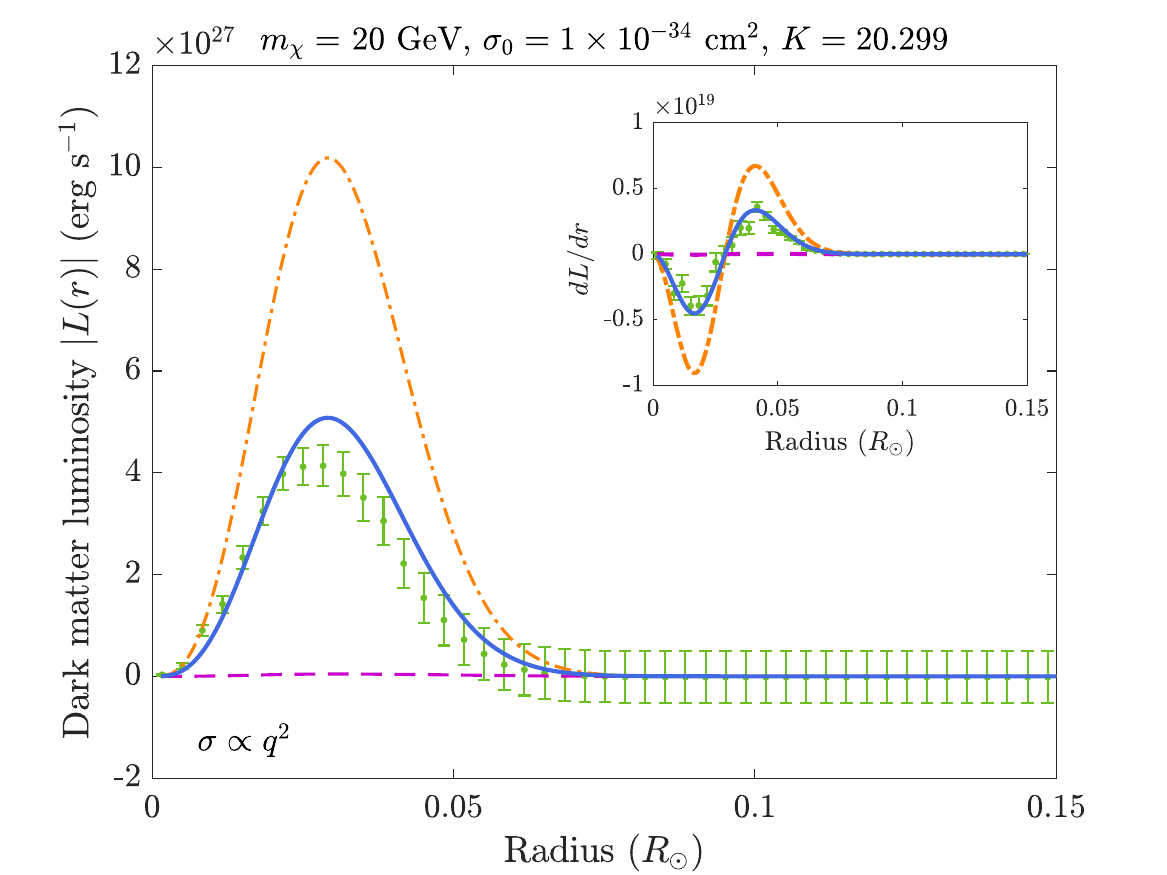}\\
    \includegraphics[width=0.8\textwidth]{figures/legend.pdf}
    \caption{
    Transported heat in the \textit{realistic} solar model (simple harmonic oscillator potential, using SSM profiles for target density and temperature profiles) for elastic scattering cross sections that depend on $v^{2}$ (left) or $q^{2}$ (right), for different masses than those shown in Fig.\ \ref{fig:realconduction_constant}. Top: $m_\chi = 5$\,GeV; bottom: $m_\chi = 20$\,GeV. Lines as in previous figures.}
    \label{fig:realconduction_v2q2mass}
\end{figure}

\afterpage{\clearpage}

Fig.\ \ref{fig:realconduction_vnqn} shows an example for each of the $v^{2n}$ and $q^{2n}$ cases. As in the \textit{idealized} case (Fig.\ \ref{fig:idealconduction_vnqn}), after being adapted to heat transport by more general nuclear scattering cross-sections the SP formalism provides a very good description of the shape of the luminosity and energy transport curves.  The radially and Knudsen-corrected GR scheme fares even worse than in the constant case, underpredicting $L(r)$ by several orders of magnitude for positive $n$, and overpredicting it for negative $n$.

The maximum luminosity for each model is shown as a function of the reference cross section $\sigma_0$ in Fig.\ \ref{fig:maxlumReal}, where each data point corresponds to a single \textit{realistic} simulation. As in the \textit{idealized} case, the maximum luminosity achievable is independent of the interaction type. The Knudsen transition occurs at very different values of $\sigma_0$ due to the (arbitrary) choices of $v_0$ and $q_0$. In terms of Knudsen number, the peak locations range from $K \sim 0.1$ for $v^{-2}$ scattering to $K \sim 1.7$ for $q^{4}$ scattering. We will discuss these in more detail in the following section.

We finally show some examples of heat transport with different masses in Fig.\ \ref{fig:realconduction_v2q2mass}, for the $v^2$ (left) and $q^2$ (right) cases.  Here we see that even with $m_\chi =5 $\,GeV (top) or $m_\chi =20 $\,GeV (bottom), the behaviour is largely similar to the $m_\chi =10 $\,GeV case that we have considered so far: at large $K$, the SP prediction is too large by about a factor of 2, but the radially and Knudsen-corrected GR prediction is too low by orders of magnitude.

\section{A universal dark matter heat transport model}
\label{sec:discussion}

In the previous section, we examined the heat transported radially by interacting DM particles in two very different setups: the \textit{idealized} constant density sphere with a linear temperature gradient and a scale radius of similar order to the stellar radius, and the \textit{realistic} case, using a reference SSM, but retaining a harmonic oscillator gravitational potential in order to allow trajectories to be parameterised analytically.

In both setups, we found that although the GR scheme was fairly robust for constant cross sections, the shape of the $L(r)$ curve was generally better described for all interaction types by the SP formalism.

Based on the above observation, we propose a simple, phenomenologically-calibrated expression for the transported luminosity.  This expression works reasonably well across all models (constant, $q^{2n}$, $v_{rel}^{2n}$), masses, and cross section values that we have tested.  It is also easy to implement into stellar simulations thanks to the absence of numerical derivatives.

We first take the Spergel \& Press luminosity $L_\mathrm{SP}$ \eqref{eq:SPint}, and rescale by a factor of $0.5$ in order to obtain sensible results in the limit $K\gg1$.  We then apply a modification factor in order to obtain the correct scaling of the overall luminosity with $K$.  The bare SP expression scales as $L \varpropto \sigma \varpropto K^{-1}$, which is appropriate in the isothermal regime.  In the LTE regime, we instead require that the luminosity scales as $L\varpropto K$.  We therefore apply a further correction factor that multiplies the luminosity by $K^2$ at small $K$.  The turnover point from the LTE to the isothermal behaviour is given by $K_0$, which is a model-dependent constant determined empirically from our Monte Carlo simulations.  The overall expression is then
\begin{equation}
\boxed{    L(r) =  \frac{0.5}{1+(K_0/K)^2}L_\mathrm{SP} (r).}
    \label{eq:SPrescale}
\end{equation}

The values of $K_0$ that we find from our simulations are tabulated in Tab.\ \ref{tab:K0real}. As noted previously, the deviation from a Knudsen transition at the expected $K \sim 1/3$ is likely due to the definition of $K$ at the centre of the star, which is not necessarily representative of the full radial interaction rate.

Predictions using this `calibrated SP' parameterisation are shown with solid blue lines in Figs.\ \ref{fig:idealconduction_constant}, \ref{fig:idealconduction_vnqn}, \ref{fig:realconduction_constant}, \ref{fig:realconduction_vnqn} and \ref{fig:realconduction_v2q2mass}. As seen in these figures, this parameterisation provides a far more accurate prediction of the transported luminosity and deposited energy than either of the existing conduction schemes. It works well in both \textit{idealized} and \textit{realistic} setups, and across interaction types and masses.

The range of applicability in terms of stellar mass and evolutionary stage of these tabulated values of $K_0$ has yet to be determined. For modelling the Sun, the \textit{realistic} values should apply directly, as our Monte Carlo setup closely matches the structure of the present-day Sun. Other main sequence stars with similar masses should also be robustly modeled with the same set of $K_0$. The possibility exists that significantly different structures, including convective or degenerate cores could modify these values of $K_0$. However, if the DM distribution is significantly compact such that the value of $K$ evaluated at the centre of the star is representative, we do not anticipate large differences. Conversely, this means that caution may be warranted for DM masses that approach the evaporation threshold in stars or planets. The fact that the Knudsen transitions follow similar patterns, leading only to O(1) differences between idealized and realistic simulations despite their extreme differences (1~K versus $10^{6}$~K temperature scales, kg versus GeV DM masses, 2.5 m versus $10^7$~m radii) does indicate to us that this method is quite robust, and any deviations from our results may not be too severe.

An exact investigation of this formalism in the context of a wider range of stellar structures and temperature profiles would require a significant additional investment in computational time, is left for future work. 

\begin{table}[tb]
    \centering
    \begin{tabular}{l|c | c} \hline
    Cross section     &  $K_0$ (\textit{idealized} simulations) & $K_0$ (\textit{realistic} simulations)\\ \hline
         $Const.$ & 0.31 & 0.40  \\
         $v^{-2}$ & 0.16 & 0.11 \\
       $v^2$ & 0.39 & 0.73 \\
       $v^4$ & 0.47 &1.20 \\
       $q^{-2}$ & 0.33 & 0.21 \\
       $q^2$ & 0.52 &  1.05 \\
       $q^4$ & 0.73 &1.72 \\ \hline
    \end{tabular}
    \caption{Values of $K_0$ in Eq.\ \eqref{eq:SPrescale} found in our simulations.}
    \label{tab:K0real}
\end{table}

Figs.\ \ref{fig:maxlumGR} and \ref{fig:maxlumReal} respectively show the predicted maximum luminosities for the \textit{idealized} and \textit{realistic} setups, using Eq.\ \eqref{eq:SPrescale} across a large range of $K$, for all the models that we have considered in this paper. Although this model does not capture the LTE limit all that well in \textit{idealized} simulations, it provides an excellent description of $L_{max}(K)$ over many decades in the \textit{realistic} simulations.

\section{Conclusions}
\label{sec:conclusions}

We have developed a first-principles Monte-Carlo program to investigate energy transport by gravitationally trapped dark matter in stars. We validated the technique for the case of constant DM-nucleus couplings, by comparing to both an earlier numerical study by Gould and Raffelt \cite{Gould1990}, and to analytical predictions for the limiting LTE and Knudsen regimes. The excellent agreement found in both cases provides strong support to our approach.

We used our simulation program to provide the first  numerical study of energy transport by DM particles with non-constant couplings across all regimes of energy transport. We performed simulations in an \textit{idealized} setup, following the simplified approach of Ref.\ \cite{Gould1990} with a constant density and linear temperature gradient, and a \textit{realistic} stellar profile, using a simple harmonic oscillator potential, but with realistic target nuclei densities and temperatures. We compared our results with the predictions of the two existing standard schemes for heat transport: the Spergel \& Press (SP) isothermal approximation, and the corrected Gould \& Raffelt (GR) LTE scheme. Based on our simulations, we provided in Sec.\ \ref{sec:discussion} a recommendation for energy transport modelling in future work.

We found that both GR and SP correctly predict the radius at which the luminosity carried by DM in stars peaks. In the Knudsen regime (large mean free path, $K \gg 1$), SP systematically overestimates heat transport by a factor of two. This was also observed by Gould \& Raffelt \cite{Gould1990CosmionLimit}; we have been unable to locate the source of this discrepancy in the derivation of the SP formalism, but our prescription builds in a simple factor of $1/2$ in our correction factor in order to account for it.

The GR diffusivity coefficients $\alpha(\mu)$ as computed in Ref.\ \cite{Gould1990} and extended to include general interactions \cite{Vincent13} accurately predict the radial DM distributions $n_\chi(r)$ for constant and non-constant cross sections. Unsurprisingly, GR does well in predicting the luminosity profile in the LTE regime, for constant cross sections. However, when extended to $v$ or $q$-dependent cross sections as done in Ref.\ \cite{Vincent13}, it proves unreliable. Conversely, SP is slightly better at predicting the \textit{shape} of the energy injection function, especially at radii where heat is deposited, as GR slightly overpredicts the range of heat transport. By performing the collision operator integrals in the isothermal regime for non-constant cross sections in Sec.\ \ref{sec:SP}, and thus generalizing the luminosity computed in Ref.\ \cite{Spergel1985EffectInterior}, we found that SP continues to robustly reproduce the shape of $L(r)$ for non-constant cross-sections, over the full range of $K$ that we have considered.

 A short comment on how these conclusions affect previous results by Vincent \textit{et al.} is in order. As we have shown, the $q$ and $v$-dependent LTE computations of Ref. \cite{Vincent13} implemented in \cite{Vincent14,Vincent2015,Vincent2016UpdatedMatter} do not correctly capture the magnitude of heat transport, underestimating it for positive $n$, and overestimating it for negative $n$. However, the location of the Knudsen peak remains unaffected---this can be readily seen via a direct comparison between Fig.\ \ref{fig:maxlumReal} and Fig. 2 of \cite{Vincent2016UpdatedMatter}. This suggests that, for positive $n$, a larger amount of parameter space may be available to address the solar composition problem via transport by asymmetric spin-dependent DM than we previously thought. Conversely, $v^{-2}$ and $q^{-2}$ models are likely less promising. A more quantitative statement will require detailed stellar evolution modeling.

We suspect that breakdown of GR is heavily influenced by the Knudsen correction $\mathfrak{f}(K)$ and the radial suppression $\mathfrak{h}(r)$, which were introduced to parameterise the deviations observed in the simulations of Refs.\ \cite{Gould1990,Gould1990CosmionLimit}. In particular, the GR approach relies on heat transport being purely dipolar, and the radial  suppression attempts to compensate for the breakdown of this approximation near the stellar centre. The approximation also breaks down when transport becomes non-local at $K>1$. Even if the majority of the Maxwell-Boltzmann distribution has a low Knudsen number, particles in the tail of the distribution can acquire a long mean free path if $\sigma$ is proportional to $v^{2n}$ or $q^{2n}$. Because the isothermal SP approach is by definition non-local, it is more robust to these considerations. 

We thus recommend a modified version of the SP heat transport scheme, given in Eq.\ \eqref{eq:SPrescale}.  This `calibrated SP' scheme includes the modifications for $v^{2n}$, $q^{2n}$ cross sections in Sec.\ \ref{sec:SP}, as well as a simple Knudsen correction that changes the scaling in the LTE regime and relies on a value of the Knudsen transition $K_0$ determined from our simulations.  These values are tabulated in Tab.\ \ref{tab:K0real}. The SP approach has the further advantage of being more stable when implemented in stellar evolution codes such as GARSTEC and MESA, due to the ease with which one can obtain analytical derivatives of $\epsilon(r)$; with GR, one must compute derivatives numerically. We leave a full comparison between the effects of these schemes in such evolution codes for future work \cite{SimsInProgress}.

\acknowledgments
Simulations were performed on cx1 (Helen) at Imperial College London. HB is partially supported by STFC Consolidated HEP grants ST/P000681/1 and ST/T000694/1 and thanks other members of the Cambridge Pheno Working Group for helpful discussions. ACV is supported by the Arthur B.~McDonald Canadian Astroparticle Physics Research Institute and NSERC, with equipment funded by the Canada Foundation for Innovation and the Province of Ontario, and housed at the Queen's Centre for Advanced Computing. Research at Perimeter Institute is supported by the Government of Canada through the Department of Innovation, Science, and Economic Development, and by the Province of Ontario. PS is supported by the Australian Research Council (ARC) under grant FT190100814. 

\appendix
\section{Orbital trajectories in the simple harmonic oscillator potential}
\label{sec:app}
Under the potential given by Eq.\ \ref{eq:pot}, the Cartesian components of the DM position vector independently undergo simple harmonic motion with an angular frequency $\Omega = \sqrt{(4/3) \pi G \rho_\mathrm{sho}}$. Given an initial position and velocity, the DM position and velocity vectors at any later time $t$ (i.e.\ prior to the next collision) can be expressed as
\begin{equation}
\label{eq:traj}
\vec{r}(t) =
\begin{pmatrix}
r_x(t) \\
r_y(t) \\
r_z(t)
\end{pmatrix}
=
\begin{pmatrix}
A_{x}\cos(\Omega t + a_x)  \\
A_{y}\cos(\Omega t + a_y)  \\
A_{z}\cos(\Omega t + a_z)
\end{pmatrix}
\end{equation}
and
\begin{equation}
\label{eq:velt}
\vec{v} =
\begin{pmatrix}
v_x(t)\\
v_y(t)\\
v_z(t)\\
\end{pmatrix}
=
\begin{pmatrix}
-A_{x}\Omega \sin(\Omega t + a_x)  \\
-A_{y}\Omega \sin(\Omega t + a_y) \\
-A_{z}\Omega \sin(\Omega t + a_z)
\end{pmatrix}
\end{equation}
where the constants $A_{i}$ and $a_i$ are set by the initial position and velocity according to
\begin{equation}
a_i = \arctan\Big(\frac{-v_i(0)}{\Omega x_i(0)}\Big)
\end{equation}
and
\begin{equation}
A_i = \frac{x_i(0)}{\cos(a_i)}.
\end{equation}

\bibliographystyle{JHEP}
\bibliography{bib}

\providecommand{\href}[2]{#2}\begingroup\raggedright\begin{thebibliography}{10}

\bibitem{Bertone2005ParticleConstraints}
G.~Bertone, D.~Hooper and J.~Silk, \emph{{Particle dark matter: evidence,
  candidates and constraints}},
  \href{https://doi.org/10.1016/j.physrep.2004.08.031}{\emph{Physics Reports}
  {\bfseries 405} (2005) 279--390},
  [\href{https://arxiv.org/abs/hep-ph/0404175}{{\ttfamily hep-ph/0404175}}].

\bibitem{BertoneBook}
G.~{Bertone}, ed., \emph{{Particle Dark Matter: Observations, Models and
  Searches}}.
\newblock Cambridge University Press, 2010.

\bibitem{Lisanti2016LecturesPhysics}
M.~{Lisanti}, \emph{{Lectures on Dark Matter Physics}},  in \emph{New Frontiers
  in Fields and Strings (TASI 2015)}, pp.~399--446, 2017,
  \href{https://arxiv.org/abs/1603.03797}{{\ttfamily 1603.03797}}.

\bibitem{Press1985CaptureParticles}
W.~H. Press and D.~N. Spergel, \emph{{Capture by the sun of a galactic
  population of weakly interacting, massive particles}},
  \href{https://doi.org/10.1086/163485}{\emph{\apj} {\bfseries 296} (1985)
  679--684}.

\bibitem{Gould1987}
A.~Gould, \emph{{Resonant enhancements in weakly interacting massive particle
  capture by the earth}}, \href{https://doi.org/10.1086/165653}{\emph{\apj}
  {\bfseries 321} (1987) 571--585}.

\bibitem{Gould1987a}
A.~Gould, \emph{{Weakly interacting massive particle distribution in and
  evaporation from the sun}}, \href{https://doi.org/10.1086/165652}{\emph{\apj}
  {\bfseries 321} (1987) 560--570}.

\bibitem{Steigman78}
G.~{Steigman}, H.~{Quintana}, C.~L. {Sarazin} and J.~{Faulkner},
  \emph{{Dynamical interactions and astrophysical effects of stable heavy
  neutrinos}}, {\emph{\aj} {\bfseries 83} (1978) 1050--1061}.

\bibitem{Spergel1985EffectInterior}
D.~N. Spergel and W.~H. Press, \emph{{Effect of hypothetical, weakly
  interacting, massive particles on energy transport in the solar interior}},
  \href{https://doi.org/10.1086/163336}{\emph{\apj} {\bfseries 294} (1985)
  663--673}.

\bibitem{Faulkner1985}
J.~Faulkner and R.~L. Gilliland, \emph{{Weakly interacting, massive particles
  and the solar neutrino flux}},
  \href{https://doi.org/10.1086/163766}{\emph{\apj} {\bfseries 299} (1985)
  994--1000}.

\bibitem{Gould1990}
A.~Gould and G.~Raffelt, \emph{{Thermal conduction by massive particles}},
  \href{https://doi.org/10.1086/168568}{\emph{\apj} {\bfseries 352} (1990)
  654--668}.

\bibitem{Lopes2001}
I.~P. Lopes, J.~Silk and S.~H. Hansen, \emph{{Helioseismology as a New
  Constraint on SUSY Dark Matter}},
  \href{https://doi.org/10.1046/j.1365-8711.2002.05238.x}{\emph{Monthly Notices
  of the Royal Astronomical Society} {\bfseries 331} (2001) 361--368},
  [\href{https://arxiv.org/abs/astro-ph/0111530}{{\ttfamily
  astro-ph/0111530}}].

\bibitem{Lopes2002SolarMatter}
I.~P. Lopes, G.~Bertone and J.~Silk, \emph{{Solar seismic model as a new
  constraint on supersymmetric dark matter}},
  \href{https://doi.org/10.1046/j.1365-8711.2002.05835.x}{\emph{Monthly Notices
  of the Royal Astronomical Society} {\bfseries 337} (2002) 1179--1184},
  [\href{https://arxiv.org/abs/astro-ph/0205066}{{\ttfamily
  astro-ph/0205066}}].

\bibitem{Bottino2002DoesParticles}
A.~Bottino, G.~Fiorentini, N.~Fornengo, B.~Ricci, S.~Scopel and F.~L. Villante,
  \emph{{Does solar physics provide constraints to weakly interacting massive
  particles?}}, \href{https://doi.org/10.1103/PhysRevD.66.053005}{\emph{\prd}
  {\bfseries 66} (2002) 053005},
  [\href{https://arxiv.org/abs/hep-ph/0206211v1}{{\ttfamily
  hep-ph/0206211v1}}].

\bibitem{Lopes2012}
I.~Lopes and J.~Silk, \emph{{Solar Constraints on Asymmetric Dark Matter}},
  \href{https://doi.org/10.1088/0004-637X/757/2/130}{\emph{\apj} {\bfseries
  757} (2012) 130}, [\href{https://arxiv.org/abs/1209.3631}{{\ttfamily
  1209.3631}}].

\bibitem{Lopes2014}
I.~{Lopes}, K.~{Kadota} and J.~{Silk}, \emph{{Constraint on Light Dipole Dark
  Matter from Helioseismology}},
  \href{https://doi.org/10.1088/2041-8205/780/2/L15}{\emph{\apjl} {\bfseries
  780} (2014) 15}, [\href{https://arxiv.org/abs/1310.0673}{{\ttfamily
  1310.0673}}].

\bibitem{Pantsy}
I.~{Lopes}, P.~{Panci} and J.~{Silk}, \emph{{Helioseismology with Long-range
  Dark Matter-Baryon Interactions}},
  \href{https://doi.org/10.1088/0004-637X/795/2/162}{\emph{\apj} {\bfseries
  795} (2014) 162}, [\href{https://arxiv.org/abs/1402.0682}{{\ttfamily
  1402.0682}}].

\bibitem{Geytenbeek2016EffectInterior}
B.~{Geytenbeek}, S.~{Rao}, P.~{Scott}, A.~{Serenelli}, A.~C. {Vincent},
  M.~{White} et~al., \emph{{Effect of electromagnetic dipole dark matter on
  energy transport in the solar interior}},
  \href{https://doi.org/10.1088/1475-7516/2017/03/029}{\emph{\jcap} {\bfseries
  3} (2017) 029}, [\href{https://arxiv.org/abs/1610.06737}{{\ttfamily
  1610.06737}}].

\bibitem{Frandsen2010}
M.~T. Frandsen and S.~Sarkar, \emph{{Asymmetric Dark Matter and the Sun}},
  \href{https://doi.org/10.1103/PhysRevLett.105.011301}{\emph{\prl} {\bfseries
  105} (2010) 011301}, [\href{https://arxiv.org/abs/1003.4505}{{\ttfamily
  1003.4505}}].

\bibitem{Cumberbatch2010LightHelioseismology}
D.~T. Cumberbatch, J.~A. Guzik, J.~Silk, L.~S. Watson and S.~M. West,
  \emph{{Light WIMPs in the Sun: Constraints from helioseismology}},
  \href{https://doi.org/10.1103/PhysRevD.82.103503}{\emph{\prd} {\bfseries 82}
  (2010) 103503}, [\href{https://arxiv.org/abs/1005.5102}{{\ttfamily
  1005.5102}}].

\bibitem{Taoso2010EffectSun}
M.~Taoso, F.~Iocco, G.~Meynet, G.~Bertone and P.~Eggenberger, \emph{{Effect of
  low mass dark matter particles on the Sun}},
  \href{https://doi.org/10.1103/PhysRevD.82.083509}{\emph{\prd} {\bfseries 82}
  (2010) 083509}, [\href{https://arxiv.org/abs/1005.5711}{{\ttfamily
  1005.5711}}].

\bibitem{Vincent14}
A.~C. {Vincent}, P.~{Scott} and A.~{Serenelli}, \emph{{Possible Indication of
  Momentum-Dependent Asymmetric Dark Matter in the Sun}},
  \href{https://doi.org/10.1103/PhysRevLett.114.081302}{\emph{\prl} {\bfseries
  114} (2015) 081302}, [\href{https://arxiv.org/abs/1411.6626}{{\ttfamily
  1411.6626}}].

\bibitem{Vincent2015}
A.~C. Vincent, A.~Serenelli and P.~Scott, \emph{{Generalised form factor dark
  matter in the Sun}},
  \href{https://doi.org/10.1088/1475-7516/2015/08/040}{\emph{\jcap} {\bfseries
  8} (2015) 040}, [\href{https://arxiv.org/abs/1504.04378}{{\ttfamily
  1504.04378}}].

\bibitem{Vincent2016UpdatedMatter}
A.~C. {Vincent}, P.~{Scott} and A.~{Serenelli}, \emph{{Updated constraints on
  velocity and momentum-dependent asymmetric dark matter}}, {\emph{\jcap}
  {\bfseries 11} (2016) 007},
  [\href{https://arxiv.org/abs/1605.06502}{{\ttfamily 1605.06502}}].

\bibitem{Gould1990CosmionLimit}
A.~Gould and G.~Raffelt, \emph{{Cosmion energy transfer in stars - The Knudsen
  limit}}, \href{https://doi.org/10.1086/168569}{\emph{\apj} {\bfseries 352}
  (1990) 669}.

\bibitem{Scott2009}
P.~Scott, M.~Fairbairn and J.~Edsj\"o, \emph{{Dark stars at the Galactic Centre
  - The main sequence}},
  \href{https://doi.org/10.1111/j.1365-2966.2008.14282.x}{\emph{Monthly Notices
  of the Royal Astronomical Society} {\bfseries 394} (2009) 82--104},
  [\href{https://arxiv.org/abs/0809.1871}{{\ttfamily 0809.1871}}].

\bibitem{Iocco:2012wk}
F.~Iocco, M.~Taoso, F.~Leclercq and G.~Meynet, \emph{{Main sequence stars with
  asymmetric dark matter}},
  \href{https://doi.org/10.1103/PhysRevLett.108.061301}{\emph{Phys. Rev. Lett.}
  {\bfseries 108} (2012) 061301}.

\bibitem{Lopes:2019jca}
J.~Lopes and I.~Lopes, \emph{{Asymmetric Dark Matter Imprint on Low-mass
  Main-sequence Stars in the Milky Way Nuclear Star Cluster}},
  \href{https://doi.org/10.3847/1538-4357/ab2392}{\emph{Astrophys. J.}
  {\bfseries 879} (2019) 50}.

\bibitem{Raen:2020qvn}
T.~J. Raen, H.~Mart\'\i{}nez-Rodr\'\i{}guez, T.~J. Hurst, A.~R. Zentner,
  C.~Badenes and R.~Tao, \emph{{The Effects of Asymmetric Dark Matter on
  Stellar Evolution I: Spin-Dependent Scattering}},
  \href{https://doi.org/10.1093/mnras/stab865}{\emph{Mon. Not. Roy. Astron.
  Soc.} {\bfseries 503} (2021) 5611--5623}.

\bibitem{Rato:2021tfc}
J.~a. Rato, J.~Lopes and I.~Lopes, \emph{{On asymmetric dark matter constraints
  from the asteroseismology of a subgiant star}}, .

\bibitem{Lopes:2021jcy}
J.~Lopes and I.~Lopes, \emph{{Dark matter capture and annihilation in stars:
  Impact on the red giant branch tip}},
  \href{https://doi.org/10.1051/0004-6361/202140750}{\emph{Astron. Astrophys.}
  {\bfseries 651} (2021) A101}.

\bibitem{Lopes:2018wgp}
I.~Lopes and J.~Silk, \emph{{Dark matter imprint on $^8$B neutrino spectrum}},
  \href{https://doi.org/10.1103/PhysRevD.99.023008}{\emph{Phys. Rev. D}
  {\bfseries 99} (2019) 023008}.

\bibitem{Asplund2009TheSun}
M.~Asplund, N.~Grevesse, A.~J. Sauval and P.~Scott, \emph{{The Chemical
  Composition of the Sun}},
  \href{https://doi.org/10.1146/annurev.astro.46.060407.145222}{\emph{Annual
  Reviews of Astronomy {\&} Astrophysics} {\bfseries 47} (2009) 481--522},
  [\href{https://arxiv.org/abs/0909.0948v1}{{\ttfamily 0909.0948v1}}].

\bibitem{Serenelli2009NewRevisited}
A.~M. Serenelli, S.~Basu, J.~W. Ferguson and M.~Asplund, \emph{{New Solar
  Composition: The problem with Solar Models Revisited}},
  \href{https://doi.org/10.1088/0004-637X/705/2/L123}{\emph{\apj} {\bfseries
  705} (2009) L123--L127}, [\href{https://arxiv.org/abs/0909.2668}{{\ttfamily
  0909.2668}}].

\bibitem{Bergemann2014}
M.~{Bergemann} and A.~{Serenelli}, \emph{Solar abundance problem},  in
  \emph{Determination of Atmospheric Parameters of B-, A-, F- and G-Type Stars}
  (E.~Niemczura, B.~Smalley and W.~Pych, eds.), GeoPlanet: Earth and Planetary
  Sciences, pp.~245--258.
\newblock Springer International Publishing (Cham), 2014.
\newblock \href{https://arxiv.org/abs/1403.3097}{{\ttfamily 1403.3097}}.

\bibitem{Vincent13}
A.~C. {Vincent} and P.~{Scott}, \emph{{Thermal conduction by dark matter with
  velocity and momentum-dependent cross-sections}},
  \href{https://doi.org/10.1088/1475-7516/2014/04/019}{\emph{\jcap} {\bfseries
  4} (2014) 19}, [\href{https://arxiv.org/abs/1311.2074}{{\ttfamily
  1311.2074}}].

\bibitem{PICO:2019vsc}
{\scshape PICO} collaboration, C.~Amole et~al., \emph{{Dark Matter Search
  Results from the Complete Exposure of the PICO-60 C$_3$F$_8$ Bubble
  Chamber}}, \href{https://doi.org/10.1103/PhysRevD.100.022001}{\emph{Phys.
  Rev. D} {\bfseries 100} (2019) 022001}.

\bibitem{Kosovichev2011}
A.~G. Kosovichev, \emph{{Advances in Global and Local Helioseismology: an
  Introductory Review}},  in \emph{Lecture Notes in Physics}, vol.~832,
  pp.~3--84.
\newblock Springer International Publishing (Cham), 2011.
\newblock \href{https://arxiv.org/abs/1103.1707}{{\ttfamily 1103.1707}}.

\bibitem{CtoO}
C.~{Allende Prieto}, D.~L. {Lambert} and M.~{Asplund}, \emph{{A Reappraisal of
  the Solar Photospheric C/O Ratio}}, {\emph{\apjl} {\bfseries 573} (2002)
  L137--L140}, [\href{https://arxiv.org/abs/astro-ph/0206089}{{\ttfamily
  astro-ph/0206089}}].

\bibitem{AspIV}
M.~{Asplund}, N.~{Grevesse}, A.~J. {Sauval}, C.~{Allende Prieto} and
  D.~{Kiselman}, \emph{{Line formation in solar granulation. IV. [O I], O I and
  OH lines and the photospheric O abundance}}, {\emph{Astronomy {\&}
  Astrophysics} {\bfseries 417} (2004) 751--768},
  [\href{https://arxiv.org/abs/astro-ph/0312290}{{\ttfamily
  astro-ph/0312290}}].

\bibitem{AspVI}
M.~{Asplund}, N.~{Grevesse}, A.~J. {Sauval}, C.~{Allende Prieto} and
  R.~{Blomme}, \emph{{Line formation in solar granulation. VI. [C I], C I, CH
  and C$_{2}$ lines and the photospheric C abundance}}, {\emph{Astronomy {\&}
  Astrophysics} {\bfseries 431} (2005) 693--705},
  [\href{https://arxiv.org/abs/astro-ph/0410681}{{\ttfamily
  astro-ph/0410681}}].

\bibitem{AspARAA}
M.~{Asplund}, \emph{{New Light on Stellar Abundance Analyses: Departures from
  LTE and Homogeneity}}, {\emph{Annual Reviews of Astronomy {\&} Astrophysics}
  {\bfseries 43} (2005) 481--530}.

\bibitem{AGS05}
M.~{Asplund}, N.~{Grevesse} and A.~J. {Sauval}, \emph{{The Solar Chemical
  Composition}},  in \emph{Cosmic Abundances as Records of Stellar Evolution
  and Nucleosynthesis} (T.~G. {Barnes III} and F.~N. {Bash}, eds.), vol.~336,
  p.~25, Astron. Soc. Pac., San Francisco, 2005.

\bibitem{ScottVII}
P.~{Scott}, M.~{Asplund}, N.~{Grevesse} and A.~J. {Sauval}, \emph{{Line
  formation in solar granulation. VII. CO lines and the solar C and O isotopic
  abundances}},
  \href{https://doi.org/10.1051/0004-6361:20064986}{\emph{Astronomy {\&}
  Astrophysics} {\bfseries 456} (2006) 675--688},
  [\href{https://arxiv.org/abs/astro-ph/0605116}{{\ttfamily
  astro-ph/0605116}}].

\bibitem{Scott09Ni}
P.~{Scott}, M.~{Asplund}, N.~{Grevesse} and A.~J. {Sauval}, \emph{{On the Solar
  Nickel and Oxygen Abundances}},
  \href{https://doi.org/10.1088/0004-637X/691/2/L119}{\emph{\apjl} {\bfseries
  691} (2009) L119--L122}, [\href{https://arxiv.org/abs/0811.0815}{{\ttfamily
  0811.0815}}].

\bibitem{AGSS_NaCa}
P.~{Scott}, N.~{Grevesse}, M.~{Asplund}, A.~J. {Sauval}, K.~{Lind}, Y.~{Takeda}
  et~al., \emph{{The elemental composition of the Sun. I. The intermediate mass
  elements Na to Ca}},
  \href{https://doi.org/10.1051/0004-6361/201424109}{\emph{Astronomy {\&}
  Astrophysics} {\bfseries 573} (2015) A25},
  [\href{https://arxiv.org/abs/1405.0279}{{\ttfamily 1405.0279}}].

\bibitem{AGSS_FePeak}
P.~{Scott}, M.~{Asplund}, N.~{Grevesse}, M.~{Bergemann} and A.~J. {Sauval},
  \emph{{The elemental composition of the Sun. II. The iron group elements Sc
  to Ni}}, \href{https://doi.org/10.1051/0004-6361/201424110}{\emph{Astronomy
  {\&} Astrophysics} {\bfseries 573} (2015) A26},
  [\href{https://arxiv.org/abs/1405.0287}{{\ttfamily 1405.0287}}].

\bibitem{AGSS_heavy}
N.~{Grevesse}, P.~{Scott}, M.~{Asplund} and A.~J. {Sauval}, \emph{{The
  elemental composition of the Sun. III. The heavy elements Cu to Th}},
  \href{https://doi.org/10.1051/0004-6361/201424111}{\emph{Astronomy {\&}
  Astrophysics} {\bfseries 573} (2015) A27},
  [\href{https://arxiv.org/abs/1405.0288}{{\ttfamily 1405.0288}}].

\bibitem{Catena2015FormTheories}
R.~{Catena} and B.~{Schwabe}, \emph{{Form factors for dark matter capture by
  the Sun in effective theories}},
  \href{https://doi.org/10.1088/1475-7516/2015/04/042}{\emph{\jcap} {\bfseries
  4} (2015) 042}, [\href{https://arxiv.org/abs/1501.03729}{{\ttfamily
  1501.03729}}].

\bibitem{Gould1992}
A.~{Gould}, \emph{{Big Bang Archeology: WIMP Capture by the Earth at Finite
  Optical Depth}}, \href{https://doi.org/10.1086/171057}{\emph{\apj} {\bfseries
  387} (Mar., 1992) 21}.

\bibitem{Busoni2017EvaporationSun}
G.~{Busoni}, A.~{De Simone}, P.~{Scott} and A.~C. {Vincent}, \emph{{Evaporation
  and scattering of momentum- and velocity-dependent dark matter in the Sun}},
  \href{https://doi.org/10.1088/1475-7516/2017/10/037}{\emph{\jcap} {\bfseries
  10} (2017) 037}, [\href{https://arxiv.org/abs/1703.07784}{{\ttfamily
  1703.07784}}].

\bibitem{Bramante:2017xlb}
J.~Bramante, A.~Delgado and A.~Martin, \emph{{Multiscatter stellar capture of
  dark matter}}, \href{https://doi.org/10.1103/PhysRevD.96.063002}{\emph{Phys.
  Rev. D} {\bfseries 96} (2017) 063002}.

\bibitem{Ilie:2020vec}
C.~Ilie, J.~Pilawa and S.~Zhang, \emph{{Comment on
  \textquotedblleft{}Multiscatter stellar capture of dark
  matter\textquotedblright{}}},
  \href{https://doi.org/10.1103/PhysRevD.102.048301}{\emph{Phys. Rev. D}
  {\bfseries 102} (2020) 048301}.

\bibitem{Clementz2015AsymmetricSun}
S.~Clementz and M.~Blennow, \emph{{Asymmetric capture of Dirac dark matter by
  the Sun}}, {\emph{\jcap} {\bfseries 2015} (2015) 36}.

\bibitem{Garani:2021feo}
R.~Garani and S.~Palomares-Ruiz, \emph{{Evaporation of dark matter from
  celestial bodies}}, .

\bibitem{Nauenberg1987}
M.~Nauenberg, \emph{{Energy transport and Evaporation of weakly interacting
  particles in the sun}}, {\emph{\prd} {\bfseries 36} (1987) 1080--1087}.

\bibitem{gilliland1986solar}
R.~L. Gilliland, J.~Faulkner, W.~H. Press and D.~N. Spergel, \emph{Solar models
  with energy transport by weakly interacting particles}, {\emph{The
  Astrophysical Journal} {\bfseries 306} (1986) 703--709}.

\bibitem{Akrami11DD}
Y.~{Akrami}, C.~{Savage}, P.~{Scott}, J.~{Conrad} and J.~{Edsj{\"o}},
  \emph{{How well will ton-scale dark matter direct detection experiments
  constrain minimal supersymmetry?}},
  \href{https://doi.org/10.1088/1475-7516/2011/04/012}{\emph{\jcap} {\bfseries
  4} (Apr., 2011) 12}.

\bibitem{Fitzpatrick2013TheDetection}
A.~L. Fitzpatrick, W.~Haxton, E.~Katz, N.~Lubbers and Y.~Xu, \emph{{The
  effective field theory of dark matter direct detection}},
  \href{https://doi.org/10.1088/1475-7516/2013/02/004}{\emph{\jcap} {\bfseries
  2013} (2013) 004}, [\href{https://arxiv.org/abs/1203.3542}{{\ttfamily
  1203.3542}}].

\bibitem{Kozar:2021iur}
N.~A. Kozar, A.~Caddell, L.~Fraser-Leach, P.~Scott and A.~C. Vincent,
  \emph{{Capt'n General: A generalized stellar dark matter capture and heat
  transport code}},  in \emph{{Tools for High Energy Physics and Cosmology}},
  5, 2021.

\bibitem{Krstic}
P.~S. Krsti\ifmmode~\acute{c}\else \'{c}\fi{} and D.~R. Schultz,
  \emph{Consistent definitions for, and relationships among, cross sections for
  elastic scattering of hydrogen ions, atoms, and molecules},
  \href{https://doi.org/10.1103/PhysRevA.60.2118}{\emph{\pra} {\bfseries 60}
  (Sep, 1999) 2118--2130}.

\bibitem{Tulin:2013teo}
S.~{Tulin}, H.-B. {Yu} and K.~M. {Zurek}, \emph{{Beyond collisionless dark
  matter: Particle physics dynamics for dark matter halo structure}},
  \href{https://doi.org/10.1103/PhysRevD.87.115007}{\emph{\prd} {\bfseries 87}
  (June, 2013) 115007}.

\bibitem{SimsInProgress}
H.~Banks, L.~Fraser-Leach, P.~Scott and A.~Vincent (2022), in prep.

\end{thebibliography}\endgroup

\end{document}